\pdfoutput=1
\documentclass[twocolumn,prd,superscriptaddress,preprintnumbers,nofootinbib]{revtex4}
\usepackage{graphicx}
\usepackage{epsfig}
\usepackage{lipsum}
\usepackage{enumitem}
\usepackage{bm}
\usepackage{latexsym,amssymb,amsmath,amsfonts,amssymb,txfonts,pxfonts,wasysym,float}
\usepackage{color}
\usepackage{xspace} 

\newcommand{\beq}[1]{\begin{equation}\label{#1}}
\newcommand{\eeq}{\end{equation}}
\newcommand{\bea}[1]{\begin{eqnarray} \label{#1}}
\newcommand{\eea}{\end{eqnarray}}
\newcommand{\ba}{\begin{array}}
\newcommand{\ea}{\end{array}}

\def\be{\begin{equation}}
\def\ee{\end{equation}}
\def\gs{\mathrel{
   \rlap{\raise 0.511ex \hbox{$>$}}{\lower 0.511ex \hbox{$\sim$}}}}
\def\ls{\mathrel{
   \rlap{\raise 0.511ex \hbox{$<$}}{\lower 0.511ex \hbox{$\sim$}}}}

\newcommand{\CAMB}{\texttt{CAMB}\xspace}
\newcommand{\CosmoMC}{\texttt{CosmoMC}\xspace}

\newcommand{\comment}[1]{}

\usepackage[usenames,dvipsnames]{xcolor}
\definecolor{orange}{cmyk}{0,0.5,1,0}
\definecolor{rossoCP3}{cmyk}{0,.88,.77,.40}
\definecolor{graa}{rgb}{0.8,0.8,0.8}
\definecolor{blaa}{rgb}{0.2,0.2,0.6}

\begin{document}

\title{\color{rossoCP3}{Dissecting the $\bm{H_0}$ and $\bm{S_8}$ tensions with Planck + BAO + supernova type Ia\\ in multi-parameter cosmologies}}

\author{Luis A. Anchordoqui}
\email{luis.anchordoqui@gmail.com}
\affiliation{Department of Physics and Astronomy,  Lehman College, City University of New York, NY 10468, USA}

\affiliation{Department of Physics,
 Graduate Center, City University of New York,  NY 10016, USA
}

\affiliation{Department of Astrophysics,
 American Museum of Natural History, NY 10024, USA}

\author{Eleonora Di Valentino}
\email{eleonora.di-valentino@durham.ac.uk}
\affiliation{Institute for Particle Physics Phenomenology, Department of Physics, Durham University, Durham DH1 3LE, UK}

\author{Supriya Pan}
\email{supriya.maths@presiuniv.ac.in}
\affiliation{Department of Mathematics, Presidency University, 86/1 College Street, Kolkata 700073, India}

\author{Weiqiang Yang}
\email{d11102004@163.com}
\affiliation{Department of Physics, Liaoning Normal University, Dalian, 116029, P. R. China}

\begin{abstract}
\vskip 2mm \noindent The mismatch between the locally measured expansion rate of the universe and the one inferred from observations of the cosmic microwave background (CMB) assuming the canonical $\Lambda$CDM model has become the new cornerstone of modern cosmology, and many new-physics set ups are rising to the challenge. Concomitant with the so-called $H_0$ problem, there is evidence of a growing tension between the CMB-preferred value and the local determination of the
weighted amplitude of matter fluctuations $S_8$. It would be appealing and compelling if both the $H_0$ and $S_8$ tensions were resolved at once, but as yet none of the proposed new-physics models have done so to a satisfactory degree. Herein, we adopt a systematic approach to investigate the possible interconnection among the free parameters in several classes of models that typify the main theoretical frameworks tackling the tensions on the universe expansion rate and the clustering of matter. Our calculations are carried out using the publicly available Boltzmann solver \CAMB in combination with the sampler \CosmoMC. We show that even after combining the leading classes of models sampling modifications of both the early and late-time universe a simultaneous solution to the $H_0$ and $S_8$ tensions remains elusive.
\end{abstract}
\maketitle

\section{Introduction}

The standard $\Lambda$-cold dark matter ($\Lambda$CDM) cosmological
model provides an exceptional description of a wide range of astrophysical
and astronomical observations~\cite{Zyla:2020zbs}. The parameters
governing the $\Lambda$CDM cosmology have been constrained with unprecedented accuracy using measurements of galaxy clusters~\cite{Alam:2016hwk,Beutler:2011hx,Ross:2014qpa}, weak
lensing~\cite{DES:2021wwk,Asgari:2020wuj}, supernovae type Ia~\cite{Scolnic:2017caz}, and anisotropies in the cosmic microwave
background (CMB) temperature and polarization fields~\cite{Aghanim:2018eyx,Aghanim:2019ame}. However, the enhanced
precision of the various steps in the local distance-ladder
measurements of the Hubble constant, $H_0 \equiv 100~h~{\rm km/s/Mpc}$, have recently opened a crack in the $\Lambda$CDM model. Indeed,
a combination of the latest Supernovae $H0$ for the Equation of State (SH0ES)
measurements with constraints of medium-to-high redshift
probes have revealed a $4.2\sigma$ discrepancy between the
$\Lambda$CDM inferred $H_0 = 67.27 \pm 0.60~{\rm
km/s/Mpc}$ using data from the Planck satellite~\cite{Aghanim:2018eyx,Aghanim:2019ame} and the local measured value $H_0 =73.2 \pm 1.3~{\rm km/s/Mpc}$~\cite{Riess:2019cxk,Riess:2020fzl}.  Moreover, these measurements are supported by other early and late time observables, as shown in~\cite{Verde:2019ivm,DiValentino:2020zio,Riess:2019qba,DiValentino:2020vnx,Freedman:2021ahq} and references therein.

Adding fuel to fire,
the $\Lambda$CDM inferred value of 
the amplitude of mass fluctuations $\sigma_8$
has been consistently higher than the one
measured in gravitational
lensing~\cite{DES:2021wwk,Asgari:2020wuj}. This leads to a tension that is
quantified in terms of the $S_8 \equiv \sigma_8 \sqrt{\Omega_m/0.3}$
parameter, where $\Omega_m$ is the present
day value of the nonrelativistic matter
density; see e.g.~\cite{DiValentino:2020vvd}.  Strictly speaking, on the
assumption of $\Lambda$CDM the Planck Collaboration estimated $S_8 =
0.834 \pm 0.016$, which is in more than $3\sigma$ tension with the result
reported by KiDS-1000: $S_8 =
0.766^{+0.020}_{-0.014}$~\cite{Asgari:2020wuj}. The tension becomes
$3.4\sigma$ if we consider a combination of BOSS and KV450: $S_8 =
0.728 \pm 0.026$~\cite{Troster:2019ean}. However, some data sets point
to higher values of $S_8$, e.g. KiDS-450+GAMA for which $S_8 =
0.800^{+0.029}_{-0.027}$~\cite{vanUitert:2017ieu} or HSC SSP finding $S_8
=0.804^{+0.032}_{-0.029} $~\cite{Hamana:2019etx}.

Models addressing the $H_0$ tension either reduce the size of the sound horizon at recombination modifying the expansion rate in the
early-universe, or else shift the matter-dark energy equality to
earlier times than it otherwise would in $\Lambda$CDM with new physics
in the post-recombination universe. Then, to keep the locations of the
peaks in the CMB angular power spectrum fixed, $H_0$ increases
diminishing the tension. Models addressing the $S_8$ tension are either based on late-universe
physics processes that yield a suppression of the linear matter power
spectrum, or else decrease the CMB-predicted value of
$\Omega_m$. It would be appealing and compelling if both the $H_0$
and $\sigma_8$ tensions were resolved simultaneously, but as yet none
of the extant new physics models on this front have done so to a
satisfactory degree~\cite{DiValentino:2021izs}.

In the recent $H_0$
olympics of~\cite{Schoneberg:2021qvd}, spanning both early- and
late-time modifications of the universe expansion rate,  the ``gold medal'' for the best
scenario has been given to a varying effective electron mass in a
curved universe~\cite{Sekiguchi:2020teg}. However, the so-called
``interacting dark energy - dark matter (IDE) models'' did not
participate in this competition.  It has long been suspected that IDE is a
compelling framework settling several cosmological
issues~(see e.g., \cite{Wetterich:1994bg,Amendola:1999er,Comelli:2003cv,Franca:2003zg,Anchordoqui:2007sb,Chimento:2012aea,Bolotin:2013jpa,Wang:2016lxa}) and
recently it gained significant attention for ameliorating the $H_0$
and $S_8$ tensions (see
e.g.~\cite{Kumar:2016zpg,Kumar:2017dnp,DiValentino:2017iww,Yang:2018euj,Yang:2018uae,Kumar:2019wfs,Pan:2019jqh,Pan:2019gop,DiValentino:2019ffd,DiValentino:2019jae,Yang:2019uog,Gao:2021xnk,Pan:2020bur,Lucca:2020zjb,Wang:2021kxc,Kumar:2021eev,Lucca:2021dxo}). In
this paper we adopt a systematic approach to study the possible
interconnection among the free parameters in 24 combinations that
typify the various theoretical frameworks tackling the $H_0$ and $S_8$
tensions, including both early- and late-time modifications of the
universe expansion rate, as well as IDE and curved-space
models. Hence, the results presented herein  are complementary to those reported in~\cite{Schoneberg:2021qvd}.

The layout of the paper is as follows. We begin in Sec.~\ref{sec:2} by
introducing the classes of cosmological models to be explored and
discussing the relevant phenomenology. In Sec.~\ref{sec:3} we first
describe the observational data sets used in our study together with
the method of data analysis and the priors imposed on the cosmological
parameters. After that, for each class of models, we use the Boltzmann
solver \CAMB~\cite{Lewis:1999bs} in combination with \CosmoMC~\cite{Lewis:2002ah,Lewis:2013hha} to establish which regions
of parameter space are empirically viable to resolve the $H_0$ and
$S_8$ tensions. Armed with our findings, in Sec.~\ref{sec:4} we investigate the
crosscorrelation between parameters. The paper wraps up with some
conclusions in Sec.~\ref{sec:5}.

\section{$\bm{\Lambda}$CDM and beyond}
\label{sec:2}

Experiments show that the distribution of matter and radiation in the
observable universe is almost homogeneous and isotropic. Thus, the
evolution of the universe is well-described by the maximally-symmetric Friedmann-Lama\^{i}tre-Robertson-Walker (FLRW) line
element
\begin{equation}
ds^2 = dt^2 - a^2(t) \left[\frac{dr^2 }{1 -k r^2}+ r^2 \ (d \theta^2 +
  \sin^2 \phi \ d\phi^2)\right] \,,   
\label{RW}
\end{equation}
where $(t,r,\theta,\phi$) are comoving coordinates, $a(t)$ is the
cosmic scale factor, and $k \ (= -1,0,1)$  parametrizes the curvature of the homogeneous and
isotropic spatial sections~\cite{Kolb:1990vq}. It is often convenient
to express the line element as 
\begin{equation}
ds^2 =  a^2(\eta) \left\{ d\eta^2 - \left[\frac{dr^2 }{1 -k r^2}+ r^2  (d \theta^2 +
  \sin^2 \phi  d\phi^2)\right] \right\},   
\label{RW2}
\end{equation}
where $\eta = \int_0^t dt'/a(t')$ is the ``conformal time.'' Small
deviations from homogeneity and isotropy are generally modeled as
perturbations over the background metric given in (\ref{RW2}).

The cosmic expansion is driven by the first Friedmann equation for the Hubble parameter $H$,
\begin{equation}
  H^2(a) = \frac{8 \pi G}{3} \left\{\sum_i \rho_i(a) \right\} - \frac{k}{a^2} \,,
\label{Friedmann}  
\end{equation}
where $G$ is the gravitational constant and the sum runs over  the
energy densities $\rho_i$ of the various components of the cosmic fluid:  dark energy
(DE), CDM ($c$), baryons ($b$), photons ($\gamma$), and three flavors of one helicity state neutrinos (left-handed $\nu_L$ along with their right-handed $\overline
\nu_R$, that we denote $\nu$ for short). Equation~(\ref{Friedmann}) can be rewritten as
\begin{eqnarray}
  H^2(z) & = & H_0^2 \Bigg[(\Omega_c + \Omega_b) (1 +z)^3 + \Omega_\gamma
    (1+z)^4 + \Omega_k (1+z)^2 \nonumber \\
    & + &  \Omega_{\rm DE} \exp \left( 3 \; \int_{0}^{z} \frac{1+w}{1+z^\prime} dz^\prime \right)  +
     \frac{\rho_\nu (z)}{\rho_{\rm crit,0}}\; \Bigg]\; ,
\label{Hdez}
\end{eqnarray}
where $z = a_0/a -1$ is the redshift, $\rho_{\rm crit,0} = 3 H_0^2/(8\pi G)$ is the present day value of
the critical density,  and  $\Omega_i = \rho_{i,0}/\rho_{\rm crit,0}$ denotes the present-day
density parameters. Throughout the article we use the subscript $0$ to indicate the 
quantities evaluated today. Since we always refer to the present day
density  parameters, we omit the subscript $0$ in this case. The
energy densities of non-relativistic matter and radiation scale as
$a^{-3}$ and $a^{-4}$ respectively, and set the scalings with $(1 + z)$. The scaling of $\Omega_{\rm DE}$ is usually described by an ``equation-of-state''
parameter $w \equiv p_{\rm DE}/\rho_{\rm DE}$, the ratio of the spatially-homogeneous
dark energy pressure to its energy density $\rho_{\rm DE}$. To
accommodate the observed cosmic acceleration we should have $w < - 1/3$. The most
economic explanation for dark energy is the cosmological constant
$\Lambda$, for which $w=-1$. An alternative possibility is to consider
a cosmic scalar field  slowly rolling to the minimum of its
potential~\cite{Wetterich:1987fm,Peebles:1987ek,Ratra:1987rm}, the so-called  ``quintessence filed''~\cite{Caldwell:1997ii}. For this class of models, $-1 < w
< -1/3$ and the dark-energy density decreases as $\rho_q \propto
a^{-3 (1+w)}$. Yet another possibility is to consider  ``phantom dark
energy'' for which $w <-1$~\cite{Caldwell:1999ew}. A point worth noting at this juncture is that phantom DE models
violate the dominant energy condition~\cite{Carroll:2003st,Caldwell:2003vq,Carroll:2004hc,Sawicki:2012pz} a
cherished notion adopted to prohibit 
wormholes and time machines~\cite{Morris:1988tu}. However, it is hard
to envision how wormholes and time machines could originate from phantom energy. In (\ref{Hdez}) we have left open the
possibility for an arbitrary (albeit constant) value of $w$. For $\rho_\nu$, we could not write a simple scaling
with $z$ because the equation-of-state parameter is not constant. The
curvature density parameter is defined as $\Omega_k = -k/H_0^2$. 

For the benchmark spatially-flat $\Lambda$CDM 6-parameter model, we
have:  
$\Omega_k = 0$, $\Omega_{\rm DE} = \Omega_{\Lambda}$, and $w
=-1$. The Hubble constant is inferred from one
of these free parameters: the angular size of the sound horizon at
recombination $\theta_*$, which is given by the ratio of the comoving sound horizon to the
comoving angular diameter distance to last-scattering surface
\begin{equation}
  \theta_* = \frac{r_s(z_{\rm LS})}{D_M(z_{\rm LS})} \, .
\label{theta*}
\end{equation}
The comoving linear size  of the sound horizon and the comoving angular
diameter distance are linked to the expansion history of the universe
via
\begin{equation}
  r_s(z) = \int_z^\infty \frac{c_s(z')}{H(z')}  \ dz'
\end{equation}
and
\begin{equation}
D_M(z) =  \int_0^z  \frac{1}{H(z')} \ dz' 
\end{equation}
respectively, with $c_s$ the speed of
sound~\cite{Aghanim:2018eyx}. The set of free parameters that describe
the $\Lambda$CDM model is:
\begin{eqnarray}
\mathcal{P}_0 \equiv\Bigl\{\Omega_{b}h^2, \Omega_{c}h^2, 100\theta_{\rm MC},
  \tau, n_{s}, \ln[10^{10}A_{s}] \Bigr\}~,
\label{eq:LCDM}
\end{eqnarray}
where $\tau$ is the reionization optical depth, $n_s$ is the scalar
spectral index, $A_{s}$ is the amplitude  of the
scalar primordial power spectrum, and the $\theta_{\rm MC}$ parameter is
an approximation of $\theta_*$ (which is adopted in \CosmoMC~\cite{Lewis:2002ah,Lewis:2013hha} and is based on
fitting formulae given in~\cite{Hu:1995en}). 

A class of spatially-flat extensions to the base $\Lambda$CDM model
that can reduce the $H_0$ tension is based on the addition of  relativistic degrees-of-freedom in the early universe.  The
presence of any additional light species (such as sterile
neutrinos~\cite{Anchordoqui:2011nh,Anchordoqui:2012qu,Jacques:2013xr,DiValentino:2015sam},
axions~\cite{Giusarma:2014zza,DiValentino:2015zta,DiValentino:2015wba,Baumann:2016wac,Poulin:2018dzj,DEramo:2018vss,Giare:2020vzo}, massless
Goldstone bosons~\cite{Weinberg:2013kea}, or any other massless fields
from the dark sector~\cite{Anchordoqui:2020djl}) can be characterized by the number of ``equivalent'' light neutrino species 
\begin{equation}
N_{\rm eff} \equiv \frac{\rho_{\rm R} -
\rho_\gamma}{\rho_{\nu}} 
\label{neff}
\end{equation} 
in units of the density of a single Weyl neutrino $\rho_\nu$, where
$\rho_{\rm R}$ is the total energy density in relativistic particles
and $\rho_\gamma$ is the energy density of
photons~\cite{Steigman:1977kc}. For three families of massless
(Standard Model) neutrinos,  $N_{\rm eff} = 3.046$~\cite{Mangano:2005cc,deSalas:2016ztq,Akita:2020szl,Froustey:2020mcq,Bennett:2020zkv}. The set of free parameters describing this class of
models is given by  
\begin{eqnarray}
\mathcal{P}_1 \equiv\Bigl\{\Omega_{b}h^2, \Omega_{c}h^2, 100\theta_{\rm MC}, \tau, n_{s}, \ln[10^{10}A_{s}], \notag \\ 
N_{\rm eff} \Bigr\}~ .
\label{eq:parameter_space1}
\end{eqnarray}
Note that by adding dark relativistic degrees-of-freedom into the
early universe we are increasing the expansion rate $H(z)$, which
in turn reduces $r_s(z_{\rm LS})$. The accurate measurement of the
location of the acoustic peaks by the Planck mission sets the value of
the $\Lambda$CDM free parameter $\theta_*$, and so to maintain the
ratio in (\ref{theta*}) we must increase $H_0$ to decrease
$D_M(z_{\rm LS})$. However, adding relativistic degrees-of-freedom
into the early universe also affects the damping scale $\theta_D$ of
the CMB power spectrum, with
$\theta_D/\theta_* \propto \sqrt{H(z_{\rm
    LS})}$~\cite{Hou:2011ec}. Therefore, while we increase $N_{\rm
  eff}$ for a faster expansion rate $H(z_{\rm LS})$ at $\theta_*$
fixed we also increase  $\theta_D$, with the damping contributing
at larger scales to reduce the power in the damping tail. The main limiting factor in
constraining $N_{\rm eff}$ from CMB data
is a degeneracy with the primordial helium fraction $Y_P \equiv n_{\rm
  He}/n_b$. Namely, for fixed  $\Omega_bh^2$,
by increasing $Y_P$  at the end of big bang
nucleosynthesis (BBN) we decrease the number density of free electrons
and increase the diffusion length.  Altogether
this reduces the power
in the damping tail~\cite{Baumann:2018muz}. Using ${\cal P}_1$ to
accommodate  CMB + BAO data
and BBN observations~\cite{Aver:2015iza,Cooke:2017cwo} the Planck Collaboration reported $N_{\rm eff} =
3.12^{+0.25}_{-0.26}$ at the 95\%~CL~\cite{Aghanim:2018eyx}.  Herein
we will take this bound as an external constraint to our
numerical analysis for spatially-flat models.

Other attempts to resolve the $H_0$ tension involve tweaking
$\Lambda$CDM somehow to slow down the late-time expansion rate without
making radical changes to the early-time expansion rate. Within this
class of models the value of $r_s(z_{\rm LS})$ does not differ
appreciably from that obtained assuming $\Lambda$CDM for the same
choice of cosmological parameters. Now, a consistently lower value of $H(z)$ at low redshifts leads to a larger
value of $D_M(z_{\rm LS})$, which in turn would result in a smaller value
of $\theta_*$. Thus, we must (re)-decrease $D_M(z_{\rm LS})$  to keep
$\theta_*$ unchanged, and this can be accomplished by increasing
$H_0$.  A straightforward extension of the standard cosmology within this class 
of models emerges when considering  $w<-1$. It is easily seen in (\ref{Hdez}) that by
considering $w<-1$ we 
lower the expansion rate for $z >0$ with respect to the case
where the DE is in the form of a cosmological constant. From now
on, we denote with $w_q>-1$ and $w_p<-1$ the equation-of-state of
quintessence and phantom models, which are analyzed separately. The sets of free parameters describing these classes of
models are given by  
\begin{eqnarray}
\mathcal{P}_2 \equiv\Bigl\{\Omega_{b}h^2, \Omega_{c}h^2, 100\theta_{\rm MC}, \tau, n_{s}, \ln[10^{10}A_{s}], \notag \\ 
w_q,  \Bigr\} \, .
\label{eq:parameter_space2}
\end{eqnarray}
and 
\begin{eqnarray}
\mathcal{P}_3 \equiv\Bigl\{\Omega_{b}h^2, \Omega_{c}h^2, 100\theta_{\rm MC}, \tau, n_{s}, \ln[10^{10}A_{s}], \notag \\ 
w_p,  \Bigr\} \, .
\label{eq:parameter_space3}
\end{eqnarray}
It is worth recalling that models with $w_p<-1$ violate the null energy
condition and are typically unstable; though with optimistic
assumptions the instability time scale can be greater than the age of
the universe. Quintessence models for which $r_s(z_{\rm LS})$ is
unmodified from $\Lambda$CDM exacerbate the $H_0$ tension. This is
because for $w \geq -1$, the DE can only redshift faster than the
cosmological constant, yielding a smaller contribution to the energy
density today than it would have
been in $\Lambda$CDM with the same value of $H(z_{\rm LS})$, and so it
is easily seen from (\ref{Hdez}) that $H_0$ must also be
smaller~\cite{Raveri:2018ddi,Colgain:2019joh}. A different quintessence scenario
appears if the scalar field speeds up $H(z)$ and reduces the sound
horizon during the era leading up to
recombination~\cite{Pettorino:2013ia,Poulin:2018cxd,Agrawal:2019lmo,Gogoi:2020qif,Gomez-Valent:2021cbe}. As in the class of
models characterized by $N_{\rm eff}$, these early dark energy models
lead to larger $H_0$ values as compared to $\Lambda$CDM. However,
early dark energy models increase the tension  with large scale
structure data. It is easy to detect the source that increases  $S_8$
in this class of models, because the early dark energy slightly suppresses the growth of perturbations during the period in
which it contributes non-negligibly to the cosmic energy
density. Therefore,  to properly match the CMB data we must
increase the CDM component to compensate for the suppression in the efficiency of perturbation growth~\cite{Hill:2020osr}.

A fourth class of models incorporates a coupling between the DE and the
dark matter (DM) sectors, altering $\Lambda$CDM late-universe-based
predictions~\cite{Cai:2004dk,Barrow:2006hia,Valiviita:2008iv,Gavela:2009cy,Gavela:2010tm,Salvatelli:2013wra,Li:2013bya,Yang:2014gza,Yang:2014okp,Salvatelli:2014zta,Yang:2014hea,Valiviita:2015dfa,Pan:2012ki,Nunes:2016dlj,Abdalla:2014cla,Kumar:2016zpg,Murgia:2016ccp,Pan:2016ngu,vomMarttens:2016tdr,Sharov:2017iue,Yang:2017yme,Yang:2017zjs,Yang:2017ccc,Guo:2017deu,Pan:2017ent,Kumar:2017dnp,DiValentino:2017iww,Li:2017usw,Feng:2017usu,Yang:2018ubt,Yang:2018pej,vonMarttens:2018iav,Yang:2018euj,Pan:2019gop,Martinelli:2019dau,Paliathanasis:2019hbi,Yang:2019bpr,Yang:2019uog,Pan:2019jqh,Pan:2020bur,Yang:2019uzo,Kumar:2019wfs,Agrawal:2019dlm,DiValentino:2019ffd,DiValentino:2019jae,Anchordoqui:2019amx,Yao:2020hkw,Pan:2020mst,Pan:2020zza,Gomez-Valent:2020mqn,Lucca:2020zjb,Yao:2020pji,DiValentino:2020vnx,Anchordoqui:2020znj,Anchordoqui:2020sqo,Amirhashchi:2020qep,vonMarttens:2020apn,Sinha:2021tnr,Gao:2021xnk,Bonilla:2021dql,Kumar:2021eev,Lucca:2021dxo,Yang:2021oxc,Paliathanasis:2021egx,Mukherjee:2021ggf,Lucca:2021eqy,Nunes:2021zzi}.  At the background level, the DM-DE coupling modifies the
functional form of the continuity equation of the dark fluids as 
\begin{equation}
\dot \rho_c + 3 {\cal H} \rho_c  = Q
\end{equation}
and
\begin{equation}
 \dot \rho_{\rm DE} + 3 {\cal H} (1+w) \rho_{\rm DE} = -Q \, .
 \label{QDE}
\end{equation} 
where the dot denotes derivative with respect to $\eta$, ${\cal H}
\equiv \dot a/a$ is the conformal Hubble rate,  and $Q$ is the
interaction rate or the interaction function which characterizes the
transfer of energy or/and momentum between the dark sectors, and where $Q < 0$ and $Q > 0$ indicate energy transfer
from DE to DM and vice versa. Although the choice of the 
interaction function is not unique, a classical functional form is given by  
\begin{equation}
  Q = \xi {\cal H} \rho_{\rm DE} \,,
 \end{equation} 
where $\xi$ is a dimensionless coupling parameter quantifying the strength of the DM-DE
interaction. Since the sign of $\xi$ could be  either positive or negative, this defines two sub-classes of models 
driven by $\xi_+$ (for which $\xi >0$) and $\xi_-$ (for which $\xi
<0$). The presence of the DM-DE coupling also
modifies the evolution at the level of perturbations. Assuming the synchronous
gauge, the evolution 
of the DM and DE density perturbations as well as  velocity
divergences have been computed
in Refs.~\cite{Valiviita:2008iv,Gavela:2009cy,Gavela:2010tm}. Following~\cite{DiValentino:2019ffd},
we adopt adiabatic initial conditions in our Boltzmann system for all species. 
At this stage, it is worthwhile to note that when considering a coupling between the DE and DM sectors, the
interacting system could be unstable. Indeed, DE-DM interactions suffer from gravitational instabilities if $w =
 -1$~\cite{Valiviita:2008iv,He:2008si}. We circumvent the instability problem by taking $w = -0.999$.
This approximation is justified
because for $w \to -1$, the effect of DE perturbations is basically
unnoticeable~\cite{Salvatelli:2013wra,DiValentino:2017iww,DiValentino:2019jae}. Therefore, the interacting system essentially captures
the effect of the DM-DE coupling, while at the same time
ensuring the absence of gravitational instabilities. In addition, in
this case we require $\xi < 0$, in order to avoid the early-time
instabilities~\cite{Valiviita:2008iv,Gavela:2009cy,Clemson:2011an,Gavela:2010tm,He:2008si,Jackson:2009mz}. The set of free parameters describing this class of
models are given by  
\begin{eqnarray}
\mathcal{P}_4 \equiv\Bigl\{\Omega_{b}h^2, \Omega_{c}h^2, 100\theta_{\rm MC}, \tau, n_{s}, \ln[10^{10}A_{s}], \notag \\ 
\xi_{-},  \Bigr\}~.
\label{eq:parameter_space4}
\end{eqnarray}

Finally, we can open Pandora's box to construct another class of
models that could resolve the $H_0$ tension. By inspection of (\ref{Hdez}) we can immediately see that the increase of the
effective fractional contribution of spatial curvature to the energy
budget yields a faster expansion rate. The set of free
parameters describing this class of models is given by 
\begin{eqnarray}
\mathcal{P}_5 \equiv\Bigl\{\Omega_{b}h^2, \Omega_{c}h^2, 100\theta_{\rm MC}, \tau, n_{s}, \ln[10^{10}A_{s}], \notag \\ 
\Omega_k,  \Bigr\}
\label{eq:parameter_space5}
\end{eqnarray}
Observational data from the Planck mission point to a $3.4\sigma$ evidence for a closed universe: $-0.095 < \Omega_k <
-0.007$ at 99\%~CL~\cite{Planck:2018vyg,DiValentino:2019qzk,Handley:2019tkm,DiValentino:2020srs}. If this were the case,
we can immediately infer from
(\ref{Hdez}) that the parameter set ${\cal P}_5$ would actually exacerbate the $H_0$ tension.  

In what follows we consider the cosmological models that can be characterized by the 18 additional
possible combinations of the extra 6 free parameters discussed
above ($N_{\rm eff}$, $w_{q}$, $w_{p}$, $\xi_{\pm}$, $\Omega_k$).  In order to avoid the early-time instabilities~\cite{Valiviita:2008iv,Gavela:2009cy,Clemson:2011an,Gavela:2010tm,He:2008si,Jackson:2009mz} we will have $\xi_-$ when $w_q>-1$, and $\xi_+$ when $w_p<-1$. The set of free parameters describing these classes of models
are given by
\begin{eqnarray}
\mathcal{P}_6 \equiv\Bigl\{\Omega_{b}h^2, \Omega_{c}h^2, 100\theta_{\rm MC}, \tau, n_{s}, \ln[10^{10}A_{s}], \notag \\ 
N_{\rm eff}, w_q \Bigr\}~,
\label{eq:parameter_space6}
\end{eqnarray}
\begin{eqnarray}
\mathcal{P}_7 \equiv\Bigl\{\Omega_{b}h^2, \Omega_{c}h^2, 100\theta_{\rm MC}, \tau, n_{s}, \ln[10^{10}A_{s}], \notag \\ 
N_{\rm eff}, w_p \Bigr\}~,
\label{eq:parameter_space7}
\end{eqnarray}
\begin{eqnarray}
\mathcal{P}_{8} \equiv\Bigl\{\Omega_{b}h^2, \Omega_{c}h^2, 100\theta_{\rm MC}, \tau, n_{s}, \ln[10^{10}A_{s}], \notag \\ 
N_{\rm eff},  \xi_- \Bigr\}~,
\label{eq:parameter_space8}
\end{eqnarray}
\begin{eqnarray}
\mathcal{P}_{9} \equiv\Bigl\{\Omega_{b}h^2, \Omega_{c}h^2, 100\theta_{\rm MC}, \tau, n_{s}, \ln[10^{10}A_{s}], \notag \\ 
N_{\rm eff}, \Omega_k \Bigr\}~,
\label{eq:parameter_space9}
\end{eqnarray}
\begin{eqnarray}
\mathcal{P}_{10} \equiv\Bigl\{\Omega_{b}h^2, \Omega_{c}h^2, 100\theta_{\rm MC}, \tau, n_{s}, \ln[10^{10}A_{s}], \notag \\ 
w_q,  \xi_- \Bigr\}~,
\label{eq:parameter_space10}
\end{eqnarray}
\begin{eqnarray}
\mathcal{P}_{11} \equiv\Bigl\{\Omega_{b}h^2, \Omega_{c}h^2, 100\theta_{\rm MC}, \tau, n_{s}, \ln[10^{10}A_{s}], \notag \\ 
w_p,  \xi_+ \Bigr\}~,
\label{eq:parameter_space11}
\end{eqnarray}
\begin{eqnarray}
\mathcal{P}_{12} \equiv\Bigl\{\Omega_{b}h^2, \Omega_{c}h^2, 100\theta_{\rm MC}, \tau, n_{s}, \ln[10^{10}A_{s}], \notag \\ 
w_q, \Omega_k \Bigr\}~,
\label{eq:parameter_space12}
\end{eqnarray}
\begin{eqnarray}
\mathcal{P}_{13} \equiv\Bigl\{\Omega_{b}h^2, \Omega_{c}h^2, 100\theta_{\rm MC}, \tau, n_{s}, \ln[10^{10}A_{s}], \notag \\ 
w_p, \Omega_k \Bigr\}~,
\label{eq:parameter_space13}
\end{eqnarray}
\begin{eqnarray}
\mathcal{P}_{14} \equiv\Bigl\{\Omega_{b}h^2, \Omega_{c}h^2, 100\theta_{\rm MC}, \tau, n_{s}, \ln[10^{10}A_{s}], \notag \\ 
\xi_-, \Omega_k \Bigr\}~,
\label{eq:parameter_space14}
\end{eqnarray}
\begin{eqnarray}
\mathcal{P}_{15} \equiv\Bigl\{\Omega_{b}h^2, \Omega_{c}h^2, 100\theta_{\rm MC}, \tau, n_{s}, \ln[10^{10}A_{s}], \notag \\ 
N_{\rm eff}, w_q, \xi_- \Bigr\}~,
\label{eq:parameter_space15}
\end{eqnarray}
\begin{eqnarray}
\mathcal{P}_{16} \equiv\Bigl\{\Omega_{b}h^2, \Omega_{c}h^2, 100\theta_{\rm MC}, \tau, n_{s}, \ln[10^{10}A_{s}], \notag \\ 
N_{\rm eff}, w_p, \xi_+ \Bigr\}~,
\label{eq:parameter_space16}
\end{eqnarray}
\begin{eqnarray}
\mathcal{P}_{17} \equiv\Bigl\{\Omega_{b}h^2, \Omega_{c}h^2, 100\theta_{\rm MC}, \tau, n_{s}, \ln[10^{10}A_{s}], \notag \\ 
N_{\rm eff}, w_q,  \Omega_k \Bigr\}~,
\label{eq:parameter_space17}
\end{eqnarray}
\begin{eqnarray}
\mathcal{P}_{18} \equiv\Bigl\{\Omega_{b}h^2, \Omega_{c}h^2, 100\theta_{\rm MC}, \tau, n_{s}, \ln[10^{10}A_{s}], \notag \\ 
N_{\rm eff}, w_p,  \Omega_k \Bigr\}~,
\label{eq:parameter_space18}
\end{eqnarray}
\begin{eqnarray}
\mathcal{P}_{19} \equiv\Bigl\{\Omega_{b}h^2, \Omega_{c}h^2, 100\theta_{\rm MC}, \tau, n_{s}, \ln[10^{10}A_{s}], \notag \\
N_{\rm eff}, \xi_-,  \Omega_k \Bigr\}~,
\label{eq:parameter_space19}
\end{eqnarray}
\begin{eqnarray}
\mathcal{P}_{20} \equiv\Bigl\{\Omega_{b}h^2, \Omega_{c}h^2, 100\theta_{\rm MC}, \tau, n_{s}, \ln[10^{10}A_{s}], \notag \\
w_q, \xi_-, \Omega_k \Bigr\}~,
\label{eq:parameter_space20}
\end{eqnarray}
\begin{eqnarray}
\mathcal{P}_{21} \equiv\Bigl\{\Omega_{b}h^2, \Omega_{c}h^2, 100\theta_{\rm MC}, \tau, n_{s}, \ln[10^{10}A_{s}], \notag \\
w_p, \xi_+, \Omega_k \Bigr\}~,
\label{eq:parameter_space21}
\end{eqnarray}
\begin{eqnarray}
\mathcal{P}_{22} \equiv\Bigl\{\Omega_{b}h^2, \Omega_{c}h^2, 100\theta_{\rm MC}, \tau, n_{s}, \ln[10^{10}A_{s}], \notag \\
N_{\rm eff}, w_q, \xi_-, \Omega_k \Bigr\}~,
\label{eq:parameter_space22}
\end{eqnarray}
and
\begin{eqnarray}
\mathcal{P}_{23} \equiv\Bigl\{\Omega_{b}h^2, \Omega_{c}h^2, 100\theta_{\rm MC}, \tau, n_{s}, \ln[10^{10}A_{s}], \notag \\
N_{\rm eff}, w_p, \xi_+, \Omega_k \Bigr\}~.
\label{eq:parameter_space23}
\end{eqnarray}
 
Most of the combined models described in  (\ref{eq:parameter_space5})
to (\ref{eq:parameter_space15}) have been previously discussed in the literature, see e.g.~\cite{Vagnozzi:2019ezj,Yang:2020zuk,Yang:2020uga,Yang:2020tax,DiValentino:2020evt,DiValentino:2020naf,Yang:2020myd,Benaoum:2020qsi,DiValentino:2020leo,Vagnozzi:2020zrh,DiValentino:2020vnx,DiValentino:2020kpf,Yang:2021flj,Yang:2021hxg}. In this work we adopt a systematic approach in which we (re)examine all of these models with the goal of establishing possible intercorrelations among the free parameters. 

 For extensive reviews about possible solutions of the $H_0$ and $S_8$ tensions see~\cite{DiValentino:2020zio,DiValentino:2020vvd,DiValentino:2021izs,Jedamzik:2020zmd,Knox:2019rjx,CANTATA:2021ktz,Perivolaropoulos:2021jda,Schoneberg:2021qvd} and references therein.

\section{Observational Data, Statistical Methodology, and Numerical Analysis}
\label{sec:3}

We begin with a brief description of the cosmological data sets used
in this work.
\begin{itemize}[noitemsep,topsep=0pt]
\item \textbf{Planck 2018 CMB data}: The CMB temperature and
  polarization angular power spectra {\it plikTTTEEE+lowl+lowE} from the Planck 2018 legacy release~\cite{Aghanim:2018eyx,Aghanim:2019ame}. 
\item \textbf{BAO}: Measurements of  baryon acoustic
  oscillations (BAO) from different galaxy surveys: 6dFGS~\cite{Beutler:2011hx}, SDSS-MGS~\cite{Ross:2014qpa}, and BOSS DR12~\cite{Alam:2016hwk}. This is the same
  combination of BAO data considered by the Planck Collaboration
  in~\cite{Aghanim:2018eyx}.
\item \textbf{Pantheon}: The 1048 supernovae type Ia data points
which are distributed in the redshift interval \mbox{$0.01 \leq z \leq 2.3$}, dubbed the Pantheon sample~\cite{Scolnic:2017caz}.  
\item \textbf{R20}: A gaussian prior on the Hubble constant in agreement with the measurement obtained by the SH0ES collaboration in~\cite{Riess:2020fzl}.  
\end{itemize}

We note in passing that associated to the $H_0$ tension there is the question of the roles played by the sound horizon scale and the local expansion rate as distance anchors~\cite{Cuesta:2014asa}. In particular, the so-called ``inverse distance
ladder calibration'' of Pantheon sample~\cite{Scolnic:2017caz} based on the CMB-inferred sound-horizon scale as an anchor cannot be made compatible with the direct distance ladder 
calibration of R20~\cite{Riess:2020fzl}. This has emerged as a redefinition of the $H_0$ tension gauging the calibration of the intrinsic supernova magnitude $M_B$ adopted in the local distance ladder $H_0$ measurement, while testing the consistency of
the corresponding observed fluxes from the Pantheon catalogue with the underlying cosmological model~\cite{Camarena:2021jlr,Efstathiou:2021ocp}. However, we note that the $M_B$ tension is only referring to the mismatch observed in SH0ES vs Planck data analyses, without considering all the other probes and cannot test modifications of the expansion rate in the early universe. For this reason, herein we center the analysis on the $H_0$ tension.

For the numerical analysis, we adopt a modified version of the well
known cosmological package \CosmoMC~\cite{Lewis:2002ah,Lewis:2013hha},
which is publicly available~\cite{comomc}. This package is equipped with a convergence diagnostic based on the
Gelman-Rubin criterion~\cite{Gelman:1992zz} and includes the support
for the 2018 Planck data release~\cite{Aghanim:2019ame}. The flat priors on the free parameters of the classes of models
analyzed herein are listed in Table~\ref{table:priors}. The results of
our numerical analysis are encapsulated in Tables~\ref{tabA} to \ref{tabL} and
Figs.~\ref{fig:0} to \ref{fig:23}.

A point worth noting at this juncture is that the acoustic peaks which are prominently observed in the CMB anisotropy spectra, are also visible as BAO peaks in the galaxy power spectra and carry the footprints of a new standard ruler:  the sound horizon at the epoch of baryon decoupling,  $r_{\rm drag}$, when the photon drag on baryons becomes unimportant.  The estimated  sound horizon at the end of the baryonic-drag epoch is $r_{\rm drag} = (137 \pm 3^{\rm stat} \pm 2^{\rm syst})~{\rm Mpc}$~\cite{Arendse:2019hev}. This estimate is based on data from low-redshift probes and a set of polynomial parametrizations which are almost independent of the underlying cosmology. None of the combination of classes of models analyzed herein can accommodate the $r_{\rm drag}$ estimate of~\cite{Arendse:2019hev} at the $1\sigma$ level; see Tables~\ref{tabA} to \ref{tabL}.
\begin{table}
  \begin{center}
  \caption{Flat priors on various cosmological parameters coming from several cosmological scenarios.}
\label{table:priors}
\begin{tabular}{ccccc}
  \hline
  \hline
~~~~~~~~~~~~~~~ Parameter ~~~~~~~~~~~~~~~ & ~~~~~~~~~~~~~~ Prior ~~~~~~~~~~~~~~~ \\
\hline
~~~~$\Omega_{b} h^2$~~~~         & ~~~~$[0.005,0.1]$~~~~ \\
$\Omega_{c} h^2$         & $[0.001,0.99]$ \\
$100\theta_{\rm MC}$             & $[0.5,10]$ \\
$\tau$                       & $[0.01,0.8]$ \\
$n_s$               & $[0.8,1.2]$\\
$\ln[10^{10}A_{s}]$         & $[1.61, 3.91]$ \\
  $N_{\rm eff}$       & $[0.05,10]$ \\
$w_q$ &  $[-1, 1]$\\
$w_p$                        & $[-3, -1]$\\  
  $\xi_+$                   & $[0,1]$ \\
  $\xi_-$                               & $[-1,0]$ \\
$\Omega_{k}$                & $[-0.3,0.3]$  \\
  \hline
  \hline
\end{tabular}
\end{center}
\end{table}

\begin{table*}                                                                                                                 \resizebox{\textwidth}{!}{   
\begin{tabular}{c|cc|cc|ccc}                                                                                               
\hline\hline                                                                                    Model & $\mathcal{P}_0$ & $\mathcal{P}_0$ & $\mathcal{P}_2$ & $\mathcal{P}_2$ & $\mathcal{P}_3$ & $\mathcal{P}_3$ \\ \hline                                
 & CMB & all & CMB & all & CMB & all \\ \hline

$\Omega_{\rm b} h^2$ & $    0.02236\pm0.00015$ & $    0.02243\pm0.00013$  & $    0.02235\pm0.00015$ & $    0.02246\pm0.00014$ & $    0.02240\pm0.00015$ & $    0.02237\pm0.00014$  \\

$\Omega_{\rm c} h^2$ & $    0.1202\pm0.0014$ & $    0.11918\pm0.00097$  & $    0.1203\pm0.0014$ & $    0.1188\pm0.0010$ & $    0.1199\pm0.0013$  & $    0.1200\pm0.0011$ \\

$100\theta_{MC}$ & $    1.04090\pm0.00031$ & $    1.04102\pm0.00029$  & $    1.04088\pm0.00031$ & $    1.04106\pm0.00030$ & $    1.04095\pm0.00031$ & $    1.04093\pm0.00030$ \\

$\tau$ & $    0.0546\pm0.0078$ & $    0.0558^{+0.0070}_{-0.0078}$  & $    0.0547^{+0.0074}_{-0.0082}$  & $    0.0563^{+0.0072}_{-0.0081}$  & $    0.0540\pm0.0078$  & $    0.0545\pm0.0077$  \\

$n_s$ & $    0.9648\pm0.0043$ & $    0.9671\pm0.0038$  & $    0.9645\pm0.0044$ & $    0.9682\pm0.0038$  & $    0.9656\pm0.0043$ & $    0.9651\pm0.0040$ \\

${\rm{ln}}(10^{10} A_s)$ & $    3.045\pm0.016$ & $    3.046^{+0.014}_{-0.016}$   & $    3.046\pm0.016$ & $    3.046\pm0.016$ & $    3.044\pm0.016$ & $    3.045\pm0.016$ \\

$w$ & $   -1$ & $   -1$  & $   <-0.879 $ & $   <-0.977 $ & $   -1.60_{-    0.33}^{+    0.16}$ & $   -1.039^{+0.037}_{-0.011} $ \\

\hline

$\Omega_{m0}$ & $  0.3166\pm0.0085$ & $    0.3102\pm0.0058$   & $    0.347_{-    0.034}^{+    0.013}$ & $    0.3129\pm0.0063$ & $    0.194_{-    0.051}^{+    0.015}$  & $    0.3046\pm0.0069$  \\

$\sigma_8$ & $    0.8122\pm0.0073$ & $    0.8095\pm0.0071$   & $    0.785_{-    0.013}^{+    0.029}$ & $    0.8029\pm0.0088$ & $    0.976_{-    0.044}^{+    0.090}$ & $    0.822^{+0.010}_{-0.012}$  \\
$H_0 {\rm[km/s/Mpc]}$& $   67.27\pm0.61$ & $   67.73\pm0.43$  & $   64.4_{-    1.2}^{+    2.9}$ & $   67.35^{+0.56}_{-0.49}$  & $   >83.2$  & $ 68.53^{+0.60}_{-0.74}$   \\

$S_8$ & $    0.834\pm0.016$ &  $    0.823\pm0.012$  & $    0.842\pm0.018$ & $    0.809\pm0.012$ & $    1.29_{-    0.45}^{+    0.17}$  & $    0.828\pm0.013$  \\

$r_{\rm{drag}}$ [Mpc] & $  147.05\pm0.30$ &  $  147.24\pm0.24$  & $  147.04\pm0.30$  & $  147.32\pm0.25$  & $  147.08\pm0.29$ & $  147.10\pm0.26$  \\

\hline\hline                                               
\end{tabular}          }                                     
\caption{68\% CL constraints on various free and derived parameters of the parameter spaces $\mathcal{P}_0 \equiv\Bigl\{\Omega_{b}h^2, \Omega_{c}h^2, 100\theta_{\rm MC},
  \tau, n_{s}, \ln[10^{10}A_{s}] \Bigr\}$, $\mathcal{P}_2 \equiv\Bigl\{\Omega_{b}h^2, \Omega_{c}h^2, 100\theta_{\rm MC}, \tau, n_{s}, \ln[10^{10}A_{s}], 
w_q,  \Bigr\}$ and $\mathcal{P}_3 \equiv\Bigl\{\Omega_{b}h^2, \Omega_{c}h^2, 100\theta_{\rm MC}, \tau, n_{s}, \ln[10^{10}A_{s}],
w_p,  \Bigr\}$.   }
\label{tabA}                                              
\end{table*}  

\begin{table*}                                                                                                                 \resizebox{\textwidth}{!}{   
\begin{tabular}{c|cc|cc|ccc}                                                                                               
\hline\hline                                                                                    Model & $\mathcal{P}_1$ & $\mathcal{P}_1$ & $\mathcal{P}_4$ & $\mathcal{P}_4$ & $\mathcal{P}_8$ & $\mathcal{P}_8$ \\ \hline                                
 & CMB & all & CMB & all & CMB & all \\ \hline

$\Omega_{\rm b} h^2$ & $    0.02226\pm0.00022$ & $    0.02241\pm0.00018$  & $    0.02239\pm0.00015$ & $    0.02237\pm0.00014$ & $    0.02228\pm0.00023$ & $    0.02229\pm0.00019$  \\

$\Omega_{\rm c} h^2$ & $    0.1184\pm0.0030$ & $    0.1187\pm0.0031$  & $    <0.0634$ & $    0.106^{+0.011}_{-0.005}$ & $    <0.0657$  & $    0.103_{-    0.008}^{+    0.012}$ \\

$100\theta_{MC}$ & $    1.04110\pm0.00043$ & $    1.04108\pm0.00045$  & $    1.0458_{-    0.0021}^{+    0.0033}$ & $    1.04174_{-    0.00066}^{+    0.00045}$ & $    1.0458_{-    0.0023}^{+    0.0032}$ & $    1.04205_{-    0.00082}^{+    0.00064}$ \\

$\tau$ & $    0.0536\pm0.0078$ & $    0.0557_{-    0.0080}^{+    0.0072}$  & $    0.0541\pm0.0076$  & $    0.0545\pm0.0078$  & $    0.0528\pm0.0081$  & $    0.0538\pm0.0077$  \\

$n_s$ & $    0.9600\pm0.0085$ & $    0.9661\pm0.0071$  & $    0.9655\pm0.0043$ & $    0.9652\pm0.0039$  & $    0.9610\pm0.0088$ & $    0.9616\pm0.0074$ \\

${\rm{ln}}(10^{10} A_s)$ & $    3.039\pm0.018$ & $    3.044\pm0.018$   & $    3.044\pm0.016$ & $    3.045\pm0.016$ & $    3.037\pm0.019$ & $    3.039\pm0.018$ \\

$\xi$ & $   0$ & $   0$  & $   -0.54^{+0.12}_{-0.28} $ & $   -0.13^{+0.12}_{-0.04} $ & $   -0.53_{-    0.30}^{+    0.13}$ & $   -0.14^{+0.12}_{-0.06} $ \\

$N_{\rm eff}$ & $   2.92\pm0.19$ & $   3.02\pm0.18$  & $   3.046$  & $  3.046$  & $   2.93\pm0.20$  & $   2.94\pm0.18$  \\

\hline

$\Omega_{m0}$ & $  0.320\pm0.010$ & $    0.3106\pm0.0067$   & $    0.139_{-    0.095}^{+    0.034}$ & $    0.274^{+0.029}_{-0.017}$ & $    0.15_{-    0.10}^{+    0.04}$  & $    0.271_{-    0.019}^{+    0.029}$  \\

$\sigma_8$ & $    0.806\pm0.011$ & $    0.808\pm0.011$   & $    2.3_{-    1.4}^{+    0.4}$ & $    0.92_{-    0.10}^{+    0.04}$ & $    2.2_{-    1.4}^{+    0.4}$ & $    0.93_{-    0.11}^{+    0.05}$  \\

$H_0 {\rm[km/s/Mpc]}$ & $   66.4\pm1.4$ & $   67.6\pm1.1$  & $   72.8_{-    1.5}^{+    3.0}$ & $   68.60^{+0.62}_{-0.82}$  & $   71.8_{-    2.6}^{+    3.3}$  & $   68.1\pm 1.2$   \\

$S_8$ & $    0.833\pm0.016$ &  $    0.822\pm0.014$  & $    1.30_{-    0.44}^{+    0.17}$ & $    0.876_{-    0.047}^{+    0.025}$ & $    1.28_{-    0.45}^{+    0.17}$  & $    0.879_{-    0.050}^{+    0.027}$  \\

$r_{\rm{drag}}$ [Mpc]& $  148.3\pm1.9$ &  $  147.6\pm1.8$  & $  147.08\pm0.30$  & $  147.09^{+0.27}_{-0.24}$  & $  148.3\pm2.0$ & $  148.2\pm1.8$  \\

\hline\hline                                               
\end{tabular}          }                                     
\caption{68\% CL constraints on various free and derived parameters of the parameter spaces $\mathcal{P}_1 \equiv\Bigl\{\Omega_{b}h^2, \Omega_{c}h^2, 100\theta_{\rm MC}, \tau, n_{s}, \ln[10^{10}A_{s}], 
N_{\rm eff} \Bigr\}$,   $\mathcal{P}_4 \equiv\Bigl\{\Omega_{b}h^2, \Omega_{c}h^2, 100\theta_{\rm MC}, \tau, n_{s}, \ln[10^{10}A_{s}],
\xi_{-}  \Bigr\}$ and $\mathcal{P}_{8} \equiv\Bigl\{\Omega_{b}h^2, \Omega_{c}h^2, 100\theta_{\rm MC}, \tau, n_{s}, \ln[10^{10}A_{s}],
N_{\rm eff},  \xi_- \Bigr\}$.  }
\label{tabB}                                              
\end{table*}  


\begin{table*}                                                                                                                 \resizebox{\textwidth}{!}{   
\begin{tabular}{c|cc|cc|ccc}                                                                                               
\hline\hline                                                                                    Model & $\mathcal{P}_5$ & $\mathcal{P}_5$ & $\mathcal{P}_{12}$ & $\mathcal{P}_{12}$ & $\mathcal{P}_{13}$ & $\mathcal{P}_{13}$ \\ \hline                                
 & CMB & all & CMB & all & CMB & all \\ \hline

$\Omega_{\rm b} h^2$ & $    0.02259\pm0.00017$ & $    0.02240\pm0.00016$  & $    0.02260\pm0.00017$ & $    0.02240\pm0.00016$ & $    0.02260\pm0.00017$ & $    0.02240\pm0.00015$  \\

$\Omega_{\rm c} h^2$ & $    0.1182\pm0.0015$ & $    0.1196\pm0.0014$  & $    0.1182\pm0.0015$ & $    0.1195\pm0.0014$ & $    0.1180\pm0.0015$  & $    0.1197\pm0.0014$ \\

$100\theta_{MC}$ & $    1.04115\pm0.00033$ & $    1.04097\pm0.00030$  & $    1.04116\pm0.00032$ & $    1.04096\pm0.00031$ & $    1.04117\pm0.00033$ & $    1.04095\pm0.00030$ \\

$\tau$ & $    0.0489^{+0.0078}_{-0.0071}$ & $    0.0553^{+0.0071}_{-0.0078}$  & $    0.0483^{+0.0082}_{-0.0069}$  & $    0.0558\pm0.0079$  & $    0.0478^{+0.0087}_{-0.0076}$  & $    0.0544\pm0.0076$  \\

$n_s$ & $    0.9703\pm0.0047$ & $    0.9662\pm0.0044$  & $    0.9706\pm0.0047$ & $    0.9661\pm0.0046$  & $    0.9709\pm0.0049$ & $    0.9657\pm0.0044$ \\

${\rm{ln}}(10^{10} A_s)$ & $    3.029^{+0.017}_{-0.015}$ & $    3.046\pm0.016$   & $    3.028^{+0.017}_{-0.015}$ & $    3.046\pm0.016$ & $    3.026^{+0.019}_{-0.016}$ & $    3.044\pm0.016$ \\

$\Omega_k$ & $   -0.043^{+0.018}_{-0.014}$ & $   0.0009\pm0.0019$  & $   -0.080^{+0.041}_{-0.026} $ & $   0.0017\pm0.0020 $ & $   -0.028_{-    0.008}^{+    0.019}$ & $ -0.0004\pm0.0020 $ \\

$w$ & $   -1$ & $   -1$  & $   -0.59^{+0.24}_{-0.28} $ & $   <-0.973 $ & $   >-1.98$ & $   >-1.05 $ \\

\hline

$\Omega_{m0}$ & $  0.479^{+0.055}_{-0.065}$ & $    0.3087\pm0.0064$   & $    0.70_{-    0.19}^{+    0.14}$ & $    0.3114\pm0.0067$ & $    0.314_{-    0.15}^{+    0.07}$  & $    0.3041\pm0.0070$  \\

$\sigma_8$ & $    0.776\pm0.015$ & $    0.8115\pm0.0083$   & $    0.690_{-    0.056}^{+    0.045}$ & $    0.8050^{+0.0099}_{-0.0090}$ & $    0.926_{-    0.14}^{+    0.07}$ & $    0.822^{+0.010}_{-0.012}$  \\

$H_0 {\rm[km/s/Mpc]}$ & $   54.6^{+3.2}_{-3.8}$ & $   68.00\pm0.66$  & $   45.8_{-    6.4}^{+    4.2}$ & $   67.68\pm0.68$  & $   70^{+9}_{-20}$  & $ 68.53\pm0.77$   \\

$S_8$ & $    0.978\pm0.046$ &  $    0.823\pm0.012$  & $    1.040^{+0.060}_{-0.054}$ & $    0.820\pm0.013$ & $    0.915\pm0.069$  & $    0.827\pm0.012$  \\

$r_{\rm{drag}}$ [Mpc] & $  147.34\pm0.31$ &  $  147.15\pm0.30$  & $  147.32\pm0.31$  & $  147.19\pm0.31$  & $  147.35\pm0.31$ & $  147.14\pm0.30$  \\

\hline\hline                                               
\end{tabular}          }                                     
\caption{68\% CL constraints on various free and derived parameters of the parameter spaces $\mathcal{P}_5 \equiv\Bigl\{\Omega_{b}h^2, \Omega_{c}h^2, 100\theta_{\rm MC}, \tau, n_{s}, \ln[10^{10}A_{s}],
\Omega_k  \Bigr\}$, $\mathcal{P}_{12} \equiv\Bigl\{\Omega_{b}h^2, \Omega_{c}h^2, 100\theta_{\rm MC}, \tau, n_{s}, \ln[10^{10}A_{s}],
w_q, \Omega_k \Bigr\}$ and $\mathcal{P}_{13} \equiv\Bigl\{\Omega_{b}h^2, \Omega_{c}h^2, 100\theta_{\rm MC}, \tau, n_{s}, \ln[10^{10}A_{s}],
w_p, \Omega_k \Bigr\}$.  }
\label{tabC}                                              
\end{table*}  


\begin{table*}                                                                                                                 \resizebox{\textwidth}{!}{   
\begin{tabular}{c|cc|cc|ccc}                                                                                               
\hline\hline                                                                                    Model & $\mathcal{P}_6$ & $\mathcal{P}_6$ & $\mathcal{P}_{10}$ & $\mathcal{P}_{10}$ & $\mathcal{P}_{15}$ & $\mathcal{P}_{15}$ \\ \hline                                
 & CMB & all & CMB & all & CMB & all \\ \hline

$\Omega_{\rm b} h^2$ & $    0.02222\pm0.00022$ & $    0.02245\pm0.00019$  & $    0.02237\pm0.00015$ & $    0.02239\pm0.00014$ & $    0.02226\pm0.00022$ & $    0.02233\pm0.00020$  \\

$\Omega_{\rm c} h^2$ & $    0.1185\pm0.0030$ & $    0.1189\pm0.0030$  & $    <0.0456$ & $    <0.0599$ & $    <0.0431$  & $    <0.0577$ \\

$100\theta_{MC}$ & $    1.04109\pm0.00043$ & $    1.04107\pm0.00044$  & $    1.0467_{-    0.0014}^{+    0.0031}$ & $    1.0461_{-    0.0019}^{+    0.0034}$ & $    1.0470_{-    0.0014}^{+    0.0031}$ & $    1.0462_{-    0.0020}^{+    0.0031}$ \\

$\tau$ & $    0.0539^{+0.0072}_{-0.0081}$ & $    0.0561\pm0.0077$  & $    0.0535\pm0.0076$  & $    0.0544\pm0.0077$  & $    0.0530\pm0.0078$  & $    0.0536\pm0.0077$  \\

$n_s$ & $    0.9591\pm0.0086$ & $    0.9682\pm0.0070$  & $    0.9650\pm0.0044$ & $    0.9660\pm0.0040$  & $    0.9600\pm0.0087$ & $    0.9631\pm0.0077$ \\

${\rm{ln}}(10^{10} A_s)$ & $    3.039\pm0.019$ & $    3.045\pm0.018$   & $    3.043\pm0.016$ & $    3.044\pm0.016$ & $    3.037\pm0.019$ & $    3.039\pm0.018$ \\

$\xi$ & $   0$ & $   0$  & $   -0.62^{+0.07}_{-0.22} $ & $   -0.56^{+0.11}_{-0.27} $ & $   -0.64^{+0.07}_{-0.22}$ & $   -0.57^{+0.11}_{-0.25} $ \\

$N_{\rm eff}$ & $   2.91\pm0.19$ & $   3.05\pm0.18$ & $   3.046$  & $  3.046$  & $   2.92\pm0.19$  & $   2.96\pm0.18$  \\

$w_q$ & $   <-0.877$ & $   <-0.978$  & $   <-0.836$  & $  -0.841^{+0.093}_{-0.054}$  & $   <-0.835$  & $   -0.843^{+0.087}_{-0.054}$  \\

\hline

$\Omega_{m0}$ & $  0.353^{+0.014}_{-0.036}$ & $    0.3132\pm0.0069$   & $    0.127_{-    0.081}^{+    0.022}$ & $    0.146^{+0.037}_{-0.093}$ & $    0.125_{-    0.079}^{+    0.021}$  & $    0.144_{-    0.088}^{+    0.037}$  \\

$\sigma_8$ & $    0.779^{+0.030}_{-0.016}$ & $    0.803\pm0.012$   & $    2.56^{+0.7}_{-1.7}$ & $    2.2_{-1.3}^{+    0.5}$ & $    2.6^{+0.7}_{-1.7}$ & $    2.2_{-1.3}^{+    0.4}$  \\

$H_0 {\rm[km/s/Mpc]}$ & $   63.5^{+3.2}_{-2.0}$ & $   67.3\pm1.1$  & $   69.7^{+3.9}_{-2.6}$ & $   68.33\pm0.81$  & $   68.8^{+4.1}_{-3.0}$  & $   68.0\pm1.2$   \\

$S_8$ & $    0.843\pm0.017$ &  $    0.821\pm0.013$  & $    1.42^{+0.27}_{-0.46}$ & $    1.31_{-    0.45}^{+    0.20}$ & $    1.43^{+0.28}_{-0.47}$  & $    1.31_{-    0.42}^{+    0.19}$  \\

$r_{\rm{drag}}$ [Mpc]& $  148.4\pm1.9$ &  $  147.3\pm1.8$  & $  147.06\pm0.29$  & $  147.14\pm0.26$  & $  148.3\pm1.9$ & $  148.0\pm1.9$  \\

\hline\hline                                               
\end{tabular}          }                                     
\caption{68\% CL constraints on various free and derived parameters of the parameter spaces $\mathcal{P}_6 \equiv\Bigl\{\Omega_{b}h^2, \Omega_{c}h^2, 100\theta_{\rm MC}, \tau, n_{s}, \ln[10^{10}A_{s}],
N_{\rm eff}, w_q \Bigr\}$, $\mathcal{P}_{10} \equiv\Bigl\{\Omega_{b}h^2, \Omega_{c}h^2, 100\theta_{\rm MC}, \tau, n_{s}, \ln[10^{10}A_{s}],
w_q,  \xi_- \Bigr\}$ and $\mathcal{P}_{15} \equiv\Bigl\{\Omega_{b}h^2, \Omega_{c}h^2, 100\theta_{\rm MC}, \tau, n_{s}, \ln[10^{10}A_{s}],
N_{\rm eff}, w_q, \xi_- \Bigr\}$. }
\label{tabD}                                              
\end{table*}  


\begin{table*}                                                                                                                 \resizebox{\textwidth}{!}{   
\begin{tabular}{c|cc|cc|ccc}                                                                                               
\hline\hline                                                                                    Model & $\mathcal{P}_7$ & $\mathcal{P}_7$ & $\mathcal{P}_{11}$ & $\mathcal{P}_{11}$ & $\mathcal{P}_{16}$ & $\mathcal{P}_{16}$ \\ \hline                                
 & CMB & all & CMB & all & CMB & all \\ \hline

$\Omega_{\rm b} h^2$ & $    0.02227\pm0.00022$ & $    0.02229\pm0.00019$  & $    0.02239\pm0.00015$ & $    0.02239\pm0.00014$ & $    0.02227\pm0.00022$ & $    0.02231\pm0.00020$  \\

$\Omega_{\rm c} h^2$ & $    0.1178\pm0.0029$ & $    0.1183\pm0.0030$  & $    0.133^{+0.005}_{-0.012}$ & $    0.133^{+0.006}_{-0.012}$ & $    0.130^{+0.006}_{-0.012}$  & $    0.131_{-    0.011}^{+    0.007}$ \\

$100\theta_{MC}$ & $    1.04118\pm0.00044$ & $    1.04113\pm0.00044$  & $    1.04026_{-    0.00049}^{+    0.00065}$ & $    1.04023_{-    0.00050}^{+    0.00059}$ & $    1.04052_{-    0.00059}^{+    0.00070}$ & $    1.04043_{-    0.00058}^{+    0.00067}$ \\

$\tau$ & $    0.0528\pm0.0080$ & $    0.0540\pm0.0075$  & $    0.0539\pm0.0078$  & $    0.0548\pm0.0080$  & $    0.0529\pm0.0078$  & $    0.0543\pm0.0080$  \\

$n_s$ & $    0.9597\pm0.0083$ & $    0.9616\pm0.0074$  & $    0.9654\pm0.0043$ & $    0.9659\pm0.0040$  & $    0.9597\pm0.0085$ & $    0.9622\pm0.0076$ \\

${\rm{ln}}(10^{10} A_s)$ & $    3.035\pm0.019$ & $    3.039\pm0.018$   & $    3.043\pm0.016$ & $    3.045\pm0.016$ & $    3.036\pm0.018$ & $    3.040\pm0.019$ \\

$\xi$ & $   0$ & $   0$  & $   <0.134 $ & $   <0.181 $ & $   <0.132$ & $   <0.171 $ \\

$N_{\rm eff}$ & $   2.90\pm0.19$ & $   2.94\pm0.18$ & $   3.046$  & $  3.046$  & $   2.90\pm0.19$  & $   2.95\pm0.18$  \\

$w_p$ & $   -1.65^{+0.19}_{-0.34}$ & $   -1.044^{+0.038}_{-0.016}$  & $   -1.59^{+0.19}_{-0.34}$  & $  -1.076^{+0.049}_{-0.033}$  & $   -1.62^{+0.23}_{-0.36}$  & $   -1.081^{+0.049}_{-0.036}$  \\

\hline

$\Omega_{m0}$ & $  0.192^{+0.017}_{-0.051}$ & $    0.3056\pm0.0072$   & $    0.220_{-    0.066}^{+    0.025}$ & $    0.334^{+0.018}_{-0.024}$ & $    0.219_{-    0.068}^{+    0.027}$  & $    0.334_{-    0.023}^{+    0.017}$  \\

$\sigma_8$ & $    0.978^{+0.091}_{-0.045}$ & $    0.818\pm0.013$   & $    0.881\pm0.085$ & $    0.750_{-    0.045}^{+    0.051}$ & $    0.881\pm0.086$ & $    0.751_{-    0.040}^{+    0.050}$  \\

$H_0 {\rm[km/s/Mpc]}$ & $   >83.1$ & $   68.0\pm 1.2$  & $   >80.9$ & $   68.38\pm0.78$  & $   >80.2$  & $   67.9\pm1.2$   \\

$S_8$ & $    0.771^{+0.024}_{-0.035}$ &  $    0.826 \pm0.014$  & $    0.742\pm0.039$ & $    0.791\pm 0.026$ & $    0.740\pm0.040$  & $    0.791_{-    0.025}^{+    0.028}$  \\

$r_{\rm{drag}}$ [Mpc] & $  148.6\pm1.9$ &  $  148.2\pm1.9$  & $  147.09\pm0.29$  & $  147.14\pm0.27$  & $  148.5\pm1.9$ & $  148.1\pm1.9$  \\

\hline\hline                                               
\end{tabular}          }                                     
\caption{68\% CL constraints on various free and derived parameters of the parameter spaces $\mathcal{P}_7 \equiv\Bigl\{\Omega_{b}h^2, \Omega_{c}h^2, 100\theta_{\rm MC}, \tau, n_{s}, \ln[10^{10}A_{s}],
N_{\rm eff}, w_p \Bigr\}$, $\mathcal{P}_{11} \equiv\Bigl\{\Omega_{b}h^2, \Omega_{c}h^2, 100\theta_{\rm MC}, \tau, n_{s}, \ln[10^{10}A_{s}],
w_p,  \xi_+ \Bigr\}$ and $\mathcal{P}_{16} \equiv\Bigl\{\Omega_{b}h^2, \Omega_{c}h^2, 100\theta_{\rm MC}, \tau, n_{s}, \ln[10^{10}A_{s}],
N_{\rm eff}, w_p, \xi_+ \Bigr\}$. }
\label{tabE}                                              
\end{table*}  


\begin{table*}                                                                                                                 \resizebox{\textwidth}{!}{   
\begin{tabular}{c|cc|cc|ccc}                                                                                               
\hline\hline                                                                                    Model & $\mathcal{P}_9$ & $\mathcal{P}_9$ & $\mathcal{P}_{14}$ & $\mathcal{P}_{14}$ & $\mathcal{P}_{19}$ & $\mathcal{P}_{19}$ \\ \hline                                
 & CMB & all & CMB & all & CMB & all \\ \hline

$\Omega_{\rm b} h^2$ & $    0.02259\pm0.00025$ & $    0.02233\pm0.00023$  & $    0.02261\pm0.00017$ & $    0.02240\pm0.00016$ & $    0.02262\pm0.00025$ & $    0.02231\pm0.00023$  \\

$\Omega_{\rm c} h^2$ & $    0.1180\pm0.0031$ & $    0.1187\pm0.0030$  & $    0.077^{+0.035}_{-0.019}$ & $    0.105^{+0.012}_{-0.006}$ & $    0.074^{+0.035}_{-0.021}$  & $    0.103_{-    0.007}^{+    0.013}$ \\

$100\theta_{MC}$ & $    1.04120\pm0.00045$ & $    1.04109\pm0.00044$  & $    1.0437_{-    0.0023}^{+    0.0012}$ & $    1.04179_{-    0.00075}^{+    0.00050}$ & $    1.0439_{-    0.0023}^{+    0.0014}$ & $    1.04203_{-    0.00088}^{+    0.00059}$ \\

$\tau$ & $    0.0485\pm0.0083$ & $    0.0547\pm0.0079$  & $    0.0481^{+0.0085}_{-0.0076}$  & $    0.0546\pm0.0078$  & $    0.0468^{+0.0085}_{-0.0075}$  & $    0.0543^{+0.0072}_{-0.0082}$  \\

$n_s$ & $    0.9701\pm0.0094$ & $    0.9632\pm0.0088$  & $    0.9708\pm0.0047$ & $    0.9660\pm0.0045$  & $    0.9714\pm0.0093$ & $    0.9623\pm0.0088$ \\

${\rm{ln}}(10^{10} A_s)$ & $    3.027\pm0.020$ & $    3.042\pm0.019$   & $    3.027^{+0.017}_{-0.016}$ & $    3.044\pm0.016$ & $    3.028^{+0.020}_{-0.018}$ & $    3.040\pm0.019$ \\

$\xi$ & $   0$ & $   0$  & $   <-0.385 $ &$   >-0.171 $ & $   <-0.414$ & $   -0.15^{+0.14}_{-0.04} $ \\

$N_{\rm eff}$ & $   3.04\pm0.20$ & $   2.97\pm0.19$ & $   3.046$  & $  3.046$  & $   3.05\pm0.20$  & $   2.95\pm0.20$  \\

$\Omega_k$ & $   -0.044^{+0.020}_{-0.015}$ & $   0.0013\pm 0.0022$  & $   -0.036^{+0.017}_{-0.013}$  & $  -0.0005\pm0.0020$  & $   -0.035^{+0.018}_{-0.012}$  & $   -0.0002\pm0.0023$  \\

\hline

$\Omega_{m0}$ & $  0.484^{+0.057}_{-0.070}$ & $    0.3097\pm0.0066$   & $    0.30\pm0.11$ & $    0.273^{+0.029}_{-0.017}$ & $    0.29_{-    0.13}^{+    0.11}$  & $    0.271_{-    0.019}^{+    0.032}$  \\

$\sigma_8$ & $    0.774\pm0.017$ & $    0.808\pm0.011$   & $    1.31^{+0.10}_{-0.54}$ & $    0.93_{-    0.10}^{+    0.04}$ & $    1.36^{+0.11}_{-0.59}$ & $    0.93_{-    0.11}^{+    0.04}$  \\

$H_0 {\rm[km/s/Mpc]}$ & $   54.4^{+3.4}_{-4.0}$ & $   67.6\pm1.1$  & $   58.7^{+4.1}_{-5.2}$ & $   68.52\pm0.75$  & $   59.1^{+4.5}_{-5.1}$  & $   68.1\pm1.2$   \\

$S_8$ & $    0.980\pm0.051$ &  $    0.821\pm0.013$  & $    1.20^{+0.10}_{-0.22}$ & $    0.878^{+0.025}_{-0.050}$ & $    1.21^{+0.10}_{-0.24}$  & $    0.881_{-    0.055}^{+    0.026}$  \\

$r_{\rm{drag}}$ [Mpc] & $  147.5\pm2.0$ &  $  147.9\pm1.9$  & $  147.34\pm0.31$  & $  147.14\pm0.30$  & $  147.3\pm2.0$ & $  148.1\pm2.0$  \\

\hline\hline                                               
\end{tabular}          }                                     
\caption{68\% CL constraints on various free and derived parameters of the parameter spaces $\mathcal{P}_{9} \equiv\Bigl\{\Omega_{b}h^2, \Omega_{c}h^2, 100\theta_{\rm MC}, \tau, n_{s}, \ln[10^{10}A_{s}], 
N_{\rm eff}, \Omega_k \Bigr\}$, $\mathcal{P}_{14} \equiv\Bigl\{\Omega_{b}h^2, \Omega_{c}h^2, 100\theta_{\rm MC}, \tau, n_{s}, \ln[10^{10}A_{s}],
\xi_-, \Omega_k \Bigr\}$, and $\mathcal{P}_{19} \equiv\Bigl\{\Omega_{b}h^2, \Omega_{c}h^2, 100\theta_{\rm MC}, \tau, n_{s}, \ln[10^{10}A_{s}],
N_{\rm eff}, \xi_-,  \Omega_k \Bigr\}$.}
\label{tabF}                                              
\end{table*}  



\begin{table*}                                                                                                                 \resizebox{\textwidth}{!}{   
\begin{tabular}{c|cc|cc|ccc}                                                                                               
\hline\hline                                                                                    Model & $\mathcal{P}_{17}$ & $\mathcal{P}_{17}$ & $\mathcal{P}_{20}$ & $\mathcal{P}_{20}$ & $\mathcal{P}_{22}$ & $\mathcal{P}_{22}$ \\ \hline                                
 & CMB & all & CMB & all & CMB & all \\ \hline

$\Omega_{\rm b} h^2$ & $    0.02258\pm0.00025$ & $    0.02234\pm0.00023$  & $    0.02260\pm0.00017$ & $    0.02240\pm0.00016$ & $    0.02260\pm0.00025$ & $    0.02231\pm0.00023$  \\

$\Omega_{\rm c} h^2$ & $    0.1182\pm0.0030$ & $    0.1187\pm0.0030$  & $    0.078^{+0.036}_{-0.018}$ & $    <0.0555$ & $    0.076^{+0.036}_{-0.021}$  & $    <0.0569$ \\

$100\theta_{MC}$ & $    1.04117\pm0.00044$ & $    1.04106\pm0.00043$  & $    1.0437_{-    0.0023}^{+    0.0012}$ & $    1.0463^{+0.0031}_{-0.0019}$ & $    1.0438_{-    0.0024}^{+    0.0013}$ & $    1.0463_{-    0.0019}^{+    0.0034}$ \\

$\tau$ & $    0.0482\pm0.0082$ & $    0.0550^{+0.0073}_{-0.0083}$  & $    0.0481\pm 0.0080$  & $    0.0546\pm0.0078$  & $    0.0484\pm0.0083$  & $    0.0540\pm0.0077$  \\

$n_s$ & $    0.9703\pm0.0093$ & $    0.9633\pm0.0088$  & $    0.9706\pm0.0046$ & $    0.9661\pm0.0046$  & $    0.9708\pm0.0092$ & $    0.9623\pm0.0087$ \\

${\rm{ln}}(10^{10} A_s)$ & $    3.027\pm0.019$ & $    3.042\pm0.019$   & $    3.027\pm0.017$ & $    3.044\pm0.016$ & $    3.028\pm0.019$ & $    3.039\pm0.018$ \\

$\xi$ & $   0$ & $   0$  & $   <-0.415 $ & $   -0.58^{+0.10}_{-0.25} $ & $   <-0.426$ & $   -0.57^{+0.10}_{-0.26} $ \\

$w_q$ & $   <-0.476 $ & $   <-0.972$  & $   -0.55^{+0.26}_{-0.29} $ & $   -0.834^{+0.084}_{-0.054} $ & $   -0.55^{+0.26}_{-0.31}$ & $   -0.838^{+0.094}_{-0.056} $ \\

$N_{\rm eff}$ & $   3.04\pm0.20$ & $   2.98\pm0.19$ & $   3.046$  & $  3.046$  & $   3.05\pm0.20$  & $   2.95\pm0.19$  \\

$\Omega_k$ & $   -0.077^{+0.042}_{-0.023}$ & $   0.0021\pm 0.0023$  & $   -0.072^{+0.047}_{-0.021}$  & $  0.0000\pm0.0023$  & $   -0.072^{+0.050}_{-0.020}$  & $   0.0004\pm0.0024$  \\

\hline

$\Omega_{m0}$ & $  0.69^{+0.12}_{-0.19}$ & $    0.3123\pm0.0070$   & $    0.47^{+0.16}_{-0.23}$ & $    0.139^{+0.035}_{-0.086}$ & $    0.46_{-    0.23}^{+    0.16}$  & $    0.144_{-    0.091}^{+    0.035}$  \\

$\sigma_8$ & $    0.694\pm0.048$ & $    0.802\pm0.012$   & $    1.08^{+0.09}_{-0.43}$ & $    2.2^{+0.5}_{-1.3}$ & $    1.11^{+0.09}_{-0.45}$ & $    2.2_{-    1.4}^{+    0.5}$  \\

$H_0 {\rm[km/s/Mpc]}$ & $   46.2^{+4.6}_{-6.4}$ & $   67.4\pm1.1$  & $   48^{+6}_{-8}$ & $   68.33\pm0.84$  & $   49^{+6}_{-8}$  & $   67.9\pm1.2$   \\

$S_8$ & $    1.037\pm0.056$ &  $    0.819\pm0.014$  & $    1.21^{+0.08}_{-0.18}$ & $    1.33^{+0.20}_{-0.44}$ & $    1.22^{+0.09}_{-0.19}$  & $    1.32^{+0.20}_{-0.45}$  \\

$r_{\rm{drag}}$ [Mpc] & $  147.4\pm1.9$ &  $  147.9\pm1.9$  & $  147.33\pm0.30$  & $  147.15\pm0.30$  & $  147.3\pm1.9$ & $  148.1\pm1.9$  \\

\hline\hline                                               
\end{tabular}          }                                     
\caption{68\% CL constraints on various free and derived parameters of the parameter spaces $\mathcal{P}_{17} \equiv\Bigl\{\Omega_{b}h^2, \Omega_{c}h^2, 100\theta_{\rm MC}, \tau, n_{s}, \ln[10^{10}A_{s}],
N_{\rm eff}, w_q,  \Omega_k \Bigr\}$, $\mathcal{P}_{20} \equiv\Bigl\{\Omega_{b}h^2, \Omega_{c}h^2, 100\theta_{\rm MC}, \tau, n_{s}, \ln[10^{10}A_{s}],
w_q, \xi_-, \Omega_k \Bigr\}$ and $\mathcal{P}_{22} \equiv\Bigl\{\Omega_{b}h^2, \Omega_{c}h^2, 100\theta_{\rm MC}, \tau, n_{s}, \ln[10^{10}A_{s}],
N_{\rm eff}, w_q, \xi_-, \Omega_k \Bigr\}$.  }
\label{tabG}                                              
\end{table*}  



\begin{table*}                                                                                                                 \resizebox{\textwidth}{!}{   
\begin{tabular}{c|cc|cc|ccc}                                                                                               
\hline\hline                                                                                    Model & $\mathcal{P}_{18}$ & $\mathcal{P}_{18}$ & $\mathcal{P}_{21}$ & $\mathcal{P}_{21}$ & $\mathcal{P}_{23}$ & $\mathcal{P}_{23}$ \\ \hline                                
 & CMB & all & CMB & all & CMB & all \\ \hline

$\Omega_{\rm b} h^2$ & $    0.02259\pm0.00025$ & $    0.02229\pm0.00023$  & $    0.02262\pm0.00017$ & $    0.02239\pm0.00016$ & $    0.02261\pm0.00025$ & $    0.02231\pm0.00023$  \\

$\Omega_{\rm c} h^2$ & $    0.1178\pm0.0029$ & $    0.1182\pm0.0030$  & $    0.131^{+0.006}_{-0.011}$ & $    0.134^{+0.007}_{-0.012}$ & $    0.131^{+0.007}_{-0.011}$  & $    0.132_{-    0.011}^{+    0.008}$ \\

$100\theta_{MC}$ & $    1.04120\pm0.00044$ & $    1.04112\pm0.00044$  & $    1.04049_{-    0.00049}^{+    0.00062}$ & $    1.04021\pm0.00056$ & $    1.04052_{-    0.00058}^{+    0.00067}$ & $    1.04039_{-    0.00061}^{+    0.00067}$ \\

$\tau$ & $    0.0488^{+0.0083}_{-0.0075}$ & $    0.0540\pm0.0079$  & $    0.0477^{+0.0084}_{-0.0071}$  & $    0.0549\pm0.0076$  & $    0.0478^{+0.0084}_{-0.0074}$  & $    0.0544\pm0.0077$  \\

$n_s$ & $    0.9702\pm0.0092$ & $    0.9614\pm0.0090$  & $    0.9711\pm0.0046$ & $    0.9661\pm0.0045$  & $    0.9708\pm0.0095$ & $    0.9624\pm0.0087$ \\

${\rm{ln}}(10^{10} A_s)$ & $    3.028^{+0.019}_{-0.018}$ & $    3.039\pm0.019$   & $    3.026^{+0.017}_{-0.015}$ & $    3.045\pm0.016$ & $    3.025\pm0.019$ & $    3.040\pm0.018$ \\

$\xi$ & $   0$ & $   0$  & $   <0.288 $ & $   <0.188 $ & $   <0.287$ & $   <0.180 $ \\

$w_p$ & $   >-2.0$ & $   >-1.06$  & $   <-1.94 $ & $   -1.078^{+0.056}_{-0.035} $ & $   >-1.92$ & $   -1.079^{+0.054}_{-0.036} $ \\

$N_{\rm eff}$ & $   3.03\pm0.20$ & $   2.94\pm0.19$ & $   3.046$  & $  3.046$  & $   3.04\pm0.20$  & $   2.95^{+0.18}_{-0.20}$  \\

$\Omega_k$ & $   -0.026^{+0.018}_{-0.008}$ & $   0.0001\pm 0.0023$  & $   -0.031^{+0.020}_{-0.009}$  & $  -0.0002\pm0.0022$  & $   -0.031^{+0.020}_{-0.010}$  & $   0.0002\pm0.0022$  \\

\hline

$\Omega_{m0}$ & $  0.31^{+0.07}_{-0.14}$ & $    0.3053\pm0.0075$   & $    0.37^{+0.11}_{-0.15}$ & $    0.335^{+0.019}_{-0.025}$ & $    0.38_{-    0.15}^{+    0.11}$  & $    0.335_{-    0.024}^{+    0.018}$  \\

$\sigma_8$ & $    0.93^{+0.08}_{-0.15}$ & $    0.818^{+0.012}_{-0.014}$   & $    0.82^{+0.06}_{-0.12}$ & $    0.748\pm0.044$ & $    0.82^{+0.06}_{-0.12}$ & $    0.748_{-    0.042}^{+    0.051}$  \\

$H_0 {\rm[km/s/Mpc]}$ & $   71^{+9}_{-20}$ & $   68.0\pm1.2$  & $   68^{+8}_{-20}$ & $   68.34\pm0.82$  & $   67^{+8}_{-20}$  & $   67.9\pm1.2$   \\

$S_8$ & $    0.907\pm0.066$ &  $    0.825\pm0.014$  & $    0.879^{+0.088}_{-0.074}$ & $    0.790\pm0.026$ & $    0.883^{+0.088}_{-0.074}$  & $    0.789\pm0.025$  \\

$r_{\rm{drag}}$ [Mpc] & $  147.5\pm1.9$ &  $  148.2\pm1.9$  & $  147.37\pm0.30$  & $  147.16\pm0.30$  & $  147.5\pm2.0$ & $  148.1\pm1.9$  \\

\hline\hline                                               
\end{tabular}          }                                     
\caption{68\% CL constraints on various free and derived parameters of the parameter spaces $\mathcal{P}_{18} \equiv\Bigl\{\Omega_{b}h^2, \Omega_{c}h^2, 100\theta_{\rm MC}, \tau, n_{s}, \ln[10^{10}A_{s}],
N_{\rm eff}, w_p,  \Omega_k \Bigr\}$, $\mathcal{P}_{21} \equiv\Bigl\{\Omega_{b}h^2, \Omega_{c}h^2, 100\theta_{\rm MC}, \tau, n_{s}, \ln[10^{10}A_{s}],
w_p, \xi_+, \Omega_k \Bigr\}$ and $\mathcal{P}_{23} \equiv\Bigl\{\Omega_{b}h^2, \Omega_{c}h^2, 100\theta_{\rm MC}, \tau, n_{s}, \ln[10^{10}A_{s}],
N_{\rm eff}, w_p, \xi_+, \Omega_k \Bigr\}$. }
\label{tabH}                                              
\end{table*}  


\begin{table*}                                                                                        \resizebox{0.85\textwidth}{!}{ 

\begin{tabular}{c|c|c|c|c|ccc}                                                                                               
\hline\hline                                                                                    Model & $\mathcal{P}_3$ & $\mathcal{P}_4$ & $\mathcal{P}_{10}$ & $\mathcal{P}_{11}$ & $\mathcal{P}_{13}$ \\ \hline                                
 & CMB + R20 & CMB + R20 & CMB + R20 & CMB + R20 & CMB + R20 \\ \hline

$\Omega_{\rm b} h^2$  & $    0.02237\pm0.00015$  & $    0.02238\pm0.00014$ & $    0.02240\pm0.00015$ & $    0.02237\pm0.00015$ & $    0.02262\pm0.00017$  \\

$\Omega_{\rm c} h^2$  & $    0.1201\pm0.0013$  & $    0.039^{+0.018}_{-0.024}$ & $    <0.0276$ & $    0.133^{+0.006}_{-0.012}$  & $    0.1181\pm0.0015$ \\

$100\theta_{MC}$  & $    1.04090\pm0.00031$  & $    1.0463\pm0.0017$ & $    1.0479^{+0.0020}_{-0.0010}$ & $    1.04021^{+0.00061}_{-0.00051}$ & $    1.04115\pm0.00033$ \\

$\tau$  & $    0.0540\pm0.0080$  & $    0.0533\pm0.0080$  & $    0.0539\pm0.0077$  & $    0.0536\pm0.0077$  & $    0.0489^{+0.0082}_{-0.0072}$  \\

$n_s$  & $    0.9650\pm0.0044$  & $    0.9653\pm0.0042$ & $    0.9658\pm0.0042$  & $    0.9647\pm0.0043$ & $    0.9707\pm0.0048$ \\

${\rm{ln}}(10^{10} A_s)$ & $    3.044 \pm 0.016$   & $    3.042\pm0.016$ & $    3.043\pm0.016$ & $    3.043\pm0.016$ & $    3.029^{+0.017}_{-0.016}$ \\

$w$ & $   -1.200\pm0.048$  & $   -0.999 $ & $   -0.939^{+0.020}_{-0.054} $ & $   -1.240\pm0.055$ & $   -1.88^{+0.42}_{-0.23} $ \\

$\xi$  & $   0$  & $   -0.60^{+0.11}_{-0.16} $ & $   -0.71^{+0.05}_{-0.012} $ & $   <0.162$ & $   0 $ \\

$\Omega_k$ & $   0$  & $   0 $ & $   0 $ & $   0$ & $   -0.0203^{+0.0063}_{-0.0072} $ \\

\hline

$\Omega_{m0}$ & $    0.2660\pm0.0097$   & $    0.117_{-    0.050}^{+    0.033}$ & $    0.085^{+0.013}_{-0.040}$ & $    0.291_{-    0.023}^{+    0.015}$  & $    0.2642\pm0.0097$  \\

$\sigma_8$& $    0.867\pm0.016$   & $    2.3_{-    1.1}^{+    0.4}$ & $    3.2^{+0.9}_{-1.5}$ & $    0.794_{-    0.041}^{+    0.062}$ & $    0.953\pm0.035$  \\

$H_0 {\rm[km/s/Mpc]}$ & $   73.4\pm1.3$  & $   73.4_{-    1.1}^{+    1.4}$ & $   72.8^{+1.2}_{-1.1}$  & $   73.3 \pm 1.3$  & $ 73.2\pm1.3$   \\

$S_8$  &  $    0.816\pm0.015$  & $    1.34^{+0.15}_{-0.33}$ & $    1.58\pm0.28$ & $    0.780_{-    0.026}^{+    0.032}$  & $    0.894\pm0.031$  \\

$r_{\rm{drag}}$ [Mpc]  &  $  147.06\pm0.29$  & $  147.07\pm0.29$  & $  147.10\pm0.29$  & $  147.05\pm0.29$ & $  147.34\pm0.32$  \\

\hline\hline                                               
\end{tabular}          }                                     
\caption{68\% CL constraints on various free and derived parameters of the parameter spaces  $\mathcal{P}_3 \equiv\Bigl\{\Omega_{b}h^2, \Omega_{c}h^2, 100\theta_{\rm MC}, \tau, n_{s}, \ln[10^{10}A_{s}],
w_p,  \Bigr\}$, $\mathcal{P}_4 \equiv\Bigl\{\Omega_{b}h^2, \Omega_{c}h^2, 100\theta_{\rm MC}, \tau, n_{s}, \ln[10^{10}A_{s}],
\xi_{-},  \Bigr\}$, $\mathcal{P}_{10} \equiv\Bigl\{\Omega_{b}h^2, \Omega_{c}h^2, 100\theta_{\rm MC}, \tau, n_{s}, \ln[10^{10}A_{s}],
w_q,  \xi_- \Bigr\}$, $\mathcal{P}_{11} \equiv\Bigl\{\Omega_{b}h^2, \Omega_{c}h^2, 100\theta_{\rm MC}, \tau, n_{s}, \ln[10^{10}A_{s}],
w_p,  \xi_+ \Bigr\}$ and $\mathcal{P}_{13} \equiv\Bigl\{\Omega_{b}h^2, \Omega_{c}h^2, 100\theta_{\rm MC}, \tau, n_{s}, \ln[10^{10}A_{s}],
w_p, \Omega_k \Bigr\}$ for CMB + R20 dataset only. }
\label{tabI}                                              
\end{table*}  


\begin{table*}                                                                                                                 \resizebox{\textwidth}{!}{   
\begin{tabular}{c|cc|cc|ccc}                                                                                               
\hline\hline                                                                                    Model & $\mathcal{P}_7$ & $\mathcal{P}_7$ & $\mathcal{P}_8$ & $\mathcal{P}_8$ & $\mathcal{P}_{18}$ & $\mathcal{P}_{18}$ \\ \hline                                
 & CMB + R20 & CMB + BAO + R20 & CMB + R20 & CMB + BAO + R20 & CMB + R20 & CMB + BAO + R20 \\ \hline

$\Omega_{\rm b} h^2$ & $    0.02224\pm0.00022$ & $    0.02236\pm0.00020$  & $    0.02230\pm0.00019$ & $    0.02253\pm0.00018$ & $    0.02260\pm0.00025$ & $    0.02247\pm0.00022$  \\

$\Omega_{\rm c} h^2$ & $    0.1179\pm0.0030$ & $    0.1207\pm0.0028$  & $    <0.0421$ & $    0.0997^{+0.013}_{-0.011}$ & $    0.1180\pm0.0031$  & $    0.1203\pm0.0029$ \\

$100\theta_{MC}$ & $    1.04118\pm0.00044$ & $    1.04085\pm0.00041$  & $    1.0470^{+0.0023}_{-0.0017}$ & $    1.04198^{+0.00073}_{-0.00092}$ & $    1.04119\pm0.00045$ & $    1.04090\pm0.00042$ \\

$\tau$ & $    0.0532\pm0.0079$ & $    0.0539\pm0.0078$  & $    0.0528\pm0.0076$  & $    0.0555\pm0.0079$  & $    0.0479\pm0.0082$  & $    0.0542\pm0.0078$  \\

$n_s$ & $    0.9591\pm0.0084$ & $    0.9643\pm0.0079$  & $    0.9620\pm0.0073$ & $    0.9718\pm0.0066$  & $    0.9701\pm0.0093$ & $    0.9682\pm0.0083$ \\

${\rm{ln}}(10^{10} A_s)$ & $    3.036\pm0.019$ & $    3.045 \pm 0.018$   & $    3.037\pm0.018$ & $    3.053\pm0.018$ & $    3.026\pm0.019$ & $    3.045\pm0.018$ \\

$w$ & $  -1.238\pm0.070$ & $   -1.200\pm0.048$  & $   -1 $ & $   -1 $ & $   -1.90^{+0.41}_{-0.25}$ & $   -1.188^{+0.080}_{-0.071} $ \\

$N_{\rm eff}$ & $   2.89\pm0.19$ & $   3.06\pm0.18$ & $   2.95\pm0.16 $  & $  3.22\pm0.16$  & $   3.04\pm0.20$  & $   3.10\pm0.18$  \\

$\xi$ & $   0$ & $   0$  & $   -0.65^{+0.09}_{-0.18} $ & $   -0.20^{+0.10}_{0.09} $ & $   0$ & $   0 $ \\

$\Omega_k$ & $   0$ & $   0$  & $   0 $ & $   0 $ & $   -0.0207^{+0.0065}_{-0.0075}$ & $   -0.0028\pm0.0024 $ \\

\hline

$\Omega_{m0}$ & $  0.262^{+0.010}_{-0.012}$ & $    0.2818\pm0.0090$   & $    0.106_{-    0.057}^{+    0.022}$ & $    0.247^{+0.029}_{-0.024}$ & $    0.264_{-    0.012}^{+    0.010}$  & $    0.2782\pm0.0096$  \\

$\sigma_8$ & $    0.869\pm0.016$ & $    0.853\pm0.016$   & $    2.6_{-    1.4}^{+    0.5}$ & $    1.01^{+0.07}_{-0.13}$ & $    0.954\pm0.035$ & $    0.861\pm0.018$  \\

$H_0 {\rm[km/s/Mpc]}$ & $   73.3\pm1.3$ & $   71.4\pm1.1$  & $   73.2\pm1.2 $ & $   70.50\pm0.90$  & $   73.2 \pm 1.3$  & $ 71.8\pm1.1$   \\

$S_8$ & $    0.812\pm0.015$ &  $    0.826\pm0.013$  & $    1.42^{+0.25}_{-0.57}$ & $    0.909^{+0.036}_{-0.059}$ & $    0.895\pm0.034$  & $    0.829\pm0.014$  \\

$r_{\rm{drag}}$ [Mpc] & $  148.6\pm1.9$ &  $  146.9\pm1.8$  & $  148.1\pm 1.7$  & $  145.4\pm1.5$  & $  147.5\pm2.0$ & $  146.7\pm1.8$  \\

\hline\hline                                               
\end{tabular}          }                                     
\caption{68\% CL constraints on various free and derived parameters of the parameter spaces $\mathcal{P}_7 \equiv\Bigl\{\Omega_{b}h^2, \Omega_{c}h^2, 100\theta_{\rm MC}, \tau, n_{s}, \ln[10^{10}A_{s}],
N_{\rm eff}, w_p \Bigr\}$, $\mathcal{P}_{8} \equiv\Bigl\{\Omega_{b}h^2, \Omega_{c}h^2, 100\theta_{\rm MC}, \tau, n_{s}, \ln[10^{10}A_{s}],
N_{\rm eff},  \xi_- \Bigr\}$ and $\mathcal{P}_{18} \equiv\Bigl\{\Omega_{b}h^2, \Omega_{c}h^2, 100\theta_{\rm MC}, \tau, n_{s}, \ln[10^{10}A_{s}],
N_{\rm eff}, w_p,  \Omega_k \Bigr\}$ when R20 is considered. }
\label{tabJ}                                              
\end{table*}  


\begin{table*}                                                                                                                 \resizebox{0.85\textwidth}{!}{   
\begin{tabular}{c|cc|c|ccc}                                                                                               
\hline\hline                                                                                    Model & $\mathcal{P}_{16}$ & $\mathcal{P}_{16}$ & $\mathcal{P}_{21}$ & $\mathcal{P}_{23}$ & $\mathcal{P}_{23}$ \\ \hline                                
 & CMB + R20 & CMB + BAO + R20 & CMB + R20 & CMB + R20 & CMB + BAO + R20 \\ \hline

$\Omega_{\rm b} h^2$ & $    0.02224\pm0.00021$ & $    0.02254\pm0.00019$  & $    0.02262\pm0.00017$ & $    0.02262\pm0.00025$ & $    0.02257\pm0.00021$  \\

$\Omega_{\rm c} h^2$ & $    0.130^{+0.007}_{-0.012}$ & $    0.137^{+0.008}_{-0.012}$  & $    0.132^{+0.007}_{-0.012}$ & $    0.132^{+0.008}_{-0.012}$ & $    0.1368^{+0.008}_{-0.012}$  \\

$100\theta_{MC}$ & $    1.04051^{+0.00068}_{-0.00060}$ & $    1.03992^{+0.00067}_{-0.00060}$  & $    1.04045^{+0.00065}_{-0.00053}$ & $    1.04043^{+0.00069}_{-0.00062}$ & $    1.03992^{+0.00070}_{-0.00062}$ \\

$\tau$ & $    0.0530\pm0.0079$ & $    0.0558\pm0.0079$  & $    0.0480^{+0.0085}_{-0.0073}$  & $    0.0486\pm0.0083$  & $    0.0560\pm0.0082$   \\

$n_s$ & $    0.9590\pm0.0084$ & $    0.9724\pm0.0069$  & $    0.9709\pm0.0048$ & $    0.9713\pm0.0093$  & $    0.9729\pm0.0082$ \\

${\rm{ln}}(10^{10} A_s)$ & $    3.036\pm0.019$ & $    3.054 \pm 0.018$   & $    3.026^{+0.018}_{-0.016}$ & $    3.027\pm0.020$ & $    3.054\pm0.018$ \\

$w$ & $  -1.279\pm0.075$ & $   -1.101^{+0.047}_{-0.042}$  & $   -1.99^{+0.40}_{-0.29} $ & $   -2.00^{+0.40}_{-0.26} $ & $   -1.101^{+0.052}_{-0.042}$  \\

$N_{\rm eff}$ & $   2.89\pm0.19$ & $   3.24\pm0.16$ & $   3.046 $  & $  3.05\pm0.20$  & $   3.24\pm0.17$  \\

$\xi$ & $   <0.160$ & $   <0.176$  & $   <0.269 $ & $   <0.279 $ & $   <0.179$  \\

$\Omega_k$ & $   0$ & $   0$  & $   -0.0214^{+0.0053}_{-0.0073} $ & $   -0.0217^{+0.0058}_{-0.0069} $ & $   -0.0001\pm0.0022$ \\

\hline

$\Omega_{m0}$ & $  0.285^{+0.017}_{-0.022}$ & $    0.322^{+0.016}_{-0.025}$   & $    0.290_{-    0.023}^{+    0.017}$ & $    0.291^{+0.019}_{-0.023}$ & $    0.323_{-    0.024}^{+    0.016}$ \\

$\sigma_8$ & $    0.797^{+0.059}_{-0.043}$ & $    0.765^{+0.058}_{-0.041}$   & $    0.861_{-    0.064}^{+    0.070}$ & $    0.860\pm0.065$ & $    0.763^{+0.056}_{-0.042}$  \\

$H_0 {\rm[km/s/Mpc]}$ & $   73.3\pm1.3$ & $   70.41\pm0.91$  & $   73.1\pm1.3 $ & $   73.1\pm1.3$  & $   70.43 \pm 0.88$   \\

$S_8$ & $    0.776^{+0.032}_{-0.026}$ &  $    0.791^{+0.030}_{-0.026}$  & $    0.844\pm0.043$ & $    0.844\pm0.044$ & $    0.790^{+0.030}_{-0.026}$   \\

$r_{\rm{drag}}$ [Mpc] & $  148.6\pm1.9$ &  $  145.3\pm1.6$  & $  147.36\pm 0.31$  & $  147.4\pm1.9$  & $  145.2\pm1.6$  \\

\hline\hline                                               
\end{tabular}          }                                     
\caption{68\% CL constraints on various free and derived parameters of the parameter spaces $\mathcal{P}_{16} \equiv\Bigl\{\Omega_{b}h^2, \Omega_{c}h^2, 100\theta_{\rm MC}, \tau, n_{s}, \ln[10^{10}A_{s}],
N_{\rm eff}, w_p, \xi_+ \Bigr\}$, $\mathcal{P}_{21} \equiv\Bigl\{\Omega_{b}h^2, \Omega_{c}h^2, 100\theta_{\rm MC}, \tau, n_{s}, \ln[10^{10}A_{s}],
w_p, \xi_+, \Omega_k \Bigr\}$ and $\mathcal{P}_{23} \equiv\Bigl\{\Omega_{b}h^2, \Omega_{c}h^2, 100\theta_{\rm MC}, \tau, n_{s}, \ln[10^{10}A_{s}],
N_{\rm eff}, w_p, \xi_+, \Omega_k \Bigr\}$ when R20 is considered. }
\label{tabK}                                              
\end{table*}  


\begin{table*}                                                                                                                 \resizebox{0.7\textwidth}{!}{   
\begin{tabular}{c|cc|c|ccc}                                                                                               
\hline\hline                                                                                    Model & $\mathcal{P}_{15}$ & $\mathcal{P}_{15}$ & $\mathcal{P}_{19}$ & $\mathcal{P}_{22}$ \\ \hline                                
 & CMB + R20 & CMB + BAO + R20 & CMB + BAO + R20  & CMB + BAO + R20 \\ \hline

$\Omega_{\rm b} h^2$ & $    0.02238\pm0.00019$ & $    0.02256\pm0.00019$  & $    0.02255\pm0.00021$ & $    0.02257\pm0.00021$   \\

$\Omega_{\rm c} h^2$ & $    <0.0259$ & $    <0.0538$  & $    0.099^{+0.015}_{-0.011}$ & $    <0.0569$   \\

$100\theta_{MC}$ & $    1.0480^{+0.0020}_{-0.0009}$ & $    1.0462^{+0.0032}_{-0.0018}$  & $    1.04200^{+0.00073}_{-0.00099}$ & $    1.0460^{+0.0030}_{-0.0020}$  \\

$\tau$ & $    0.0536\pm0.0078$ & $    0.0555\pm0.0079$  & $    0.0556\pm0.0078$  & $    0.0553\pm0.0080$   \\

$n_s$ & $    0.9545\pm0.0073$ & $    0.9727\pm0.0069$  & $    0.9722\pm0.0080$ & $    0.9731\pm0.0079$  \\

${\rm{ln}}(10^{10} A_s)$ & $    3.041\pm0.018$ & $    3.054 \pm 0.018$   & $    3.053\pm0.018$ & $    3.053\pm0.018$  \\

$w$ & $  <-0.928 $ & $   -0.859^{+0.081}_{-0.050}$  & $   -1 $ & $   -0.862^{+0.080}_{-0.058} $   \\

$N_{\rm eff}$ & $   3.01\pm0.17$ & $   3.24\pm0.17$ & $   3.23\pm0.17 $  & $  3.25\pm0.18$    \\

$\xi$ & $   -0.72^{+0.05}_{-0.13}$ & $   -0.59^{+0.09}_{-0.23}$  & $   -0.20^{+0.12}_{-0.09} $ & $   -0.58^{+0.10}_{-0.24} $ \\

$\Omega_k$ & $   0$ & $   0$  & $   -0.0003\pm0.0022 $ & $   0.0000\pm0.0022 $  \\

\hline

$\Omega_{m0}$ & $  0.084^{+0.011}_{-0.039}$ & $    0.129^{+0.032}_{-0.079}$   & $    0.247_{-    0.023}^{+    0.032}$ & $    0.133^{+0.035}_{-0.081}$ \\

$\sigma_8$ & $    3.3^{+0.9}_{-1.5}$ & $    2.4^{+0.5}_{-1.4}$   & $    1.02_{-    0.15}^{+    0.07}$ & $    2.3^{+0.5}_{-1.4}$ \\

$H_0 {\rm[km/s/Mpc]}$ & $   72.7\pm1.1$ & $   70.43\pm0.91$  & $   70.47\pm0.92 $ & $   70.45\pm0.89$   \\

$S_8$ & $    1.59^{+0.34}_{-0.27}$ &  $    1.37^{+0.20}_{-0.45}$  & $    0.912^{+0.036}_{-0.067}$ & $    1.34^{+0.20}_{-0.43}$   \\

$r_{\rm{drag}}$ [Mpc] & $  147.5\pm1.7$ &  $  145.2\pm1.6$  & $  145.4\pm 1.6$  & $  145.1\pm1.6$    \\

\hline\hline                                               
\end{tabular}          }                                     
\caption{68\% CL constraints on various free and derived parameters of the parameter spaces $\mathcal{P}_{15} \equiv\Bigl\{\Omega_{b}h^2, \Omega_{c}h^2, 100\theta_{\rm MC}, \tau, n_{s}, \ln[10^{10}A_{s}],
N_{\rm eff}, w_q, \xi_- \Bigr\}$, $\mathcal{P}_{19} \equiv\Bigl\{\Omega_{b}h^2, \Omega_{c}h^2, 100\theta_{\rm MC}, \tau, n_{s}, \ln[10^{10}A_{s}],
N_{\rm eff}, \xi_-,  \Omega_k \Bigr\}$ and $\mathcal{P}_{22} \equiv\Bigl\{\Omega_{b}h^2, \Omega_{c}h^2, 100\theta_{\rm MC}, \tau, n_{s}, \ln[10^{10}A_{s}],
N_{\rm eff}, w_q, \xi_-, \Omega_k \Bigr\}$ when R20 is considered. }
\label{tabL}                                              
\end{table*}  

\section{Discussion of the Results}
\label{sec:4}

In this section we summarize the results obtained for the different extended cosmological classes of models, highlighting the indication for new physics beyond the $\Lambda$CDM model or the possible solution of the $H_0$ and/or $S_8$ tension. For each case, we show the constraints obtained from Planck alone (CMB),  CMB + BAO + Pantheon (labeled by `all'), CMB + R20, and CMB +BAO + R20.

\subsection{Planck, BAO and Pantheon}

\begin{figure*}
	\centering
	\includegraphics[width=0.75 \textwidth]{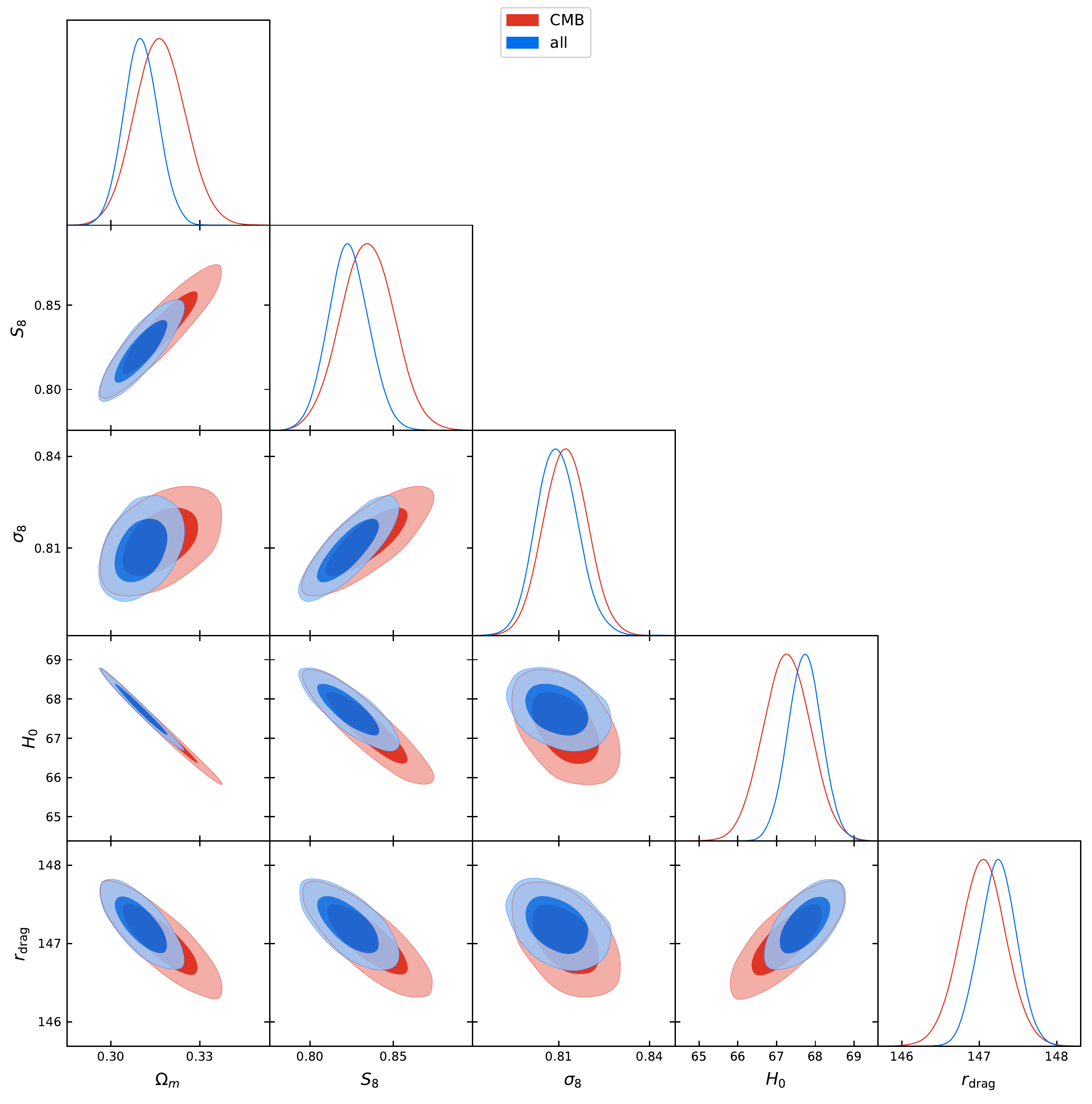}
	\caption{\textit{One dimensional posterior distributions and two dimensional joint contours for the parameter space $\mathcal{P}_0 \equiv\Bigl\{\Omega_{b}h^2, \Omega_{c}h^2, 100\theta_{\rm MC},
  \tau, n_{s}, \ln[10^{10}A_{s}] \Bigr\}$ for  CMB alone and CMB + BAO + Pantheon (referred to as `{\rm all}') dataset. The $H_0$ and $r_{\rm drag}$ parameters are respectively measured in {\rm[km/s/Mpc]} and {\rm[Mpc]} in all the triangular plots of this work. }} 
	\label{fig:0}
\end{figure*}

\begin{figure*}
	\centering
	\includegraphics[width=0.75 \textwidth]{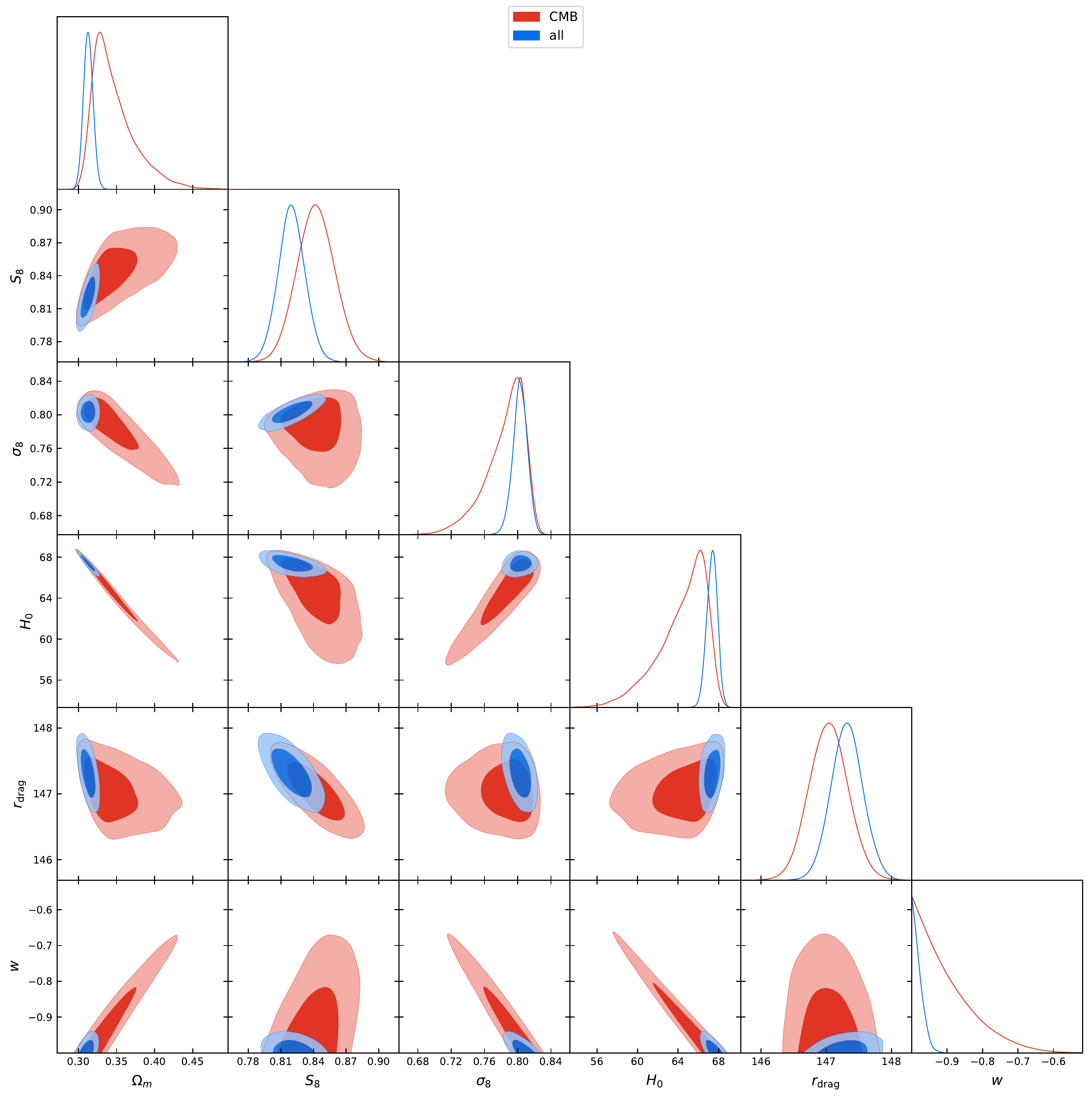}
	\caption{One dimensional posterior distributions and two dimensional joint contours for the parameter space $\mathcal{P}_2 \equiv\Bigl\{\Omega_{b}h^2, \Omega_{c}h^2, 100\theta_{\rm MC}, \tau, n_{s}, \ln[10^{10}A_{s}], 
w_q,  \Bigr\}$ for  CMB alone and CMB + BAO + Pantheon (referred to as `{\rm all}') dataset.}
	\label{fig:2}
\end{figure*}

\begin{figure*}
	\centering
	\includegraphics[width=0.75 \textwidth]{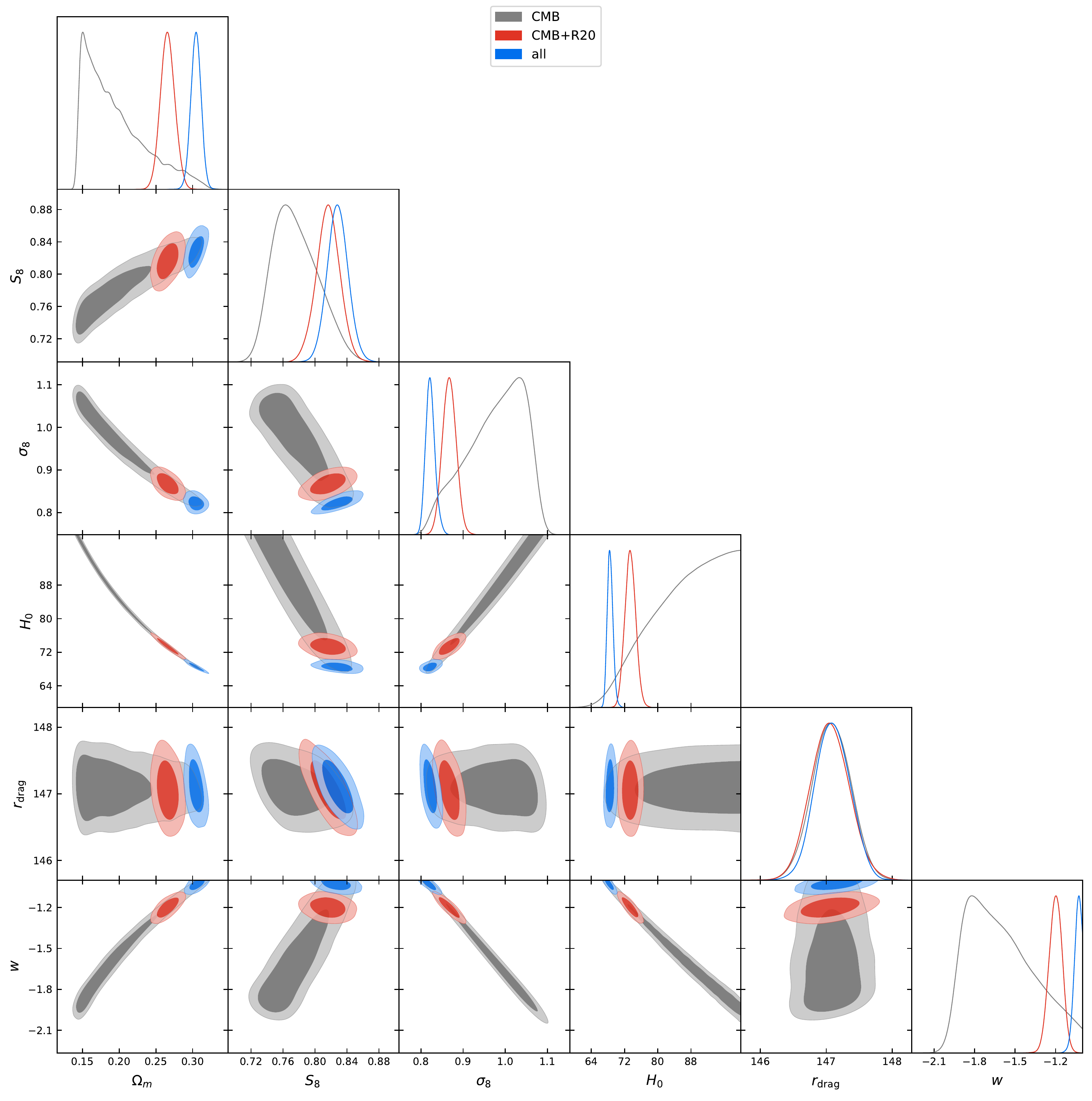}
	\caption{One dimensional posterior distributions and two dimensional joint contours for the parameter space $\mathcal{P}_3 \equiv\Bigl\{\Omega_{b}h^2, \Omega_{c}h^2, 100\theta_{\rm MC}, \tau, n_{s}, \ln[10^{10}A_{s}],
w_p,  \Bigr\}$ for  CMB alone, CMB + R20 and CMB + BAO + Pantheon (referred to as `{\rm all}') dataset.}
	\label{fig:3}
\end{figure*}

In Table~\ref{tabA} we show the results obtained for a $\Lambda$CDM model ($\mathcal{P}_0$) and two $w$CDM classes of models, divided into two different regions based on the DE equation of state parameter: when the DE equation of state lies in the quintessence regime, i.e. $w_q>-1$ ($\mathcal{P}_2$) and when the DE equation of state lies in the phantom regime, i.e. $w_p<-1$ ($\mathcal{P}_3$). We can see that, with respect to the $\Lambda$CDM model, in the $\mathcal{P}_2$ scenario the CMB + BAO + Pantheon dataset combination decreases the tension with the weak lensing data, but there is no indication for $w \neq -1$ and the  Hubble constant tension cannot be eliminated. In the $\mathcal{P}_3$ scenario, for CMB + BAO + Pantheon dataset combination the tension with the weak lensing data is restored with an indication of a phantom DE  at more than $1\sigma$. Additionally, we see a very mild increase in $H_0$ for this dataset compared to the recorded value of $H_0$ in the $\mathcal{P}_0$ for the same dataset and hence the $H_0$ tension is very mildly alleviated in this case. This increment in the Hubble constant is mainly driven by the phantom DE. In order to understand the behaviour of the parameters in these classes of models, we have displayed the one dimensional posterior distributions and the two dimensional joint contours in Figs.~\ref{fig:0} (for $\mathcal{P}_0$), \ref{fig:2} (for $\mathcal{P}_2$), and \ref{fig:3} (for $\mathcal{P}_3$). 

\begin{figure*}
	\centering
	\includegraphics[width=0.75 \textwidth]{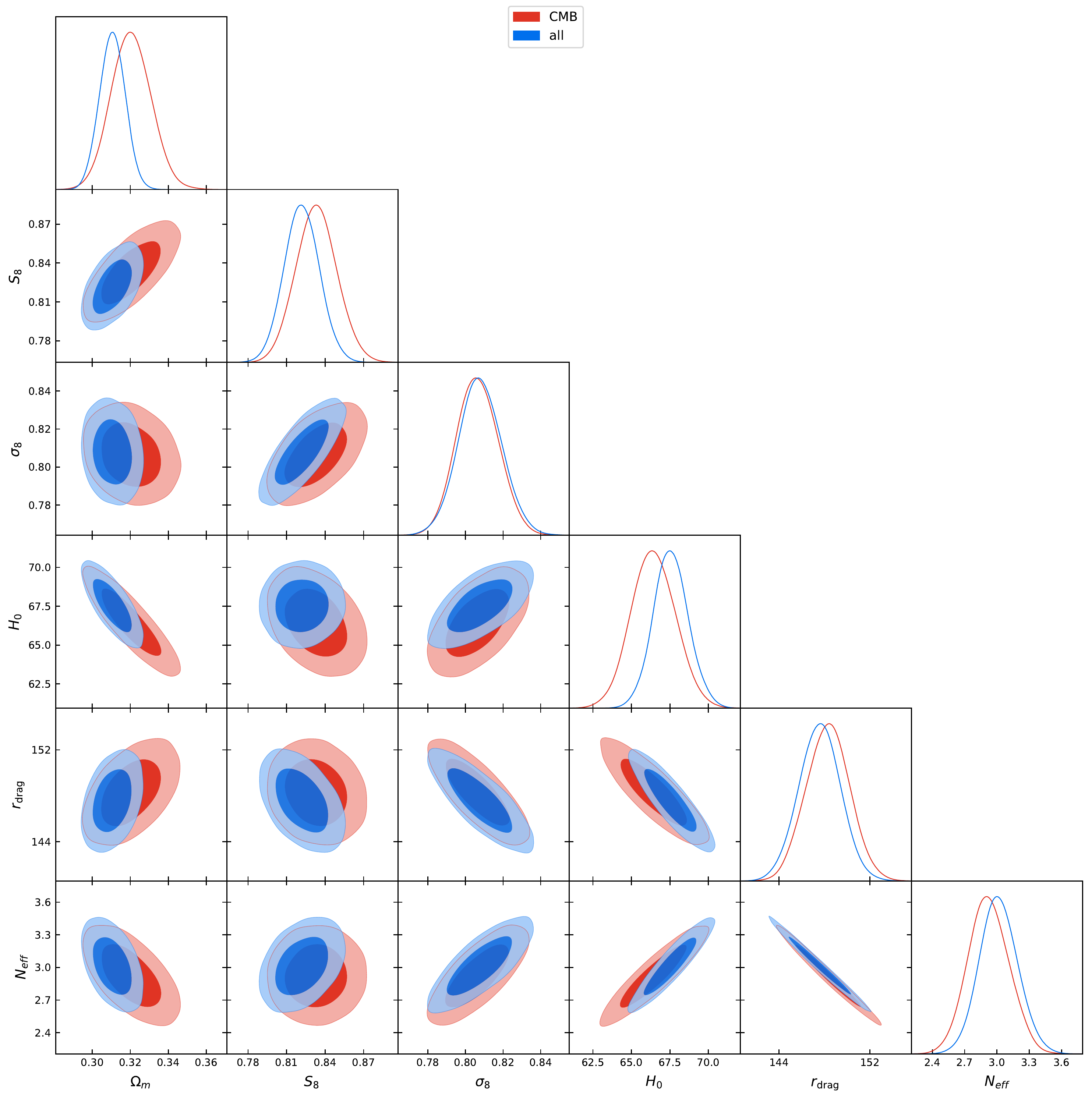}
	\caption{\textit{One dimensional posterior distributions and two dimensional joint contours for the parameter space $\mathcal{P}_1 \equiv\Bigl\{\Omega_{b}h^2, \Omega_{c}h^2, 100\theta_{\rm MC}, \tau, n_{s}, \ln[10^{10}A_{s}], 
N_{\rm eff} \Bigr\}$ for  CMB alone and CMB + BAO + Pantheon (referred to as `{\rm all}') dataset. }}
	\label{fig:1}
\end{figure*}

\begin{figure*}
	\centering
	\includegraphics[width=0.75 \textwidth]{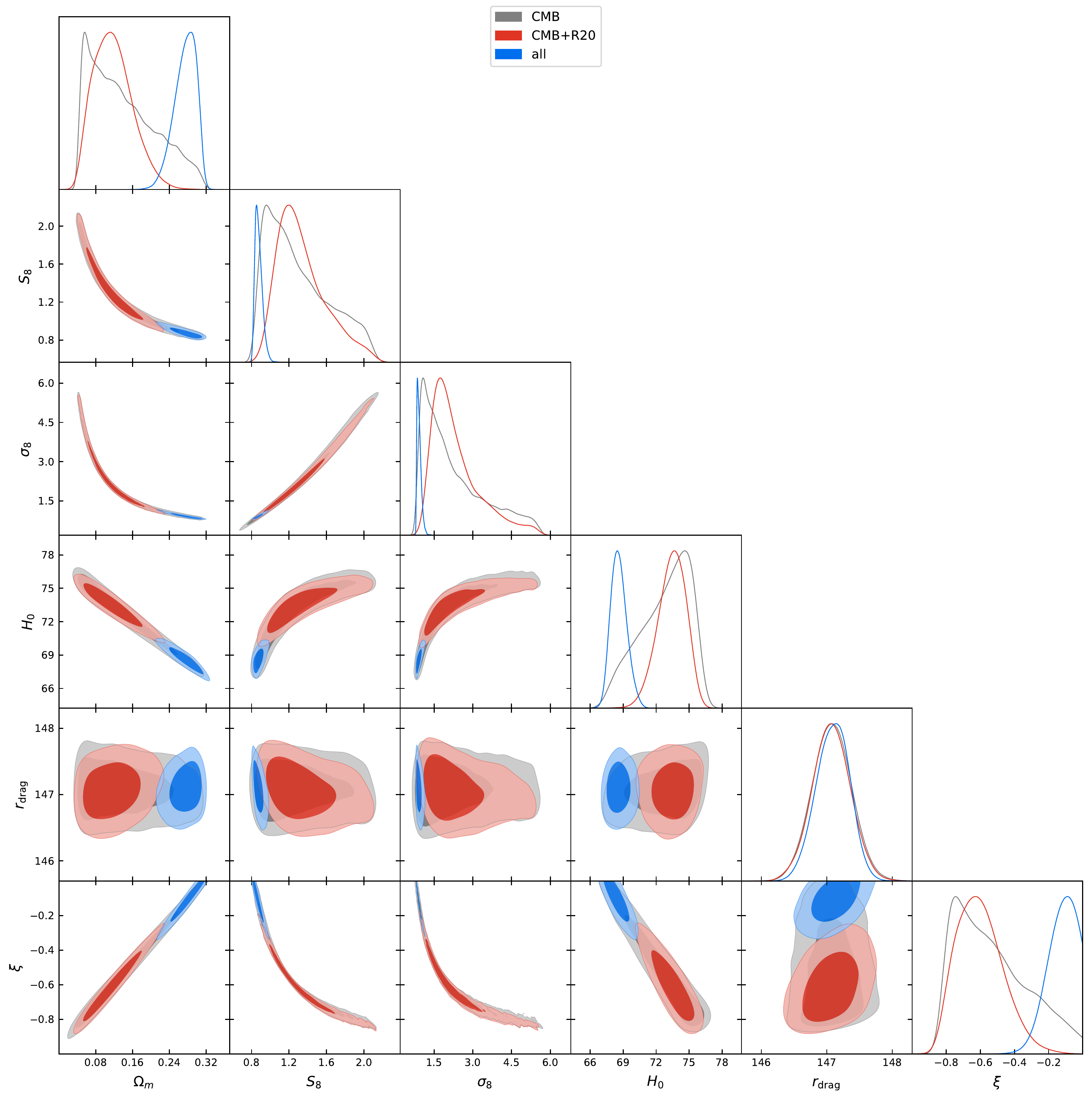}
	\caption{One dimensional posterior distributions and two dimensional joint contours for the parameter space $\mathcal{P}_4 \equiv\Bigl\{\Omega_{b}h^2, \Omega_{c}h^2, 100\theta_{\rm MC}, \tau, n_{s}, \ln[10^{10}A_{s}],
\xi_{-},  \Bigr\}$ for  CMB alone, CMB + R20 and CMB + BAO + Pantheon (referred to as `{\rm all}') dataset.}
	\label{fig:4}
\end{figure*}

\begin{figure*}
	\centering
	\includegraphics[width=0.75 \textwidth]{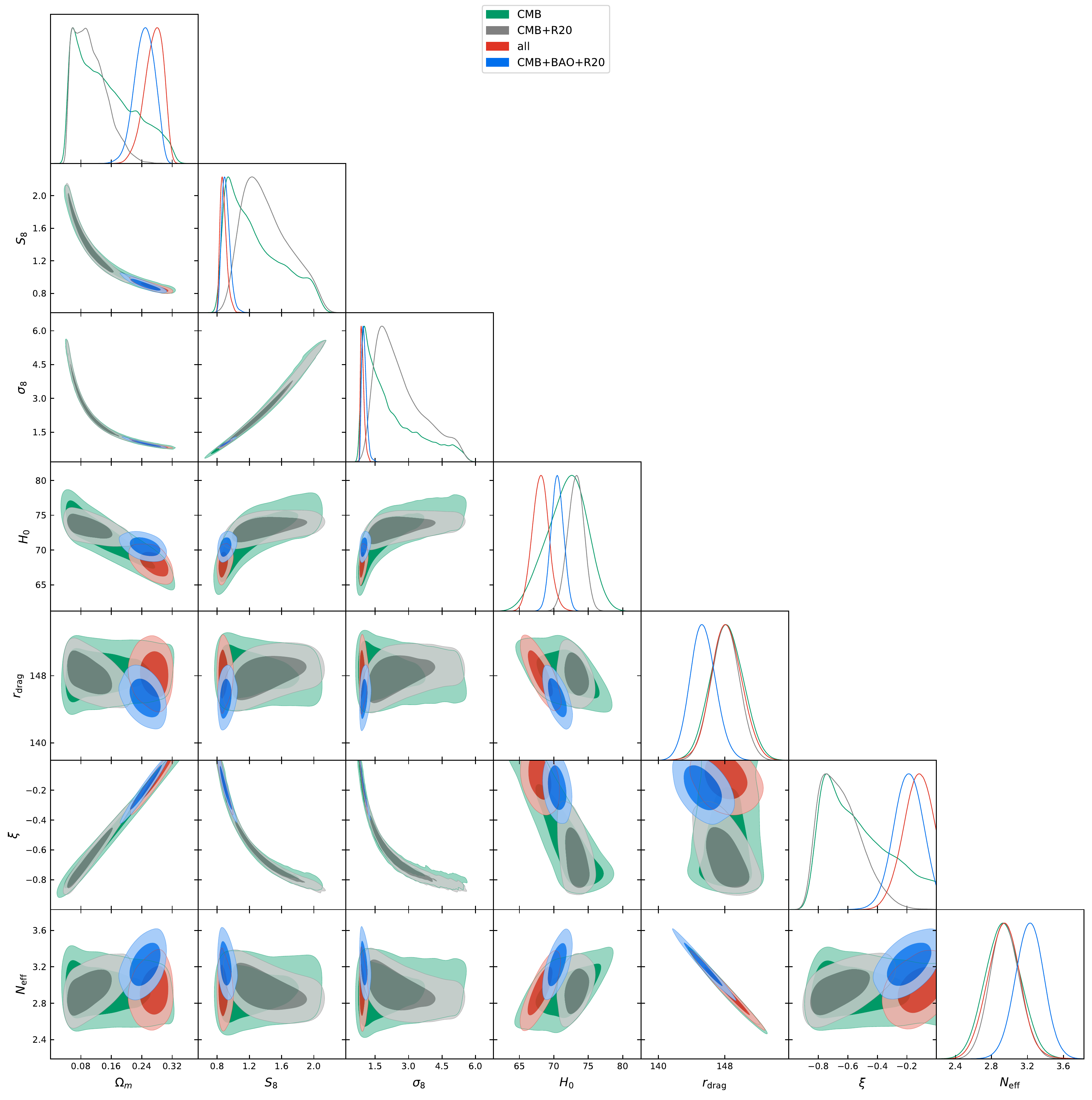}
	\caption{One dimensional posterior distributions and two dimensional joint contours for the parameter space $\mathcal{P}_{8} \equiv\Bigl\{\Omega_{b}h^2, \Omega_{c}h^2, 100\theta_{\rm MC}, \tau, n_{s}, \ln[10^{10}A_{s}],
N_{\rm eff},  \xi_- \Bigr\}$ for CMB alone, CMB + R20, CMB + BAO + Pantheon (referred to as `{\rm all}'), and CMB + BAO + R20 datasets.}
	\label{fig:8}
\end{figure*}

In Table~\ref{tabB} we show the cosmological constraints for the classes of models described by $\Lambda$CDM + $N_{\rm eff}$ model ($\mathcal{P}_1$), IDE ($\mathcal{P}_4$), and IDE + $N_{\rm eff}$ ($\mathcal{P}_8$). As already shown in the literature, extra relativistic degrees of freedom at recombination are not favored by data, and in the $\mathcal{P}_1$ framework the Hubble tension cannot be resolved. On the contrary, $H_0$ is naturally in agreement with SH0ES in the IDE scenario, at the price of a coupling between DM and DE at more than 99\% CL for Planck-data alone. The $H_0$ tension is, however, restored above $3\sigma$ when BAO and Pantheon data are included, i.e for the dataset CMB + BAO + Pantheon, and the evidence for the coupling is reduced to $1\sigma$. The combination of the two extensions in  $\mathcal{P}_8$ leaves unaltered the previous results, because $N_{\rm eff}$ is very well constrained and in agreement with the Standard Model expectation. The tension in the $S_8$ parameter is not addressed by any of these classes of models when BAO and Pantheon are included. In fact, for all these extensions, the $S_8$ parameter takes a higher value when compared to the classes of models characterized by $\mathcal{P}_1$ and $\mathcal{P}_0$ (see Table~\ref{tabA}). In a similar fashion, the one dimensional posterior distributions and the two dimensional joint contours for these classes of models are shown in Figs.~\ref{fig:1} (for $\mathcal{P}_1$), \ref{fig:4} (for $\mathcal{P}_4$), and \ref{fig:8} (for $\mathcal{P}_8$).

\begin{figure*}
	\centering
	\includegraphics[width=0.75 \textwidth]{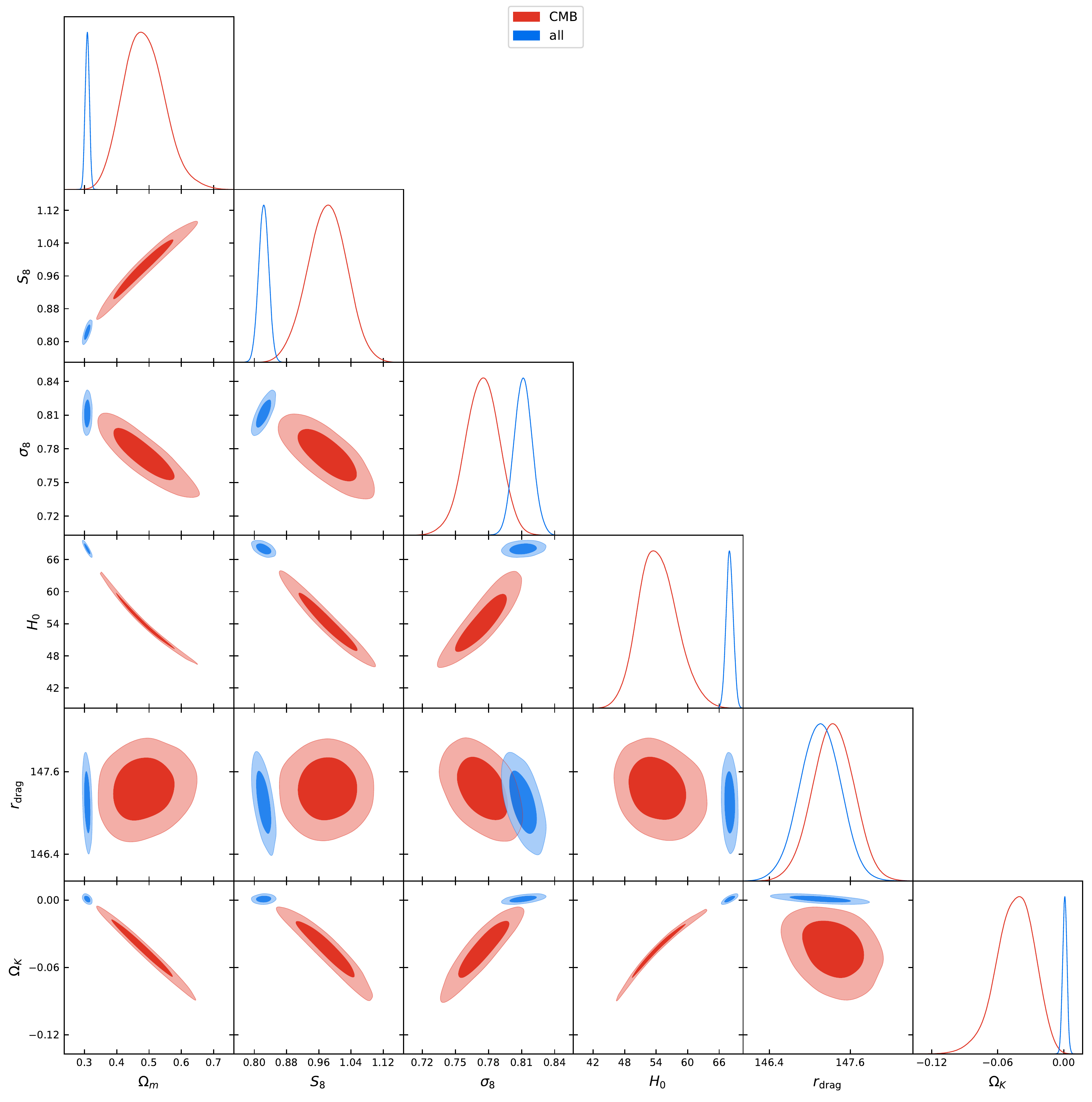}
	\caption{One dimensional posterior distributions and two dimensional joint contours for the parameter space $\mathcal{P}_5 \equiv\Bigl\{\Omega_{b}h^2, \Omega_{c}h^2, 100\theta_{\rm MC}, \tau, n_{s}, \ln[10^{10}A_{s}],
\Omega_k,  \Bigr\}$ for  CMB alone, and CMB + BAO + Pantheon (referred to as `{\rm all}') dataset.}
	\label{fig:5}
\end{figure*}

\begin{figure*}
	\centering
	\includegraphics[width=0.75 \textwidth]{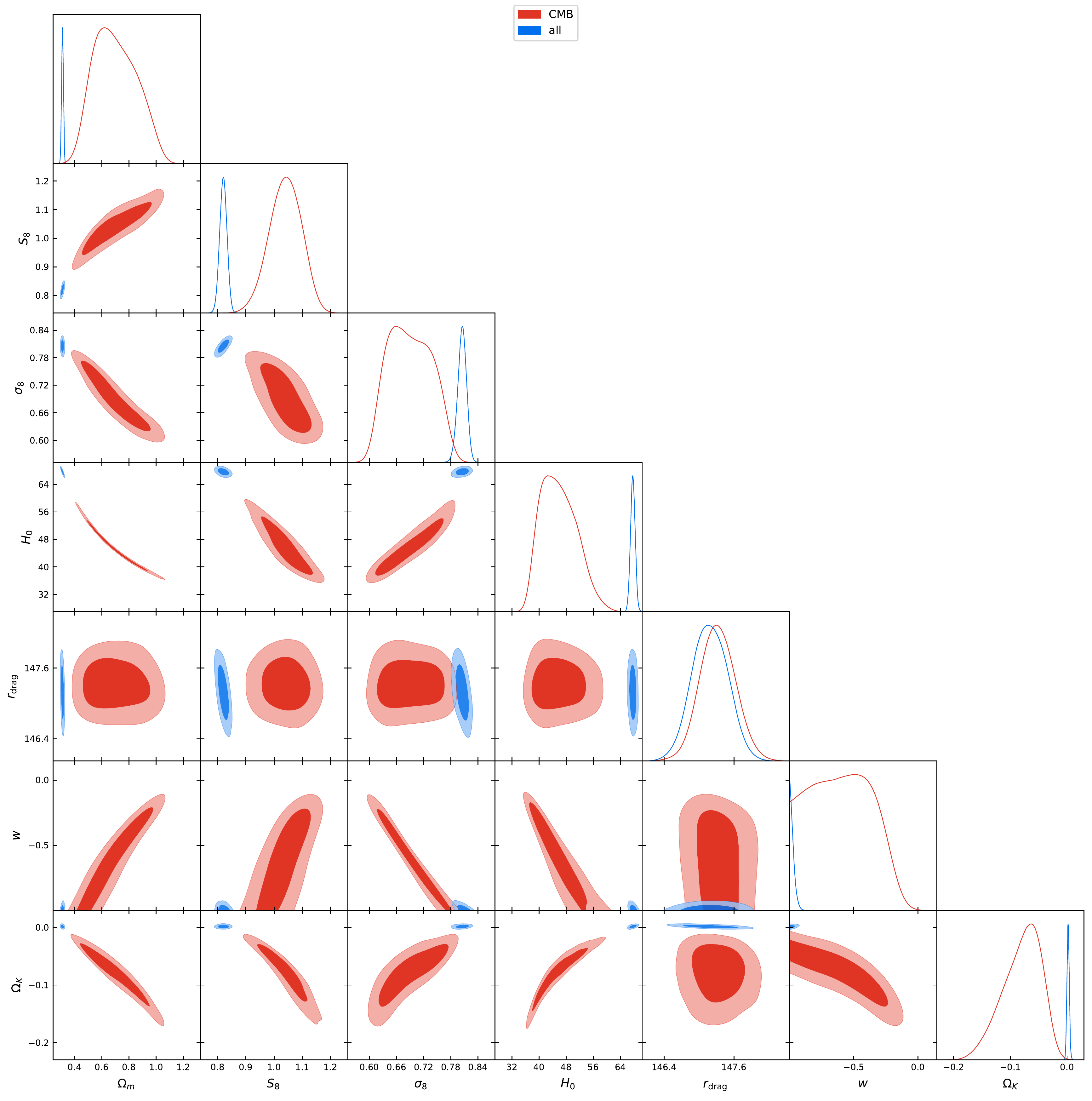}
	\caption{One dimensional posterior distributions and two dimensional joint contours for the parameter space $\mathcal{P}_{12} \equiv\Bigl\{\Omega_{b}h^2, \Omega_{c}h^2, 100\theta_{\rm MC}, \tau, n_{s}, \ln[10^{10}A_{s}],
w_q, \Omega_k \Bigr\}$ for  CMB alone and CMB + BAO + Pantheon (referred to as `{\rm all}') dataset. }
	\label{fig:12}
\end{figure*}

\begin{figure*}
	\centering
	\includegraphics[width=0.75 \textwidth]{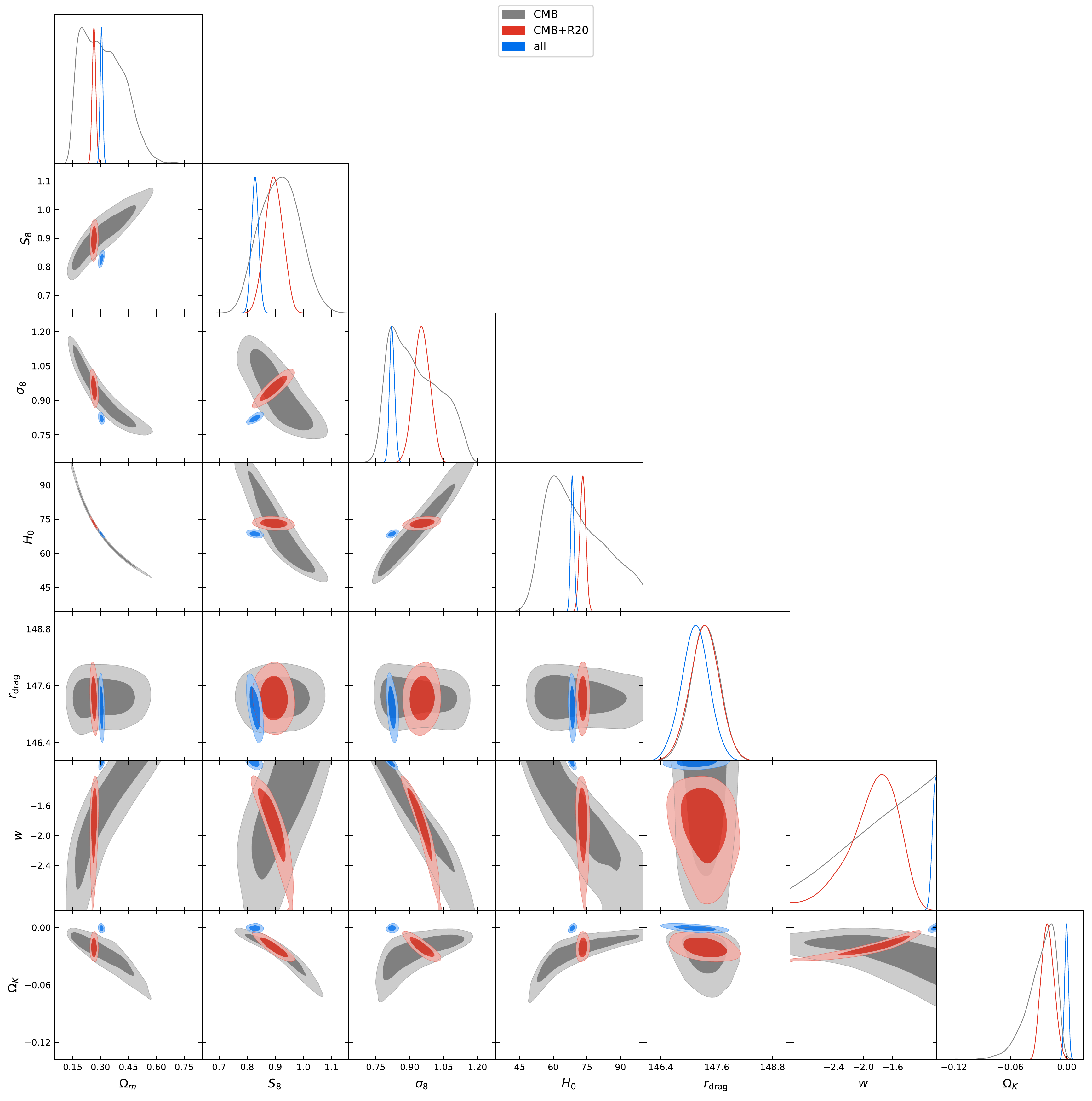}
	\caption{One dimensional posterior distributions and two dimensional joint contours for the parameter space $\mathcal{P}_{13} \equiv\Bigl\{\Omega_{b}h^2, \Omega_{c}h^2, 100\theta_{\rm MC}, \tau, n_{s}, \ln[10^{10}A_{s}],
w_p, \Omega_k \Bigr\}$ for  CMB alone, CMB + R20 and CMB + BAO + Pantheon (referred to as `{\rm all}') dataset. }
	\label{fig:13}
\end{figure*}

In Table~\ref{tabC} we show the cosmological constraints at 68\% CL for the classes of models marked out by k$\Lambda$CDM  ($\mathcal{P}_5$) and k$w$CDM, with the latter divided again into two different regions based on the nature of the DE equation of state: quintessence regime, i.e. $w_q>-1$ ($\mathcal{P}_{12}$) and phantom regime, i.e. $w_p<-1$ ($\mathcal{P}_{13}$). As already shown in the literature, a closed universe is preferred at more than $3\sigma$ by Planck-data alone~\cite{Planck:2018vyg,DiValentino:2019qzk,Handley:2019tkm}, but this trademark increases the tension with $H_0$. The evidence for a closed universe disappears completely by the inclusion of the BAO and Pantheon data samples. The inclusion of the DE equation of state free to vary in the quintessence region, i.e. $w_q>-1$ ($\mathcal{P}_{12}$) yields evidence for a closed universe at more than $3\sigma$ along  with a $w_q>-1$ at more than 68\% CL and an increased Hubble constant discrepancy with respect to the $\Lambda$CDM model, for Planck-data alone. On the contrary, if we consider a phantom DE equation of state (i.e. $w_p<-1$) free to vary in this regime ($\mathcal{P}_{13}$), we obtain $\Omega_k<0$ at more than 99\% CL, $w_p$ is in agreement with a cosmological constant, and the Hubble constant is consistent with SH0ES, for the Planck-data alone. However, for all the cases, all these features disappear for the combined dataset CMB + BAO + Pantheon data, which perfectly restores a flat $\Lambda$CDM model. Similarly, aiming to understand the correlations between the model parameters, we again show the one dimensional posterior distributions and the two dimensional joint contours for these classes of models in Figs.~\ref{fig:5} (for $\mathcal{P}_{5}$), \ref{fig:12} (for $\mathcal{P}_{12}$), and \ref{fig:13} (for $\mathcal{P}_{13}$). 

\begin{figure*}
	\centering
	\includegraphics[width=0.75 \textwidth]{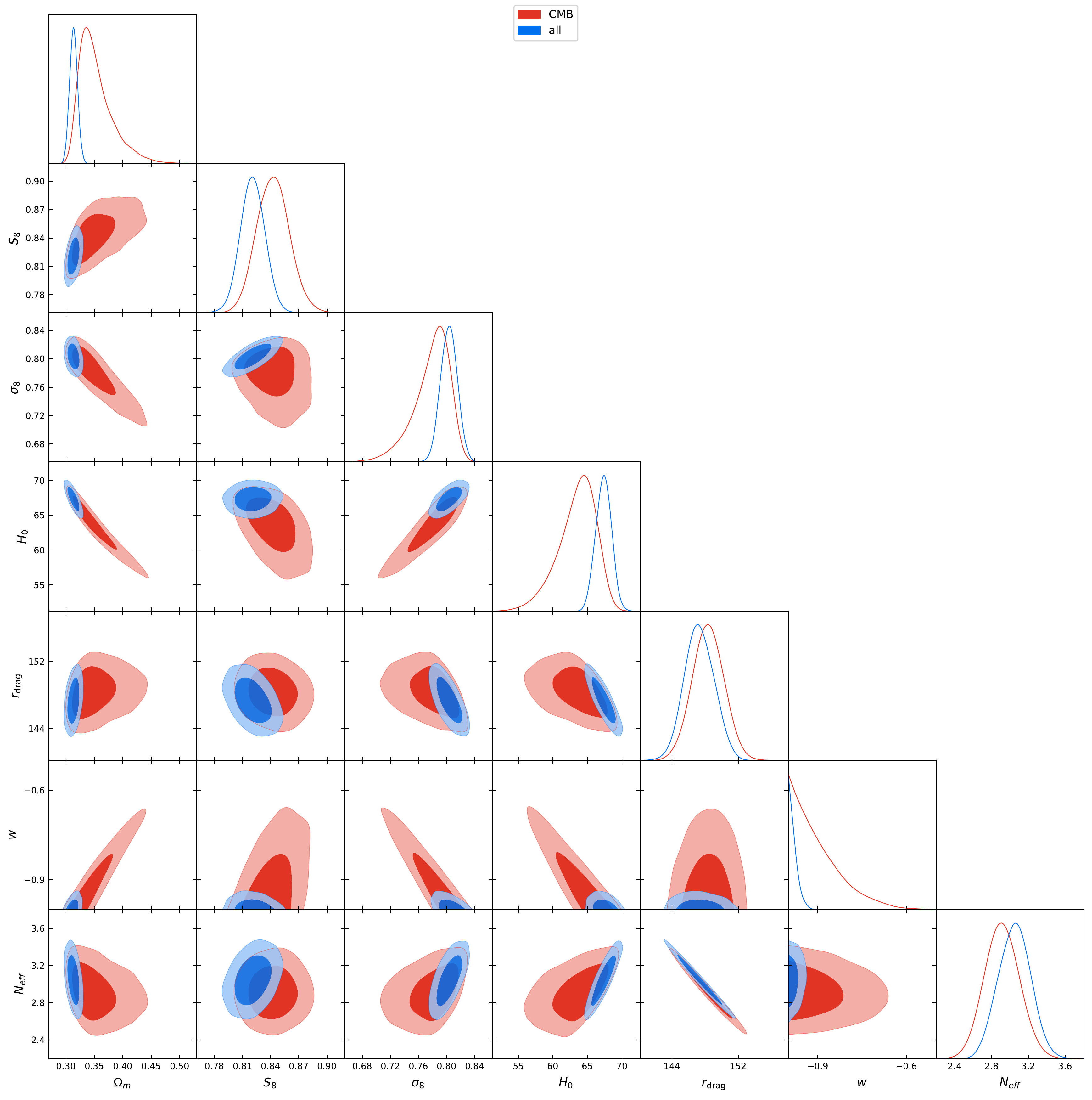}
	\caption{One dimensional posterior distributions and two dimensional joint contours for the parameter space $\mathcal{P}_6 \equiv\Bigl\{\Omega_{b}h^2, \Omega_{c}h^2, 100\theta_{\rm MC}, \tau, n_{s}, \ln[10^{10}A_{s}],
N_{\rm eff}, w_q \Bigr\}$ for  CMB alone and CMB + BAO + Pantheon (referred to as `{\rm all}') dataset. }
	\label{fig:6}
\end{figure*}
\begin{figure*}
	\centering
	\includegraphics[width=0.75 \textwidth]{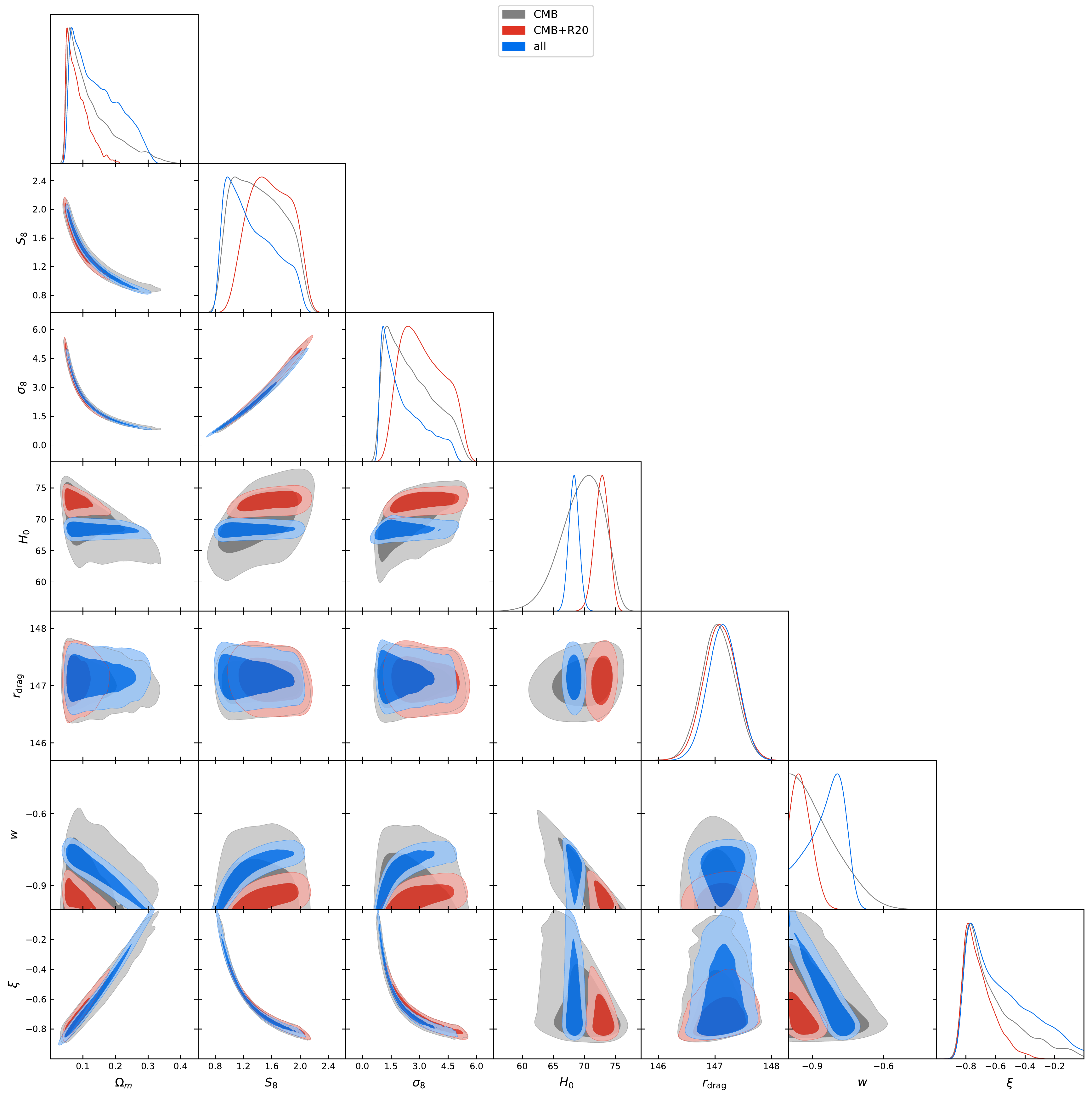}
	\caption{One dimensional posterior distributions and two dimensional joint contours for the parameter space $\mathcal{P}_{10} \equiv\Bigl\{\Omega_{b}h^2, \Omega_{c}h^2, 100\theta_{\rm MC}, \tau, n_{s}, \ln[10^{10}A_{s}],
w_q,  \xi_- \Bigr\}$ for  CMB alone, CMB + R20 and CMB + BAO + Pantheon (referred to as `{\rm all}') dataset. }
	\label{fig:10}
\end{figure*}
\begin{figure*}
	\centering
	\includegraphics[width=0.75 \textwidth]{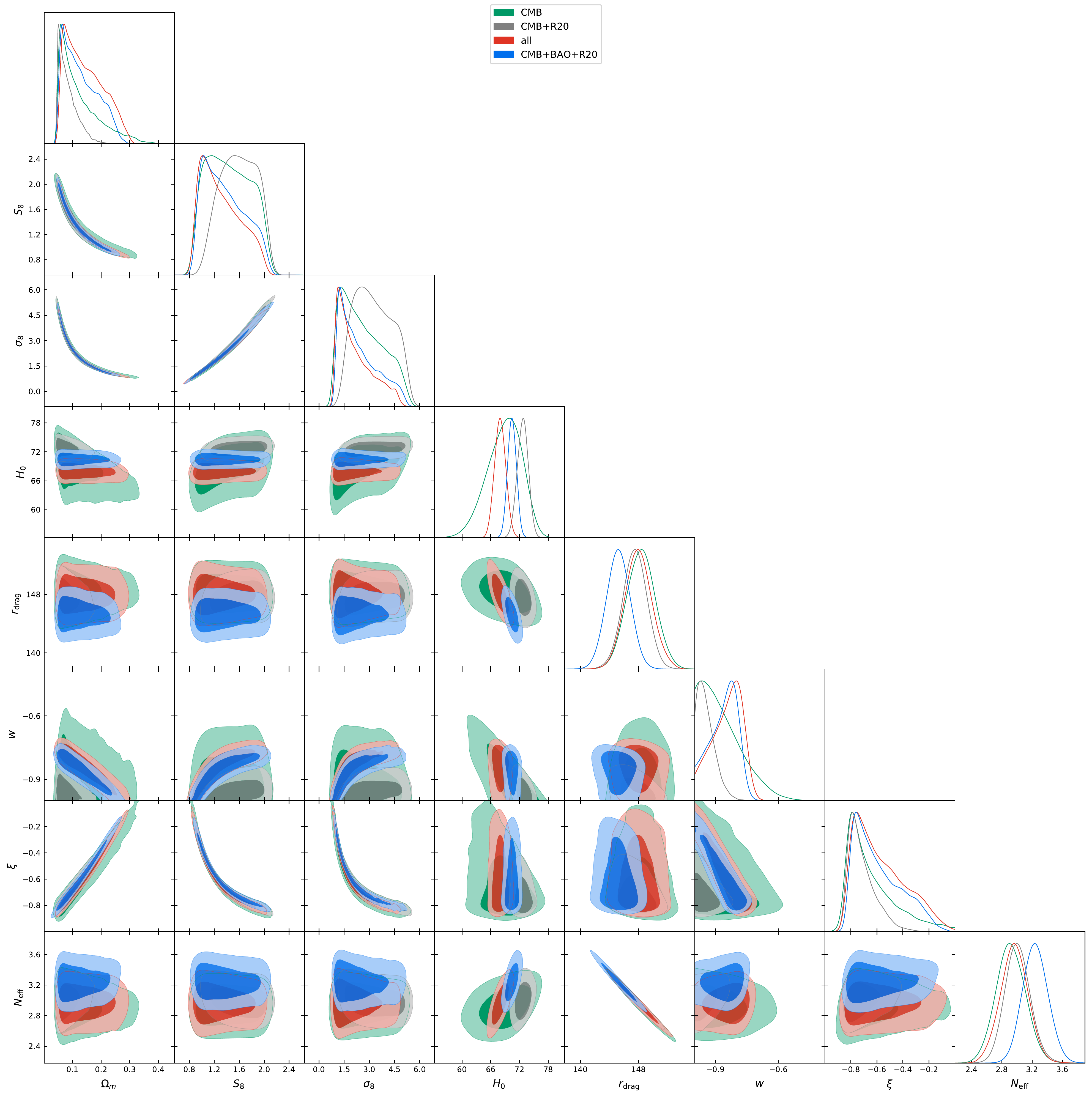}
	\caption{One dimensional posterior distributions and two dimensional joint contours for the parameter space $\mathcal{P}_{15} \equiv\Bigl\{\Omega_{b}h^2, \Omega_{c}h^2, 100\theta_{\rm MC}, \tau, n_{s}, \ln[10^{10}A_{s}],
N_{\rm eff}, w_q, \xi_- \Bigr\}$ for  CMB alone, CMB + R20, CMB + BAO + Pantheon (referred to as `{\rm all}'), and CMB + BAO + R20 datasets.}
	\label{fig:15}
\end{figure*}

In Table~\ref{tabD} we present the constraint on the cosmological parameters for the classes of models characterized by $w_q$CDM + $N_{\rm eff}$ ($\mathcal{P}_6$), $w_q$IDE  ($\mathcal{P}_{10}$), and $w_q$IDE + $N_{\rm eff}$  ($\mathcal{P}_{15}$), where the DE equation of state is always in the quintessence regime, i.e. $w_q>-1$. The $\mathcal{P}_6$ framework does not show any deviation form the standard $\Lambda$CDM scenario. However, in the $w_q$IDE scenario we find some interesting results. First of all, for the Planck-data alone case, the introduction of a $w_q$ free to vary reduces the overlap with the Hubble constant measured by SH0ES, even if this is still in agreement within $1\sigma$. Secondly, even if $w_q$ is consistent with a cosmological constant, the evidence for the coupling between DM and DE increases in significance. Lastly, the inclusion of the BAO and Pantheon data does not restore the usual $\Lambda$CDM model concordance. In fact, for the combined dataset CMB + BAO + Pantheon we have a coupling still at more than 99\% CL, a $w_q \neq -1$ above 95\% CL, but $H_0$ is still in tension with SH0ES at more than 3 standard deviations. Finally, the extended scenario $w_q$IDE + $N_{\rm eff}$ (the last two columns of Table~\ref{tabD}) does not modify the previous findings about the free parameters, because $N_{\rm eff}$ is completely consistent with the standard value of $3.046$. However, we have slightly larger error bars for an increased volume of the parameter space. The one dimensional posterior distributions and the two dimensional joint contours for these classes of models are shown in Figs.~\ref{fig:6} (for $\mathcal{P}_6$), \ref{fig:10} (for $\mathcal{P}_{10}$), and \ref{fig:15} (for $\mathcal{P}_{15}$).

\begin{figure*}
	\centering
	\includegraphics[width=0.75 \textwidth]{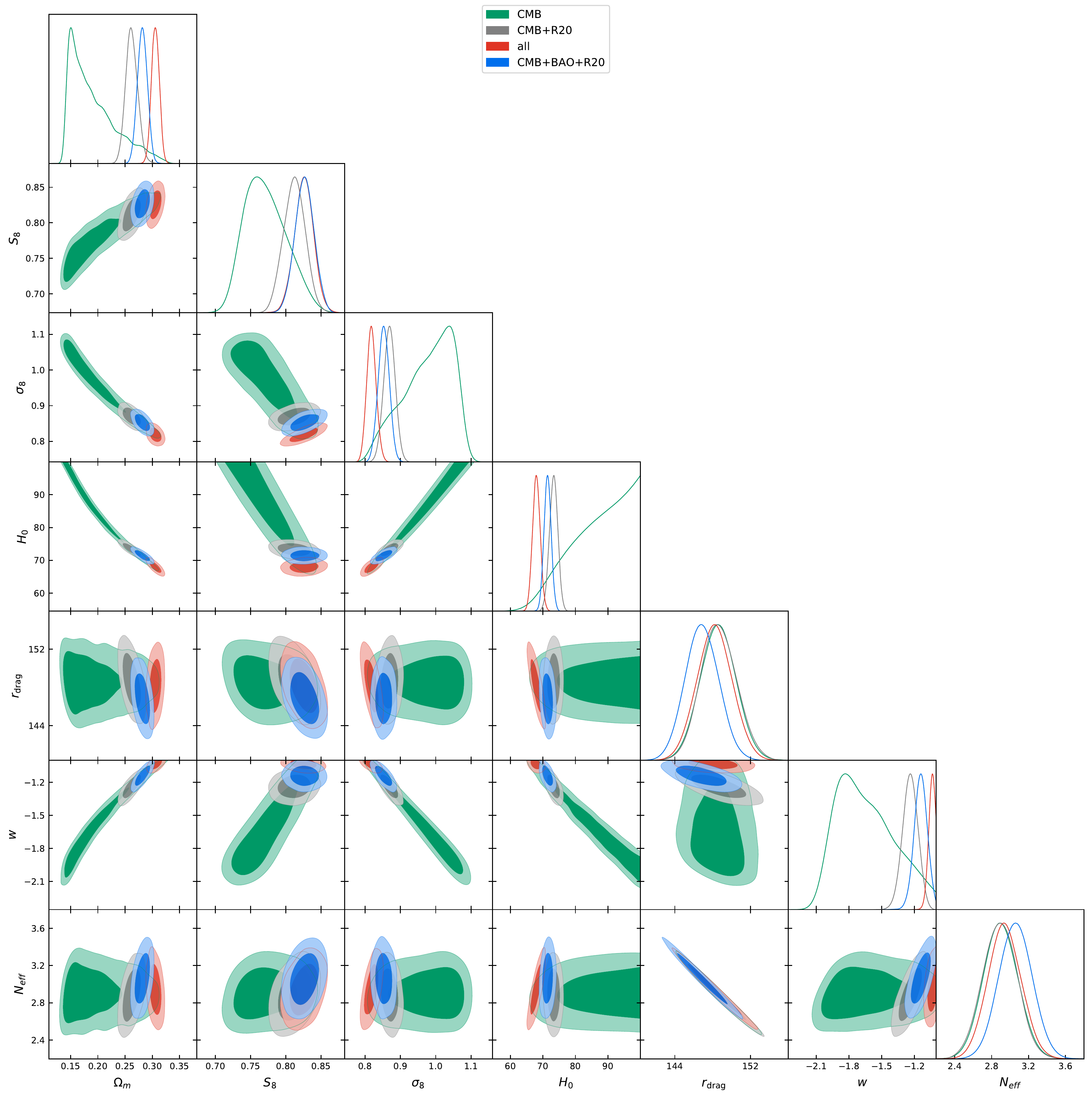}
	\caption{One dimensional posterior distributions and two dimensional joint contours for the parameter space $\mathcal{P}_7 \equiv\Bigl\{\Omega_{b}h^2, \Omega_{c}h^2, 100\theta_{\rm MC}, \tau, n_{s}, \ln[10^{10}A_{s}],
N_{\rm eff}, w_p \Bigr\}$ for CMB alone, CMB + R20, CMB + BAO + Pantheon (referred to as `{\rm all}'), and CMB + BAO + R20 datasets. }
	\label{fig:7}
\end{figure*}

\begin{figure*}
	\centering
	\includegraphics[width=0.75 \textwidth]{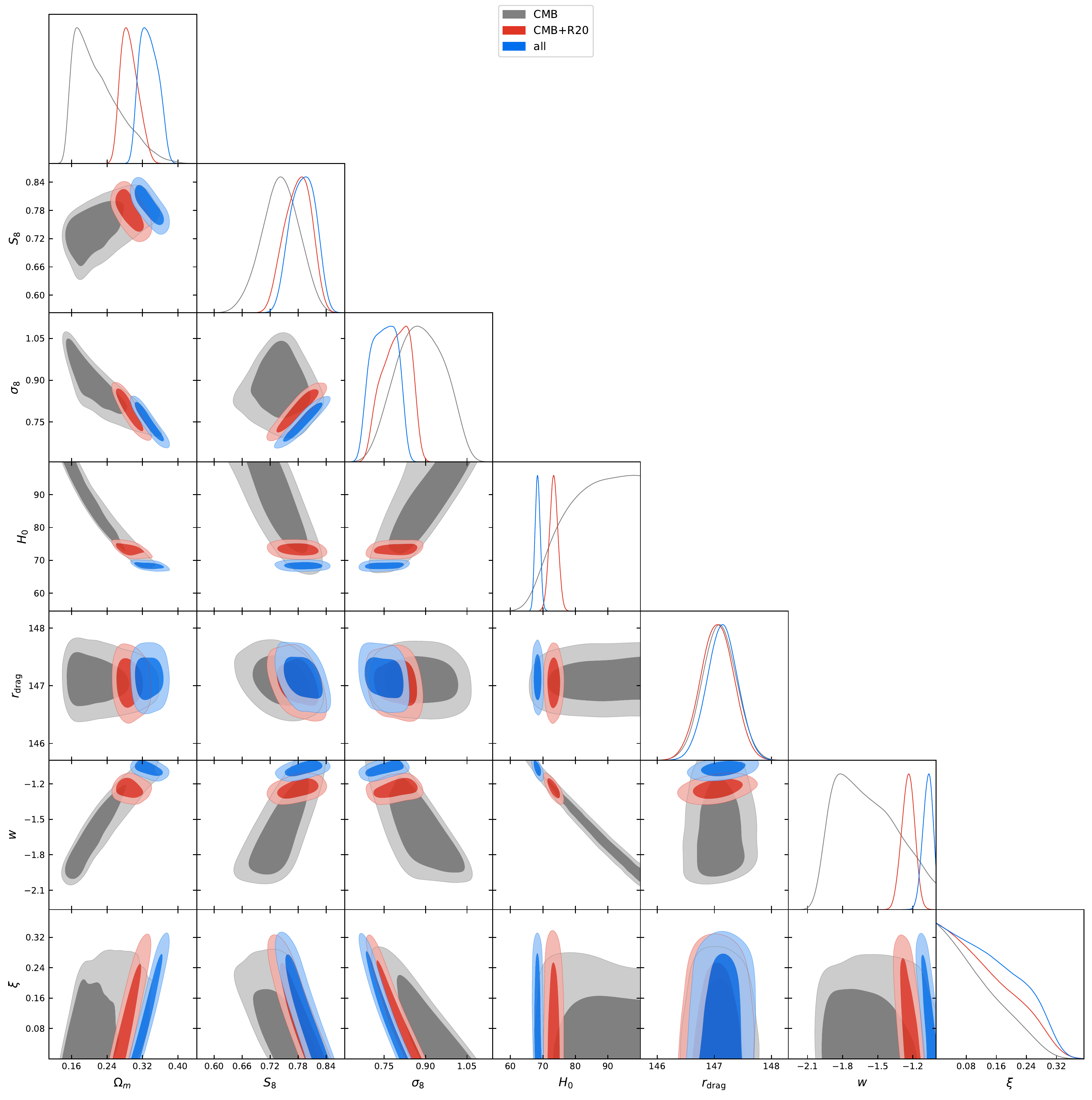}
	\caption{One dimensional posterior distributions and two dimensional joint contours for the parameter space $\mathcal{P}_{11} \equiv\Bigl\{\Omega_{b}h^2, \Omega_{c}h^2, 100\theta_{\rm MC}, \tau, n_{s}, \ln[10^{10}A_{s}],
w_p,  \xi_+ \Bigr\}$ for  CMB alone, CMB + R20 and CMB + BAO + Pantheon (referred to as `{\rm all}') dataset. }
	\label{fig:11}
\end{figure*}

Complementary, in Table~\ref{tabE} we present the constraint on the cosmological parameters for the classes of models specified by $w_p$CDM + $N_{\rm eff}$ ($\mathcal{P}_7$),  $w_p$IDE  ($\mathcal{P}_{11}$) and  $w_p$IDE + $N_{\rm eff}$  ($\mathcal{P}_{16}$), where the DE equation of state is always in the phantom regime, i.e. $w_p<-1$.
In this case, for the scenario $\mathcal{P}_7$ we find   evidence for a phantom $w_p<-1$ at more than 99\% CL when considering Planck-data alone. In addition, we find that the $S_8$ parameter is in agreement with the weak lensing experiments within $1\sigma$, and the $H_0$ tension is certainly alleviated within the 95\% CL. However,  even though the evidence for $w_p<-1$ at more than $1\sigma$ still persists when the full combination CMB + BAO + Pantheon is considered, the $H_0$ and $S_8$ tensions strike back. For the $\mathcal{P}_{11}$ scenario, we do not find any indication for a coupling $\xi \neq 0$, and our analysis confirms the same salient features present in the $\mathcal{P}_7$ classes of models for the Planck data only. On the other hand, when the combined dataset CMB + BAO + Pantheon is considered, $w_p<-1$ is still preferred at more than $1\sigma$ and the $S_8$ tension is alleviated. However, in this case the $H_0$ tension is only reduced down to $3.2\sigma$. The inclusion of the 
$N_{\rm eff}$ parameter free to vary in the last scenario $\mathcal{P}_{16}$ does not modify the previous findings, because $N_{\rm eff}$ is in complete agreement with the Standard Model value. The one dimensional posterior distributions and the two dimensional joint contours for these classes of models are shown in Figs.~\ref{fig:7} (for $\mathcal{P}_{7}$), \ref{fig:11} (for $\mathcal{P}_{11}$), and \ref{fig:16} (for $\mathcal{P}_{16}$).

\begin{figure*}
	\centering
	\includegraphics[width=0.75 \textwidth]{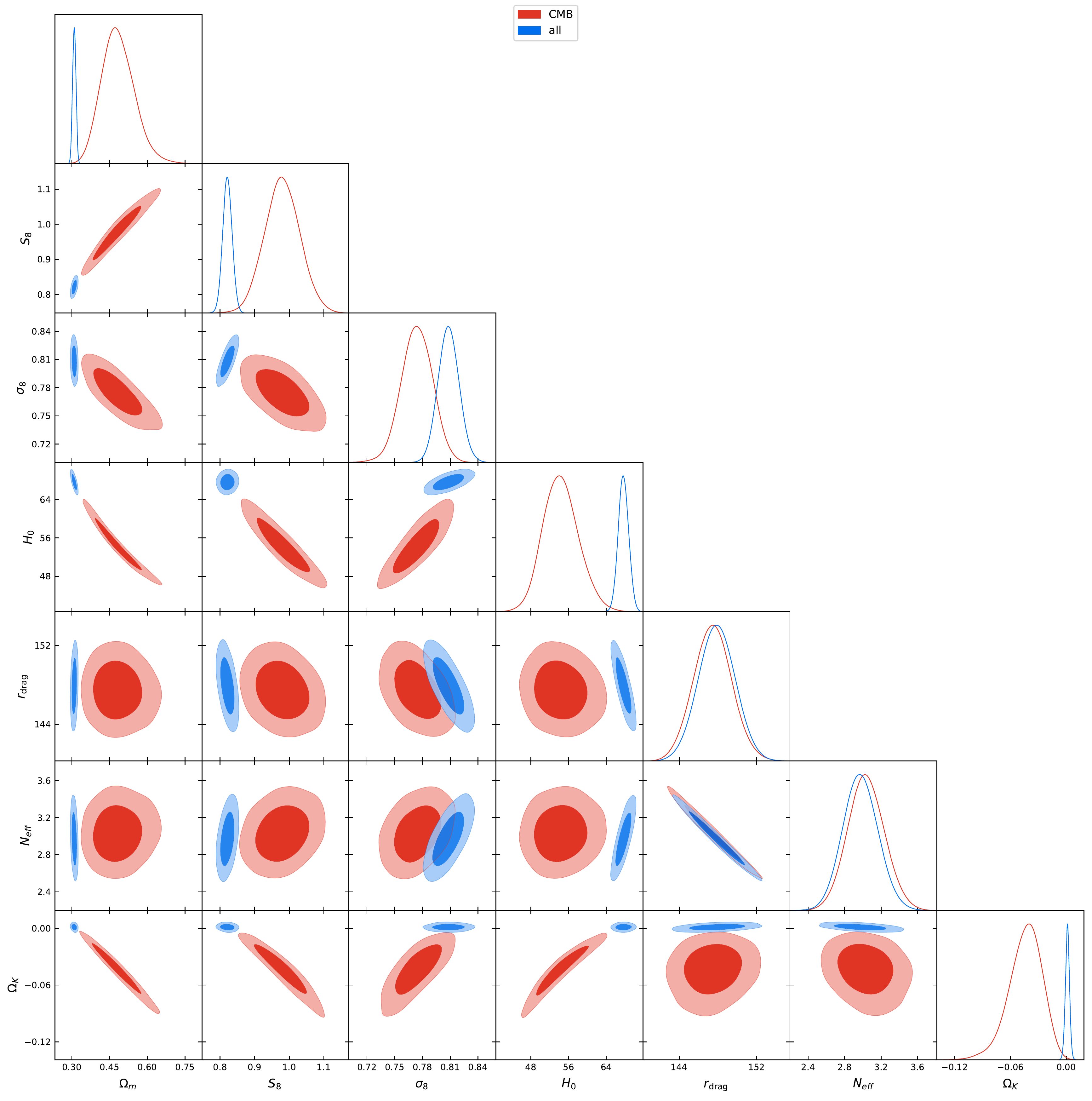}
	\caption{One dimensional posterior distributions and two dimensional joint contours for the parameter space $\mathcal{P}_{9} \equiv\Bigl\{\Omega_{b}h^2, \Omega_{c}h^2, 100\theta_{\rm MC}, \tau, n_{s}, \ln[10^{10}A_{s}], 
N_{\rm eff}, \Omega_k \Bigr\}$ for  CMB alone and CMB + BAO + Pantheon (referred to as `{\rm all}') dataset.}
	\label{fig:9}
\end{figure*}

\begin{figure*}
	\centering
	\includegraphics[width=0.75 \textwidth]{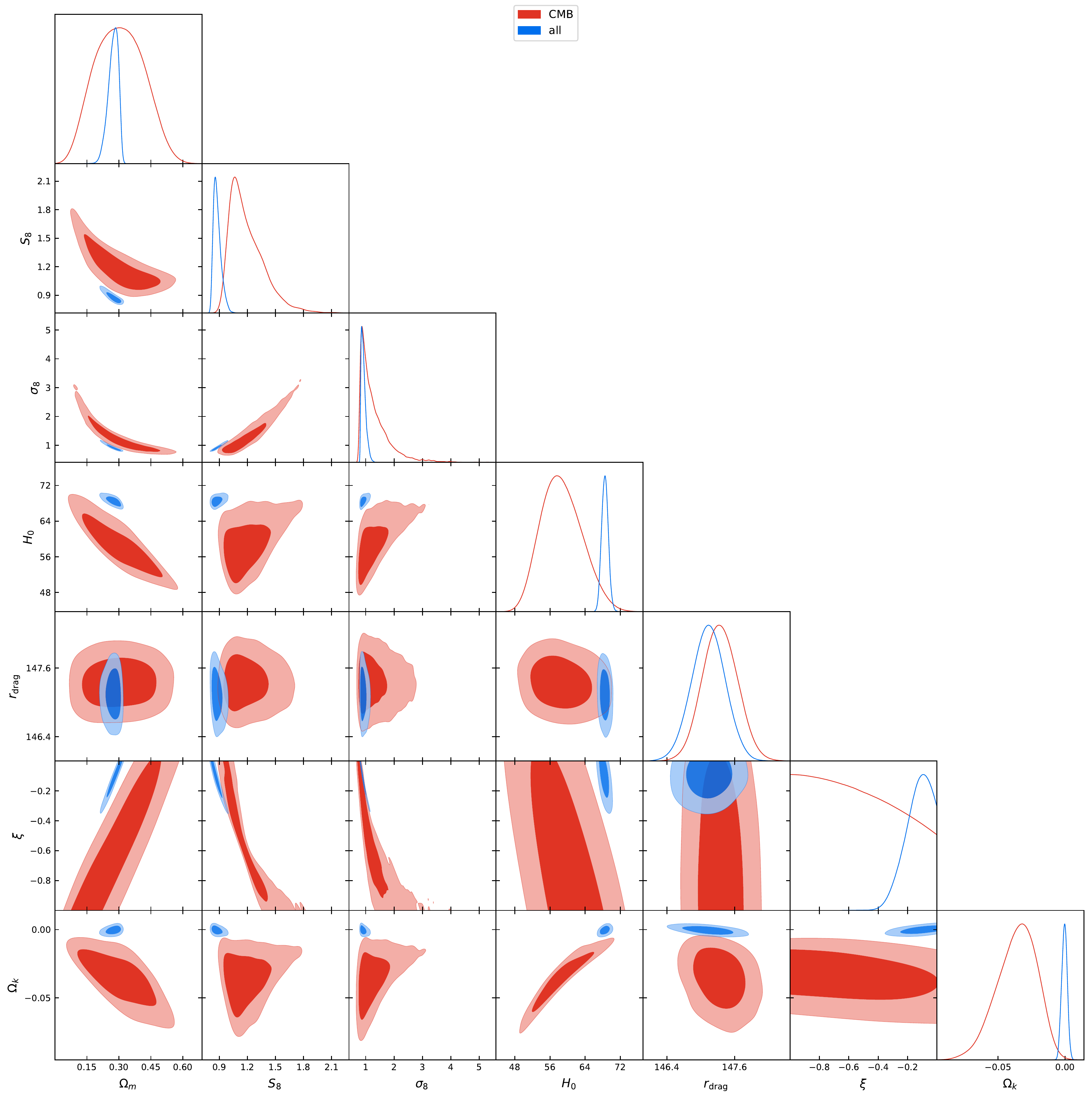}
	\caption{One dimensional posterior distributions and two dimensional joint contours for the parameter space $\mathcal{P}_{14} \equiv\Bigl\{\Omega_{b}h^2, \Omega_{c}h^2, 100\theta_{\rm MC}, \tau, n_{s}, \ln[10^{10}A_{s}],
\xi_-, \Omega_k \Bigr\}$ for  CMB alone, and CMB + BAO + Pantheon (referred to as `{\rm all}') dataset. }
	\label{fig:14}
\end{figure*}

\begin{figure*}
	\centering
	\includegraphics[width=0.75 \textwidth]{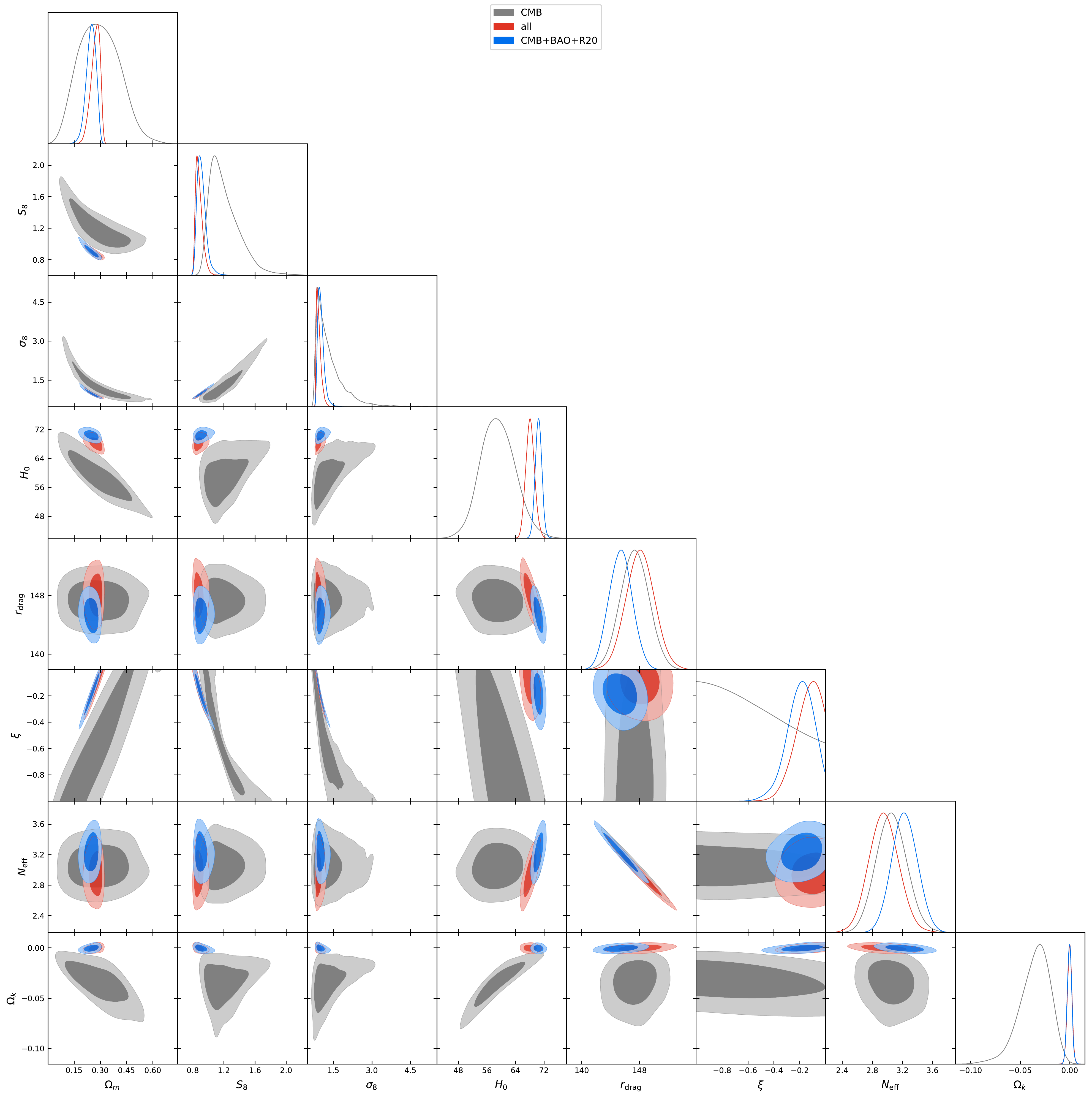}
	\caption{One dimensional posterior distributions and two dimensional joint contours for the parameter space $\mathcal{P}_{19} \equiv\Bigl\{\Omega_{b}h^2, \Omega_{c}h^2, 100\theta_{\rm MC}, \tau, n_{s}, \ln[10^{10}A_{s}],
N_{\rm eff}, \xi_-,  \Omega_k \Bigr\}$ for CMB alone, CMB + BAO + Pantheon (referred to as `{\rm all}'), and CMB + BAO + R20 datasets.}
	\label{fig:19}
\end{figure*}

In Table~\ref{tabF} we show the results for the classes of models headlining k$\Lambda$CDM + $N_{\rm eff}$  ($\mathcal{P}_9$), kIDE  ($\mathcal{P}_{14}$), and  kIDE + $N_{\rm eff}$ ($\mathcal{P}_{19}$). For the Planck-data alone,  evidence for a closed universe at more than 95\% CL is clearly visible for all three scenarios. However, such an evidence disappears when BAO and Pantheon data samples are included. Both the $H_0$ and $S_8$ tensions are exacerbated when considering just the Planck data sample, but $N_{\rm eff}$ is always found to be in agreement with the Standard Model value. In addition, when considering only the CMB-data sample for $\mathcal{P}_{14}$ and $\mathcal{P}_{19}$, we find  evidence for a coupling of DE and DM through an upper limit for $\xi$. However, for the combined dataset CMB + BAO + Pantheon, $\xi \neq 0$ remains true around $1\sigma$ for the $\mathcal{P}_{19}$ (kIDE + $N_{\rm eff}$). Further, we see that for the combined dataset the classes of models typify by $\mathcal{P}_{14}$ and $\mathcal{P}_{19}$ do not offer any alleviation of the $S_8$ tension, rather the tension in $S_8$ increases significantly. These classes of models, however, lead to a mild alleviation of the $H_0$ tension. For the $\mathcal{P}_{14}$ and $\mathcal{P}_{19}$ classes of models, the $H_0$ tension is reduced down to $3.1 \sigma$ and $2.8 \sigma$, for the combined CMB + BAO + Pantheon dataset. 
The one dimensional posterior distributions and the two dimensional joint contours for these classes of models are shown in Figs.~\ref{fig:9} (for $\mathcal{P}_{9}$), \ref{fig:14} (for $\mathcal{P}_{14}$), and \ref{fig:19} (for $\mathcal{P}_{19}$).

\begin{figure*}
	\centering
	\includegraphics[width=0.75 \textwidth]{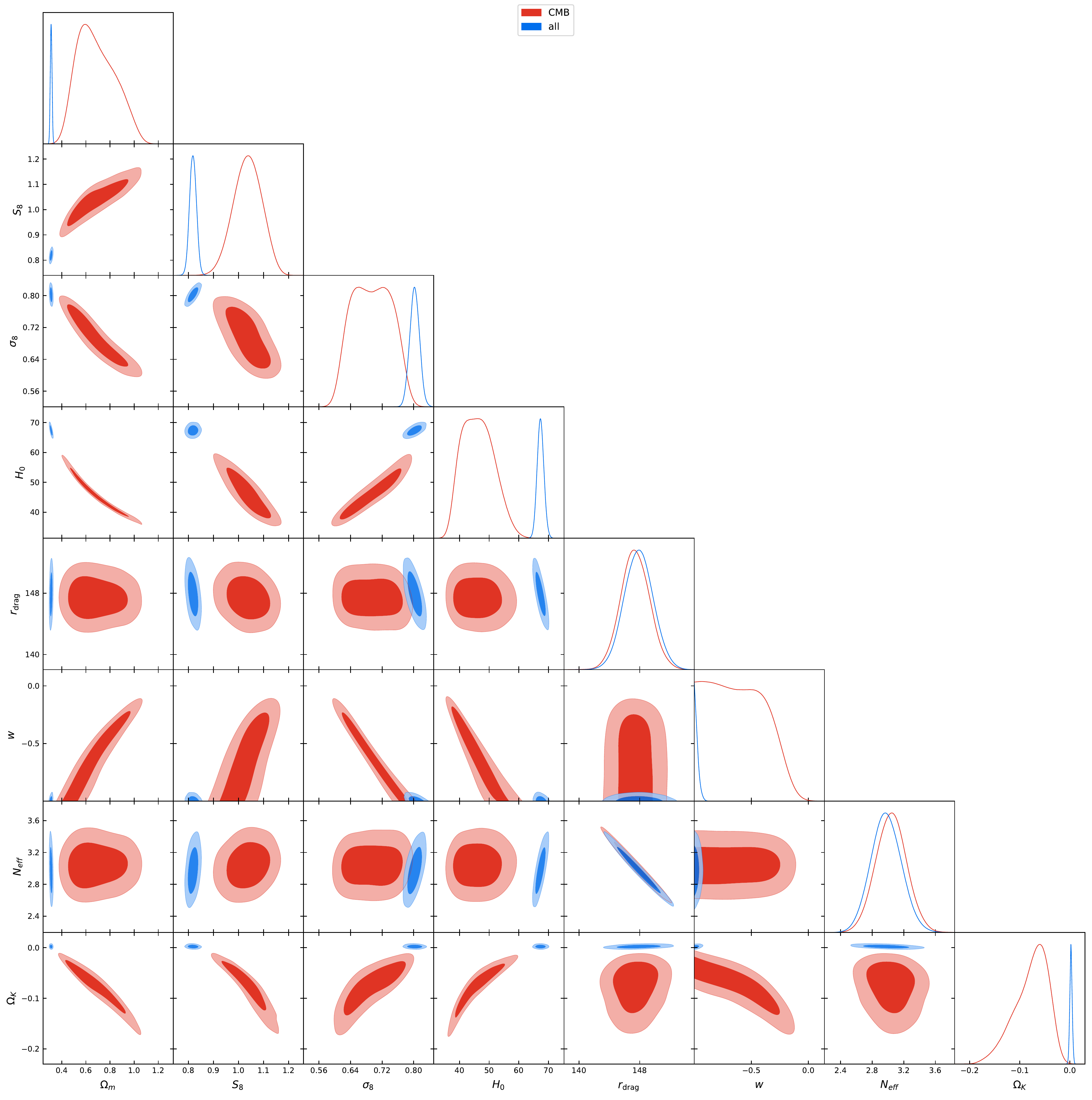}
	\caption{One dimensional posterior distributions and two dimensional joint contours for the parameter space $\mathcal{P}_{17} \equiv\Bigl\{\Omega_{b}h^2, \Omega_{c}h^2, 100\theta_{\rm MC}, \tau, n_{s}, \ln[10^{10}A_{s}],
N_{\rm eff}, w_q,  \Omega_k \Bigr\}$ for CMB alone, and CMB + BAO + Pantheon (referred to as `{\rm all}') dataset.}
	\label{fig:17}
\end{figure*}

\begin{figure*}
	\centering
	\includegraphics[width=0.75 \textwidth]{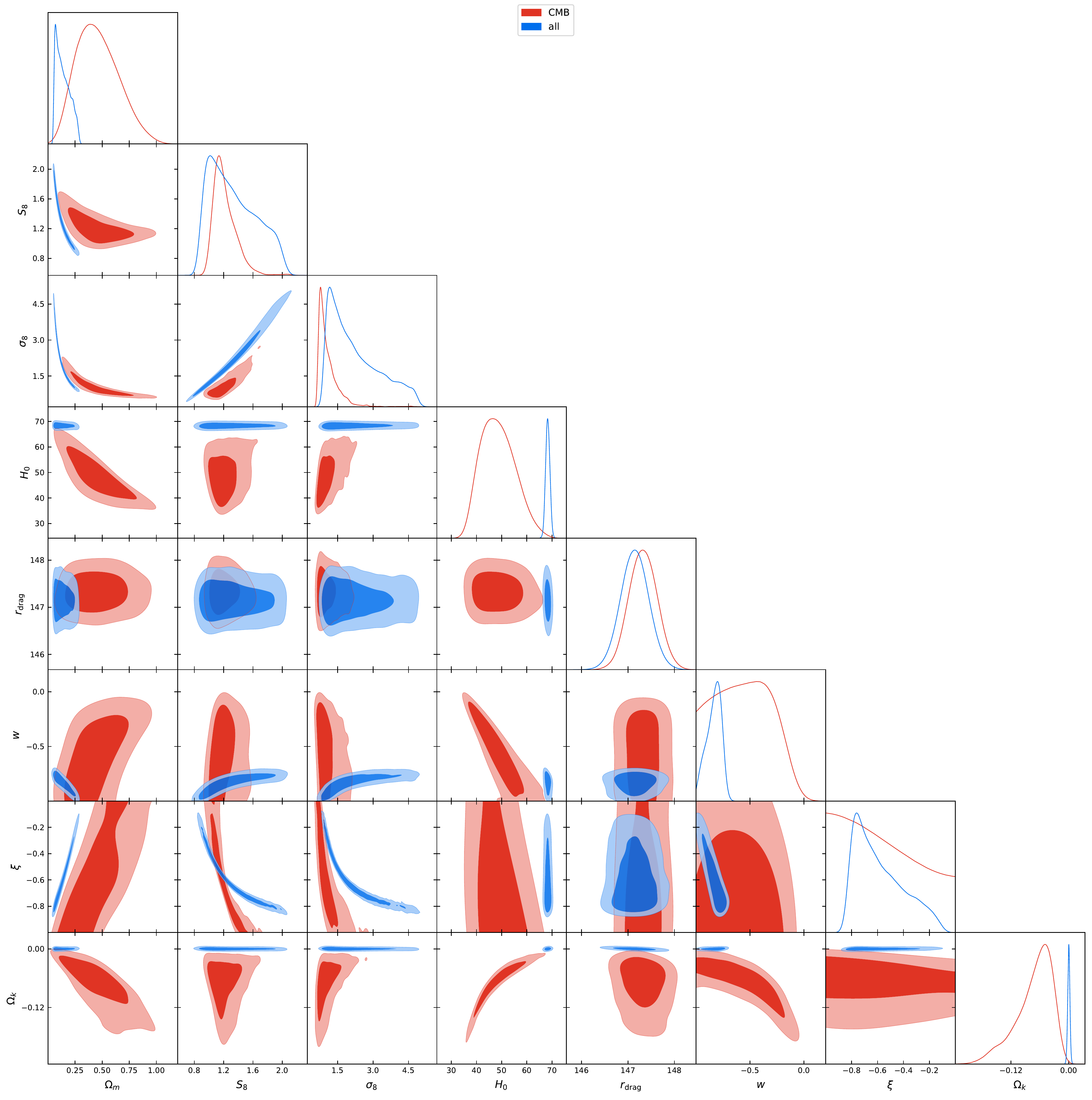}
	\caption{One dimensional posterior distributions and two dimensional joint contours for the parameter space $\mathcal{P}_{20} \equiv\Bigl\{\Omega_{b}h^2, \Omega_{c}h^2, 100\theta_{\rm MC}, \tau, n_{s}, \ln[10^{10}A_{s}],
w_q, \xi_-, \Omega_k \Bigr\}$ for CMB alone, and CMB + BAO + Pantheon (referred to as `{\rm all}') dataset.}
	\label{fig:20}
\end{figure*}

\begin{figure*}
	\centering
	\includegraphics[width=0.8 \textwidth]{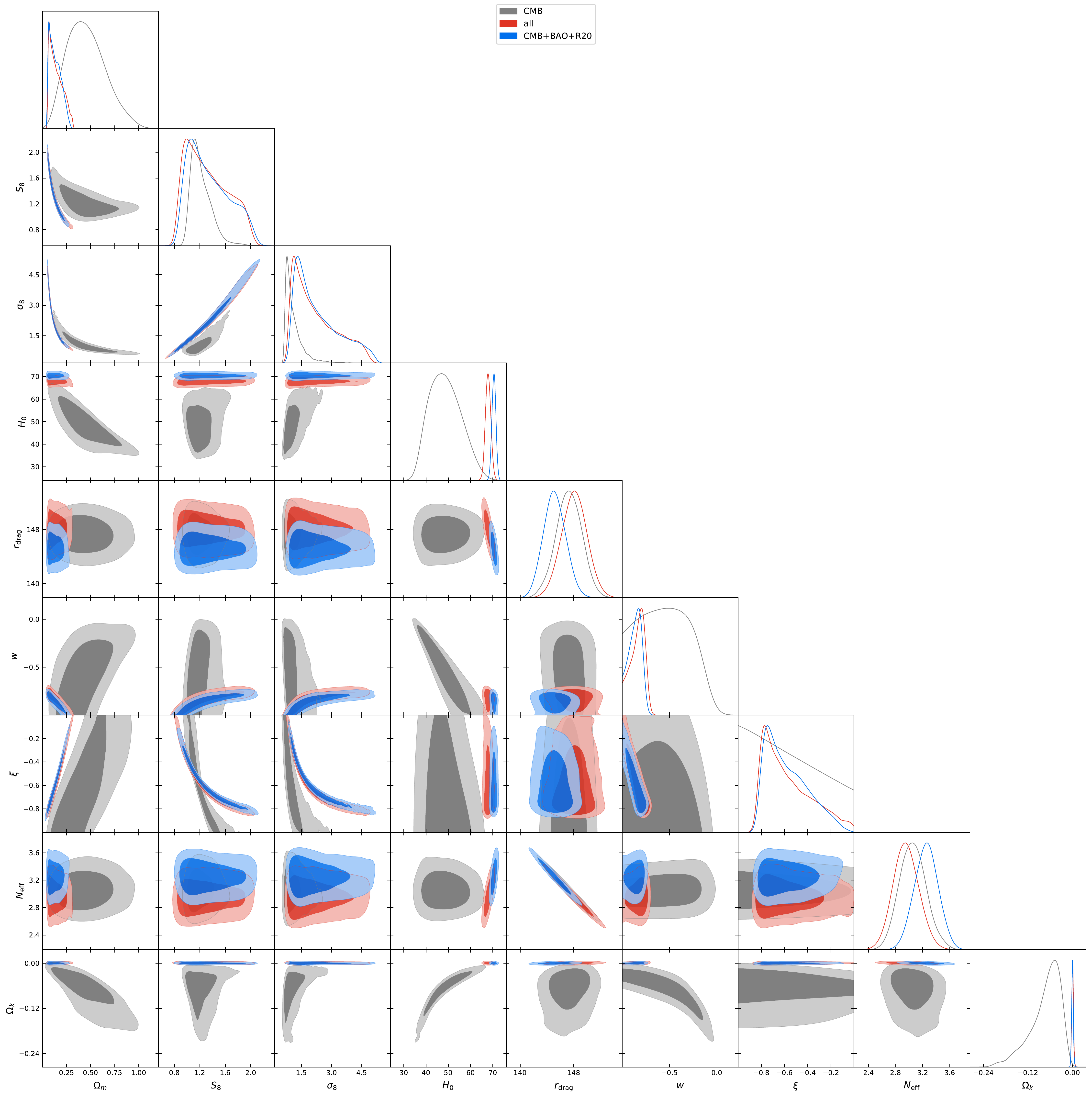}
	\caption{One dimensional posterior distributions and two dimensional joint contours for the parameter space $\mathcal{P}_{22} \equiv\Bigl\{\Omega_{b}h^2, \Omega_{c}h^2, 100\theta_{\rm MC}, \tau, n_{s}, \ln[10^{10}A_{s}],
N_{\rm eff}, w_q, \xi_-, \Omega_k \Bigr\}$ for CMB alone, CMB + BAO + Pantheon (referred to as `{\rm all}'), and CMB + BAO + R20 datasets.}
	\label{fig:22}
\end{figure*}

In Table~\ref{tabG} we show the results of the classes of models featuring k$w_q$CDM + $N_{\rm eff}$  ($\mathcal{P}_{17}$), k$w_q$IDE ($\mathcal{P}_{20}$) and the k$w_q$IDE + $N_{\rm eff}$ ($\mathcal{P}_{22}$), where the DE equation of state has been freely varying in the quintessence regime, i.e. $w_q>-1$.  For the CMB dataset, we find for all the classes of models evidence for a closed universe at more than 95\% CL. However, this evidence disappears when we combine BAO and Pantheon with CMB data. Focusing on the tensions on $H_0$ and $S_8$ parameters, we find that for both CMB and CMB + BAO + Pantheon, the $H_0$ and $S_8$ tensions cannot be resolved. In fact, for CMB alone, both the $H_0$ and $S_8$ tensions significantly increase. However, for CMB + BAO + Pantheon, we see that for $\mathcal{P}_{17}$, $\mathcal{P}_{20}$ and $\mathcal{P}_{22}$ classes of models, the tension on $H_0$ is reduced down to $3.4\sigma$, $3.1 \sigma$ and slightly below $3\sigma$ respectively. Similar to the other cases, the graphical extraction for these scenarios are shown in Figs.~\ref{fig:17} (for $\mathcal{P}_{17}$), \ref{fig:20} (for $\mathcal{P}_{20}$), and \ref{fig:22} (for $\mathcal{P}_{22}$). 

\begin{figure*}
	\centering
	\includegraphics[width=0.75 \textwidth]{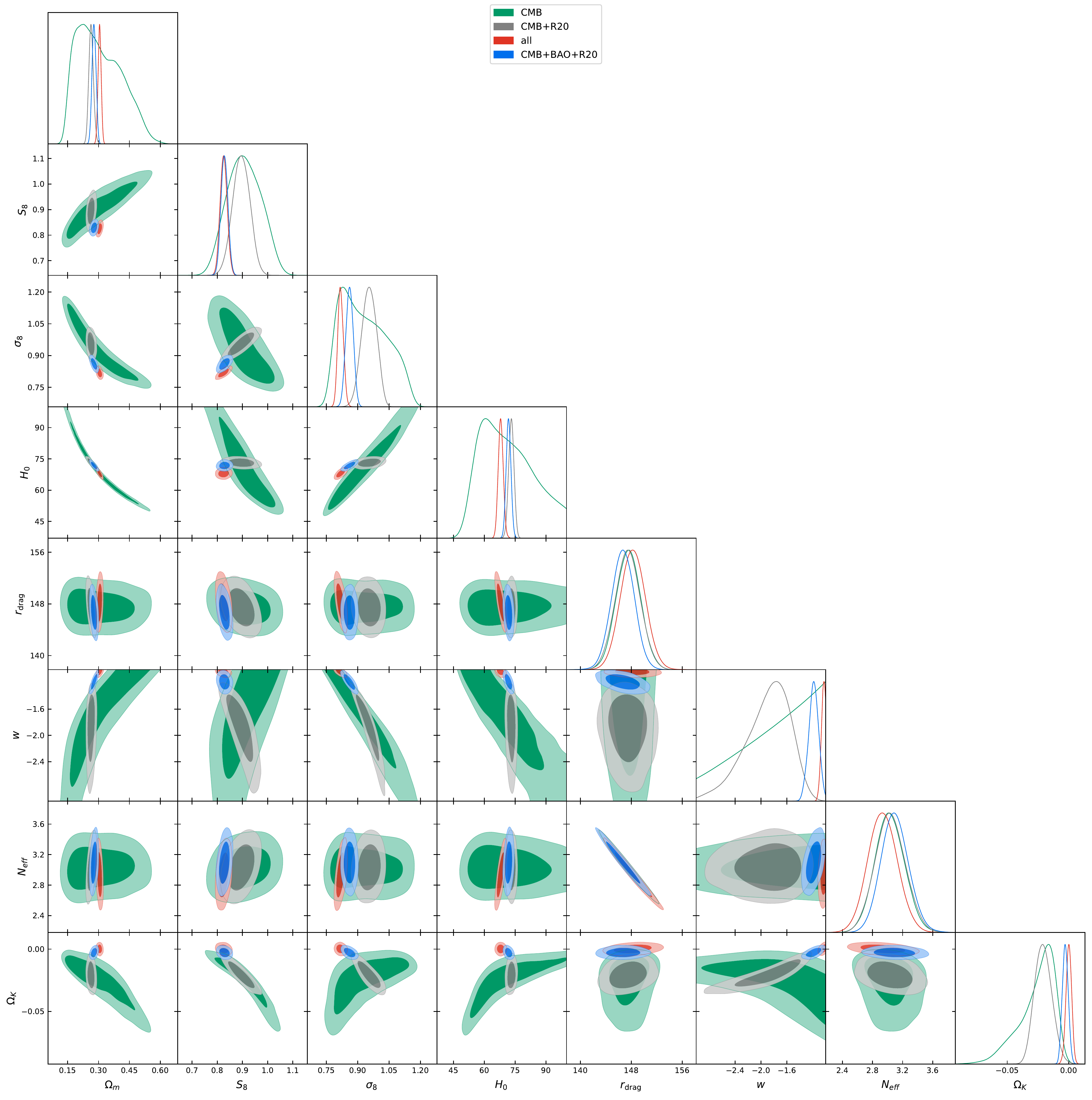}
	\caption{One dimensional posterior distributions and two dimensional joint contours for the parameter space $\mathcal{P}_{18} \equiv\Bigl\{\Omega_{b}h^2, \Omega_{c}h^2, 100\theta_{\rm MC}, \tau, n_{s}, \ln[10^{10}A_{s}],
N_{\rm eff}, w_p,  \Omega_k \Bigr\}$ for CMB alone, CMB + R20, CMB + BAO + Pantheon (referred to as `{\rm all}'), and CMB + BAO + R20 datasets.}
	\label{fig:18}
\end{figure*}

\begin{figure*}
	\centering
	\includegraphics[width=0.75 \textwidth]{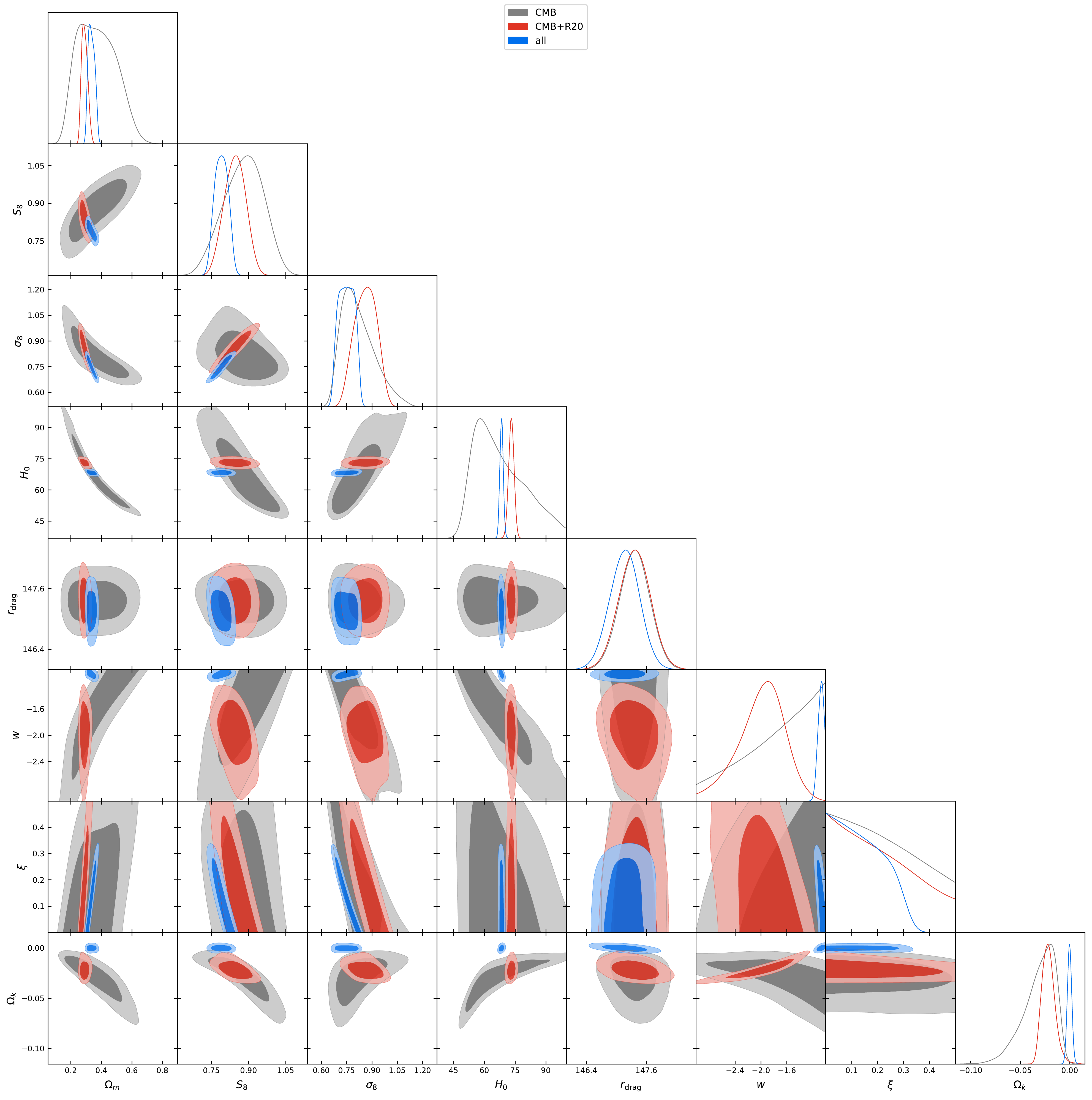}
	\caption{One dimensional posterior distributions and two dimensional joint contours for the parameter space $\mathcal{P}_{21} \equiv\Bigl\{\Omega_{b}h^2, \Omega_{c}h^2, 100\theta_{\rm MC}, \tau, n_{s}, \ln[10^{10}A_{s}],
w_p, \xi_+, \Omega_k \Bigr\}$ for CMB alone, CMB + R20 and CMB + BAO + Pantheon (referred to as `{\rm all}') datasets.}
	\label{fig:21}
\end{figure*}

\begin{figure*}
	\centering
	\includegraphics[width=0.8\textwidth]{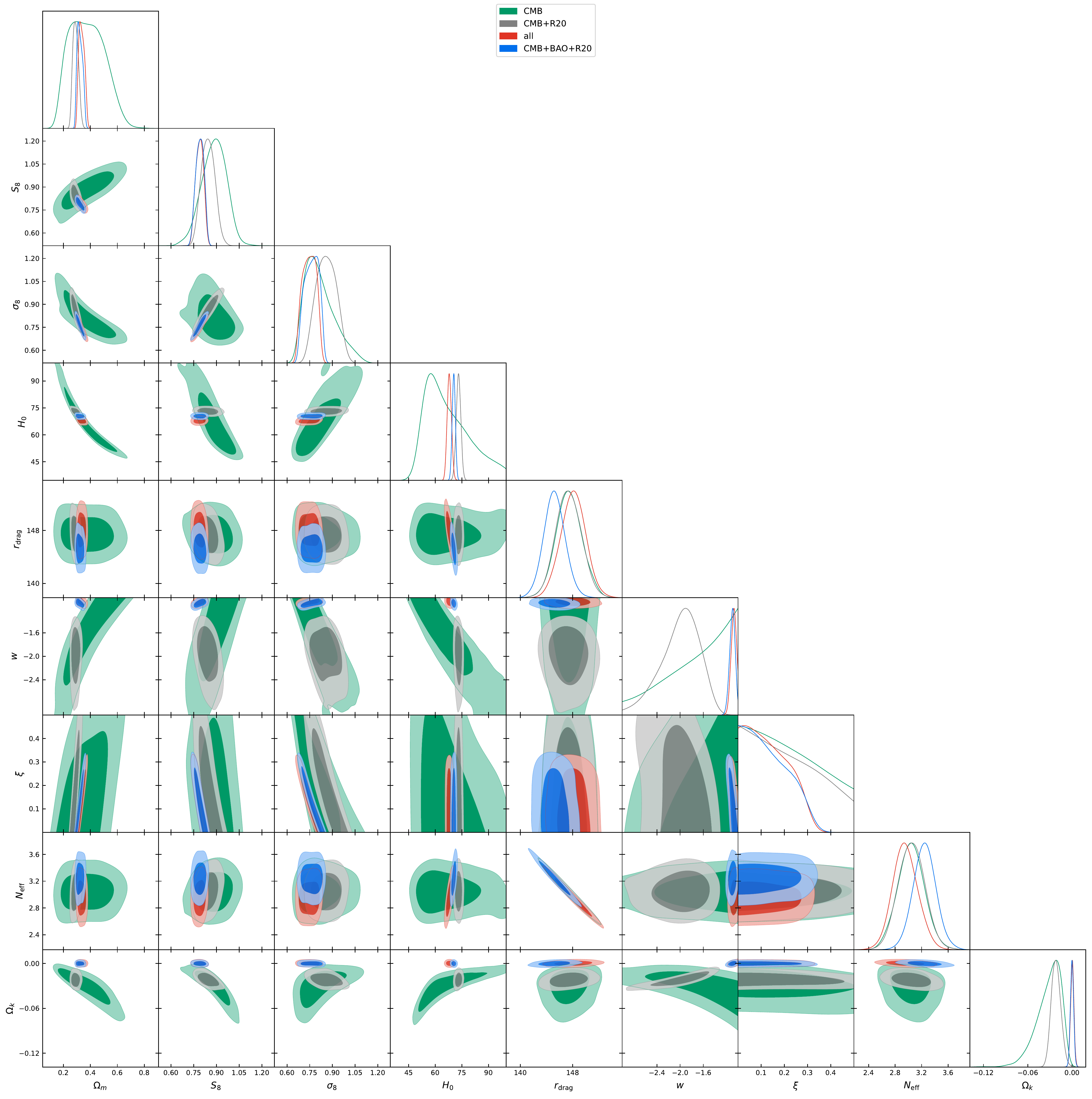}
	\caption{One dimensional posterior distributions and two dimensional joint contours for the parameter space $\mathcal{P}_{23} \equiv\Bigl\{\Omega_{b}h^2, \Omega_{c}h^2, 100\theta_{\rm MC}, \tau, n_{s}, \ln[10^{10}A_{s}],
N_{\rm eff}, w_p, \xi_+, \Omega_k \Bigr\}$ for CMB alone, CMB + R20, CMB + BAO + Pantheon (referred to as `{\rm all}'), and CMB + BAO + R20 datasets. }
	\label{fig:23}
\end{figure*}

Finally, in Table~\ref{tabH} we compare the classes of models characterized by k$w_p$CDM + $N_{\rm eff}$ ($\mathcal{P}_{18}$),  k$w_p$IDE ($\mathcal{P}_{21}$), and  k$w_p$IDE + $N_{\rm eff}$  ($\mathcal{P}_{23}$), where the DE equation of state is always in the phantom regime, i.e. $w_p<-1$. Similar to the classes of models summarized in Table~\ref{tabG}, for all the three extended cases, Planck alone suggests an indication for a closed universe and this evidence goes away when BAO and Pantheon are added to CMB, that means for the combined analysis CMB + BAO + Pantheon. However, due to the phantom behaviour of the DE, the Hubble tension is solved within $1\sigma$, even if $w_p$ is consistent with a cosmological constant. Interestingly, for the CMB + BAO + Pantheon combination and the $\mathcal{P}_{21}$ and $\mathcal{P}_{23}$ cosmological models, while the agreement with a flat universe is restored, an indication for a $w_p<-1$ appears at more than 68\% CL, and at the same time both the $S_8$ and $H_0$ tension are reduced. The one dimensional posterior distributions and the two dimensional joint contours for these models are shown in Figs.~\ref{fig:18} (for $\mathcal{P}_{18}$), \ref{fig:21} (for $\mathcal{P}_{21}$), and \ref{fig:23} (for $\mathcal{P}_{23}$).

\subsection{Including the R20 prior}

In this section we study the effects of adding a gaussian prior R20~\cite{Riess:2020fzl} on the classes of models (described in Table~\ref{tabA} $-$ Table~\ref{tabG}) which alleviate the Hubble tension below $3\sigma$ for Planck or CMB + BAO + Pantheon datasets. To make this selection we adopt a back-of-the-envelope definition of the $H_0$ tension,
\begin{equation}
T = (x_1-x_2)/\sqrt{({\rm err}\, x_1)^2 + ({\rm err} \,  x_2)^2} \,,
\end{equation}
where $x_1 \pm {\rm err} \, x_1$ is the R20 $H_0$ measurement and $x_2
\pm {\rm err} \, x_2$ is the corresponding $H_0$ prediction from model
${\cal P}_i$, with $i = 0, \cdots, 23$. To avoid double counting, when using the R20 prior we do not consider the Pantheon catalog.

For the case of study, the data samples are divided into two subgroups: CMB + R20 and CMB + BAO + R20. The results are shown in Tables~\ref{tabI} to \ref{tabL}. We can see that in all the selected cases the $H_0$ tension is reduced down to $1\sigma$, as expected when a gaussian prior is included and datasets are not in strong tension. In particular, the agreement in the CMB + R20 analyses takes place at the price of a phantom DE ($\mathcal{P}_{3}$ and $\mathcal{P}_{7,11,16,18}$), a coupling for IDE ($\mathcal{P}_{4}$ and $\mathcal{P}_{8,15}$), a quintessence DE interacting with DM ($\mathcal{P}_{10}$), and a phantom closed scenario ($\mathcal{P}_{13}$ and the extended $\mathcal{P}_{21,23}$). Alternatively, the agreement in the CMB + BAO + R20 analyses takes place at the price of a phantom DE ($\mathcal{P}_{7}$), an IDE framework with additional dark radiation ($\mathcal{P}_{8}$ and $\mathcal{P}_{19}$), an interacting quintessence DE with DM in presence of additional dark radiation ($\mathcal{P}_{15}$ and $\mathcal{P}_{22}$), a phantom DE with additional dark radiation ($\mathcal{P}_{16}$ and $\mathcal{P}_{23}$), and a phantom closed universe ($\mathcal{P}_{18}$). However, only the class of models characterized by $\mathcal{P}_{16}$ and $\mathcal{P}_{23}$ can resolve the $H_0$ tension at  $1\sigma$ level, while relaxing also the $S_8$ and the $r_{drag}$ tensions for the CMB + BAO + R20 combination. 

In closing, we note that the phantom DE can address the $H_0$ problem, but does not solve the $M_B$ tension~\cite{Camarena:2021jlr,Efstathiou:2021ocp}. On the contrary, the IDE classes of models can solve simultaneously the $H_0$ and $M_B$ tensions~\cite{Nunes:2021zzi}, but it is not supported by a full dataset combination.


\section{Conclusions}
\label{sec:5}

\begin{figure*}
	\centering
	\includegraphics[width=0.75 \textwidth]{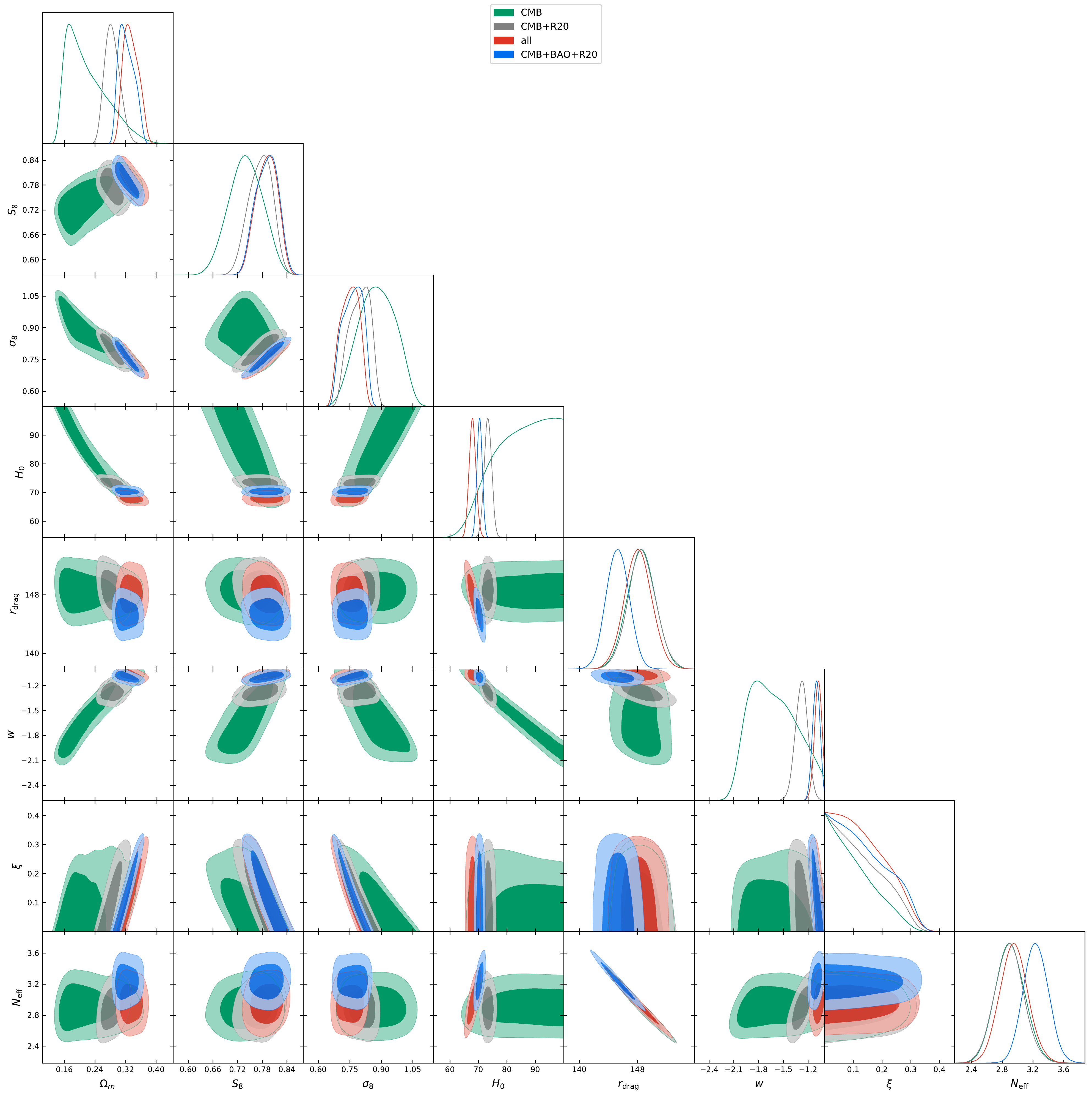}
	\caption{One dimensional posterior distributions and two dimensional joint contours for the parameter space $\mathcal{P}_{16} \equiv\Bigl\{\Omega_{b}h^2, \Omega_{c}h^2, 100\theta_{\rm MC}, \tau, n_{s}, \ln[10^{10}A_{s}],
N_{\rm eff}, w_p, \xi_+ \Bigr\}$ for CMB alone, CMB + R20, CMB + BAO + Pantheon (referred to as `{\rm all}'), and CMB + BAO + R20 datasets.}
	\label{fig:16}
\end{figure*}

We have investigated the possible interconnection among the free parameters in several classes of cosmological models that typify the main theoretical frameworks tackling the tensions on the universe expansion rate and the clustering of matter. This 
meta-analysis features interesting results on the global cosmological picture which can be summarized as follows:
\begin{description}
\item[$\bullet$] The estimate of $r_{\rm drag}$ from low-redshift probes~\cite{Knox:2019rjx,Arendse:2019hev} poses a challenge for beyond $\Lambda$CDM models trying to address the Hubble tension.
Since the baryonic-drag epoch takes place at a somewhat lower redshift than recombination, by comparing the $\Lambda$CDM value of $r_{\rm drag}$ in Table~\ref{tabA} with the estimate of Ref.~\cite{Arendse:2019hev} we can conclude that in order to accommodate $r_{\rm drag} = (137 \pm 3^{\rm stat} \pm 2^{\rm syst})~{\rm Mpc}$ we need a modification of the sound horizon at recombination. Actually, for $\Lambda$CDM
the two standard rulers are related according to $r_{\rm drag} \sim 1.0184 \,  r_*$,  and the proportionality factor is essentially the same in all analyzed classes of models introducing modifications in the expansion rate pre-recombination. In our study we have considered the latest Planck 2018 data sample  which  contains  both  temperature  and  polarization measurements, together with a new optical depth estimate which strongly correlates with $N_{\rm eff}$ keeping its value near $3.046$, and thereby the sound horizon at the epoch of baryon decoupling near the $\Lambda$CDM prediction, $r_{\rm drag} = 147.24 \pm 0.24$~{\rm Mpc}.  Altogether our conclusion points towards the need of new-physics at $z_{\rm drag} \alt z \alt z_*$, or else some unaccounted systematic effects are at play.  Note that our conclusion is complementary to the statements presented in~\cite{Jedamzik:2020zmd,Bernal:2021yli} because our study includes combination of classes of models modifying both the early and late-time expansion rate of the universe.  It should be noticed here that models involving only modifications pre-recombination of early universe physics alone are disfavored~\cite{Lin:2021sfs}.
\item[$\bullet$] For considerations of CMB + BAO + Pantheon data, string-inspired models with coupling between the DE and DM sectors characterized by ${\cal P}_{10}$ tend to fall short of fully resolving the $H_0$ tension~\cite{Agrawal:2019dlm}. The addition of extra-relativistic degrees of freedom (e.g., by considering 3 right-handed Dirac neutrinos ubiquitous in D-brane constructs~\cite{Anchordoqui:2019amx,Anchordoqui:2020znj},  or thermal axions~\cite{Giusarma:2014zza,DiValentino:2015zta,DiValentino:2015wba,Baumann:2016wac,Poulin:2018dzj,DEramo:2018vss,Giare:2020vzo}, or sterile neutrinos~\cite{Anchordoqui:2011nh,Anchordoqui:2012qu,Jacques:2013xr,DiValentino:2015sam}, or any other light species) tend to exacerbate the tension; see ${\cal P}_{15}$ in Table~\ref{tabD}.  However, when considering Planck data and the R20 prior both, the scenarios ${\cal P}_{10}$ and ${\cal P}_{15}$ can resolve the $H_0$ tension at the $1\sigma$ level; see Tables~\ref{tabI} and \ref{tabL}. One would expect that the transfer from DM to DE (a.k.a. fading DM) could ameliorate the $S_8$ tension. However, this  happens because of the larger error bars instead of a stronger overlap. In particular, for the ${\cal P}_{15}$ scenario in Table~\ref{tabL}, $N_{\rm eff}$ saturates the Planck limit. Since $N_{\rm eff}$ is correlated with $\Omega_m$, the effect of a non-negligible dark radiation  shifts the $S_8$ parameter towards higher values, even if with large errors. 
\item[$\bullet$] Frameworks featuring phantom dark energy characterize the classes of models with a potential to accommodate simultaneously the $H_0$ and $S_8$ local measurements (but not the $M_B$ tension~\cite{Camarena:2021jlr}). Classes of models with a transfer of energy from DE to DM (e.g. ${\cal P}_{16}$) keep the value of $N_{\rm eff}$ consistent with the Standard Model expectation of $3.046$, and can resolve the $S_8$ tension while ameliorating the $H_0$ tension, see Table~\ref{tabE} and Fig.~\ref{fig:16}. 
Notably, the IDE model also accommodates the $M_B$ tension~\cite{Nunes:2021zzi}. When including the R20 prior the value of $N_{\rm eff}$ saturates the Planck limit, but the $H_0$ is resolved at the $1\sigma$ level and $S_8$ remains consistent with local measurements; see Table~\ref{tabK}. 
\item[$\bullet$] The latest observations of  the Planck satellite have confirmed the presence of an enhanced lensing amplitude in CMB power spectra compared to that predicted in the standard $\Lambda$CDM model. It was noted in~\cite{DiValentino:2019qzk} that
a closed universe can provide a physical explanation for this effect, with the 2018 Planck CMB spectra preferring a positive curvature at more than 99\% CL~\cite{Planck:2018vyg,DiValentino:2019qzk,Handley:2019tkm,DiValentino:2020srs}. Altogether this motivated our consideration of $\Omega_k$ as a free parameter in the likelihood analysis. Scenarios favoring a closed universe also favor a smaller value of the expansion rate than the $H_0$ measurement by SH0ES, see e.g. Tables~\ref{tabF} and \ref{tabG}. An exception is the class of models featuring extra-relativistic degrees of freedom in the early universe  and a phantom DE, which is described by ${\cal P}_{23}$, and can simultaneously ameliorate the $H_0$ and $S_8$ tensions (i.e. a phantom closed model~\cite{DiValentino:2020hov,Shirokov:2020dwl}). 
\end{description}

 In summary, the $H_0$ and $S_8$ tensions present a daunting
challenge. In this paper we have collected some of the best insights
to extend the standard $\Lambda$CDM model and studied the interconnections among free
parameters of these classes of models. So far, all these insights have
drawbacks and herein we have shown that the extended multi-parameter
cosmologies could only help to narrow down (though not fully eliminate) the tensions. It
is crystal-clear that to unlock Pandora's box  a coordinated effort
involving theory, interpretation, and data analysis would be needed to exploit the large data sets to be collected by the next-generation experiments~\cite{DiValentino:2020vhf}.

\acknowledgments{
LAA was supported by the U.S. National Science Foundation (NSF Grant PHY-2112527).
EDV acknowledges the support of the Addison-Wheeler Fellowship awarded by the Institute of Advanced Study at Durham University.
SP acknowledges the financial supports from the Science and Engineering Research Board, Govt. of India under Mathematical Research Impact-Centric Support Scheme (File No. MTR/2018/000940) and the Department of Science and Technology (DST), Govt. of India under the Scheme  
``Fund for Improvement of S\&T Infrastructure (FIST)'' [File No. SR/FST/MS-I/2019/41].  
WY was supported by the National Natural Science Foundation of China under Grants No. 11705079 and No. 11647153.}

\clearpage

\end{document}